\documentclass[reprint,aps,prb,nofootinbib,groupedaddress,floatfix,superscriptaddress,pdflatex]{revtex4-2}

\usepackage{graphics}
\usepackage{graphicx}
\usepackage{amssymb}
\usepackage{amsmath}
\usepackage{amsfonts}
\usepackage{color}
\usepackage{setspace}
\usepackage{multirow}
\usepackage{array}
\usepackage[subrefformat=parens]{subcaption}

\def\PATHFIG{./figure}

\begin{document}

\title{Electronic band structure change with structural transition of buckled Au$_2$X monolayers induced by strain}

\author{Masahiro Fukuda}
\affiliation{Institute for Solid State Physics, The University of Tokyo,
5-1-5 Kashiwanoha, Kashiwa, Chiba 277-8581, Japan}
\author{Taisuke Ozaki}
\affiliation{Institute for Solid State Physics, The University of Tokyo,
5-1-5 Kashiwanoha, Kashiwa, Chiba 277-8581, Japan}
\email{masahiro.fukuda@issp.u-tokyo.ac.jp}

\date{\today}

\begin{abstract}
This study investigates the strain-induced structural transitions of $\eta \leftrightarrow \theta$ 
and the changes in electronic band structures of Au$_2$X (X=S, Se, Te, Si, Ge) 
and Au$_4$SSe.
We focus on Au$_2$S monolayers,
which can form multiple meta-stable monolayers theoretically, 
including $\eta$-Au$_2$S, a buckled penta-monolayer composed of a square Au lattice and S adatoms.
The $\theta$-Au$_2$S is regarded as a distorted structure of $\eta$-Au$_2$S.
Based on density functional theory (DFT) calculations using a generalized gradient approximation, the conduction and the valence bands of $\theta$-Au$_2$S intersect at the $\Gamma$ point, leading to linear dispersion, whereas $\eta$-Au$_2$S has a band gap of 1.02 eV. The conduction band minimum depends on the specific Au-Au bond distance,
while the valence band maximum depends on both Au-S and Au-Au interactions.
The band gap undergoes significant changes during the $\eta \leftrightarrow \theta$ phase transition of Au$_2$S induced by applying tensile or compressive in-plane biaxial strain to the lattice. Moreover, substituting S atoms with other elements alters the electronic band structures, resulting in a variety of physical properties without disrupting the fundamental Au lattice network. Therefore, the family of Au$_2$X monolayers holds potential as materials for atomic scale network devices.

\end{abstract}


\maketitle
\section{Introduction}

In recent years, many new structures of AB$_2$-type 2D materials, such as transition metal dichalcogenides,
have been proposed theoretically by high-throughput density functional theory (DFT) calculations~\cite{PhysRevLett.118.106101,Haastrup_2018},
and experimental material synthesis for database construction has also been actively carried out~\cite{Zhou2018}.
In our previous study, we performed high-throughput first-principles electronic structure calculations for AB$_2$-type 2D materials based on DFT 
to construct a structure map which not only assembles families of 1T and 1H structures,
such as the well-known transition metal dichalcogenides,
but also highlights intriguing structural trends through the periodic table~\cite{D0MA00999G,AB2_structure_map}.

\begin{figure}[htb]
    \centering
    \includegraphics[width=\linewidth]{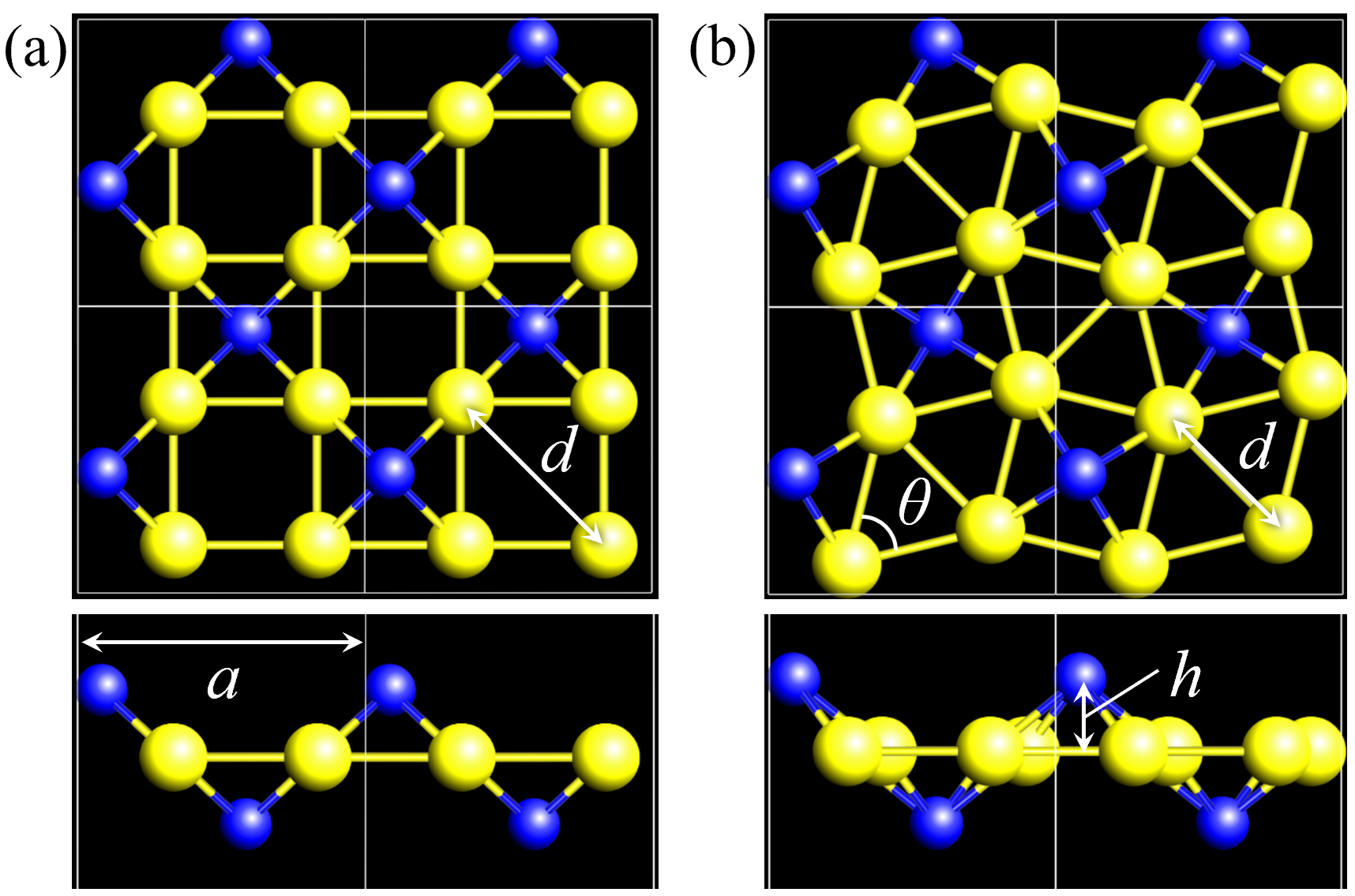}
        \caption{Structures of \textit{$\eta$}-Au$_2$S (a) and \textit{$\theta$}-Au$_2$S (b).}\label{fig:Au2S_structure}
\end{figure}

In that study, we also found some characteristic monolayers
(see Fig.10 in Ref.~\cite{D0MA00999G}),
for example, Au$_2$S [$P42_12$(90)], Au$_2$B [$P4/mbm$(127)], and GeC$_2$[$P42_1m$(113)],
which can be classified into a group of penta-monolayers.
Here, the compound name and space group are described to represent a specific phase of monolayers.
The difference among these three monolayers is how their structures are buckled.
The Au$_2$B [$P4/mbm$(127)] is a planar structure, while S atoms in the Au$_2$S [$P42_12$(90)] are buckled as shown in Fig.~\ref{fig:Au2S_structure} (b).
The GeC$_2$[$P42_1m$(113)] is in the same space group of buckled pentagonal monolayers as PdSe$_2$,
which has already been synthesized experimentally~\cite{doi:10.1021/jacs.7b04865}.
The top view of Au$_2$S [$P42_12$(90)] and Au$_2$B [$P4/mbm$(127)] are similar pentagonal monolayers.
A different perspective reveals that their structures consist of triangular and square Au lattice networks and S or B adatoms.
For Au$_2$S, there were some reports where Au$_2$S has several phases~\cite{D2NA00019A,Gao2021,doi:10.1021/acs.jpclett.9b01312}.
In particular, a structure of Au$_2$S [$P4/nmm$(129)], which consists of the square Au lattice networks and S adatoms as shown in Fig.~\ref{fig:Au2S_structure} (a),
is comprehended as a distorted structure of Au$_2$S [$P42_12$(90)].
The geometrical structure of Au$_2$S [$P42_12$(90)] and Au$_2$S [$P4/nmm$(129)] have already been reported
in the study about Cu$_2$S~\cite{ZHU2019113704, doi:10.1021/acs.jpclett.0c00613}.
However, since Au$_2$S [$P42_12$(90)] and Au$_2$S [$P4/nmm$(129)] are not the most stable structure,
researchers have not given a detailed analysis of the electronic structure of them.

In this work, we discuss how structures (phases) and electronic band structures of Au$_2$S change under strain and adatom substitution,
when Au$_2$S is considered as a monolayer consisting of Au lattice networks with S adatoms.
For simplicity, Au$_2$S [$P42_12$(90)] and Au$_2$S [$P4/nmm$(129)] are denoted as \textit{buckled}-$\eta$-Au$_2$S and \textit{buckled}-$\theta$-Au$_2$S, or more simply $\eta$-Au$_2$S and $\theta$-Au$_2$S, respectively.
This paper is organized as follows.
In Sec.~\ref{sec:Computational_details}, computational details of DFT calculations and phonon calculations are outlined.
In Sec.~\ref{sec:Au2S_structure}, geometrical structures and stability of $\eta$- and $\theta$-Au$_2$S
monolayers are discussed with a focus on $\eta \leftrightarrow \theta$ phase transitions.
In Sec.~\ref{sec:Au2S_eband_structure}, we focus on the electronic band structures of Au$_2$S,
and investigate what the origin of the band gap is and how we can control the band gap.
In Sec.~\ref{sec:Au2X_structure}, geometrical structures and stability of $\eta$- and $\theta$-Au$_2$X (X=Se, Te, Si, Ge) monolayers
are discussed in comparison to those of Au$_2$S.
Moreover, as an example of the partially atom-substituted monolayers,
we discuss $\eta$- and $\theta$-Au$_4$SSe monolayers
from the viewpoint of controlling the stability of the phase.
The electronic band structures of Au$_2$X monolayers are investigated in Sec.~\ref{sec:Au2X_eband_structure}.
The last section is devoted to the conclusions and perspectives.

\section{Computational details}
\label{sec:Computational_details}

The first-principles calculations based on DFT were carried out for Au$_2$X (X = S, Se, Te, Si, Ge) type monolayers of
$\theta$- and $\eta$-phase.
The DFT calculations within a generalized gradient approximation (GGA) \cite{PhysRev.140.A1133,PhysRevLett.77.3865} 
were performed for the geometry relaxations and variable cell optimizations using the
OpenMX code~\cite{OpenMX} based on norm-conserving pseudopotentials generated
with multireference energies \cite{PhysRevB.47.6728} and optimized pseudoatomic
basis functions~\cite{PhysRevB.67.155108}.
The basis sets we used are listed in Table~\ref{tab:basis_set}.
For example, Au7.0-s3p2d2f1 means that three, two, two, and one optimized radial functions
were allocated for the $s$, $p$, $d$, and $f$ orbitals, respectively for Au atoms,
and the cutoff radius of 7 Bohr was chosen.
The qualities of basis functions and fully relativistic pseudopotentials were carefully
benchmarked by the delta gauge method~\cite{Lejaeghere2016-gh}
to ensure accuracy of our calculations.
An electronic temperature of 700 K is used to count the number
of electrons by the Fermi-Dirac function.
The regular mesh of 240 Ry in real space was used for the numerical
integration and for the solution of the Poisson equation~\cite{PhysRevB.72.045121}.
A $5 \times 5 \times 1$ mesh of k points was adopted.
Cell vectors and internal coordinates are simultaneously optimized
without any constraint by using a combination scheme of
the rational function (RF) method~\cite{doi:10.1021/j100247a015} 
and the direct inversion iterative sub-space (DIIS) method~\cite{CSASZAR198431}
with a BFGS update~\cite{10.1093/imamat/6.1.76, 10.1093/comjnl/13.3.317, 10.2307/2004873, 10.2307/2004840} for the approximate Hessian. 
The force on each atom was relaxed to be less than 0.0003 Hartree/Bohr.
We put four Au atoms and two Si atoms in a square unit cell to form \textit{buckled}-$\eta$-phase and \textit{buckled}-$\theta$-phase as initial structures.
Calculations of phonon band structures and the density of states (DOS) were carried out by ALAMODE~\cite{Tadano_2014}.
For the force calculations for estimating force constants,
we used $2 \times 2$ supercell structures with 0.04 \AA~displaced configurations for using a $4 \times 4 \times 1$ mesh of k points and the regular mesh of about 700 Ry to improve accuracy.

\begin{table}[htb]
\centering
  \caption{List of basis sets}
\begin{tabular}{c}
\centering
  \begin{minipage}{0.95\linewidth}
    \begin{tabular}{l@{\hspace{12pt}}l@{\hspace{12pt}}l} 
    \hline
      S7.0-s2p2d1f1 & Se7.0-s3p2d2 & Te7.0-s3p2d2f1 \\
      Si7.0-s2p2d1  & Ge7.0-s3p2d2 & Au7.0-s3p2d2f1 \\ 
    \hline
    \end{tabular}
  \end{minipage}
  \end{tabular}
  \label{tab:basis_set}
\end{table}

\section{A\lowercase{u}$_2$S monolayers} \label{sec:Au2S}

\subsection{Structure and stability} \label{sec:Au2S_structure}

The structures of \textit{$\eta$}- and \textit{$\theta$}-Au$_2$S monolayers obtained by the geometry optimization are shown in Fig.~\ref{fig:Au2S_structure}.
Four Au atoms in \textit{$\eta$}-phase form four square lattice in a unit cell.
Two S atoms located $h$~\AA~away from the center of two of the square, either up or down.
The height $h=1.35$~\AA~ of $\eta$-Au$_2$S is a little bit shorter than $h=1.41$~\AA~ of $\theta$-Au$_2$S.
In the \textit{$\theta$}-Au$_2$S, two squares are distorted,
and the distorted angle $\theta \approx 60^\circ$.
Therefore, each distorted square lattice forms two almost equilateral triangles.
The lattice constant $a=5.613$\AA~of \textit{$\theta$}-Au$_2$S is shorter than $a=5.805$~\AA~of \textit{$\eta$}-Au$_2$S.

\begin{figure*}[htb]
  \begin{minipage}[b]{0.45\linewidth}
    \includegraphics[width=\linewidth]{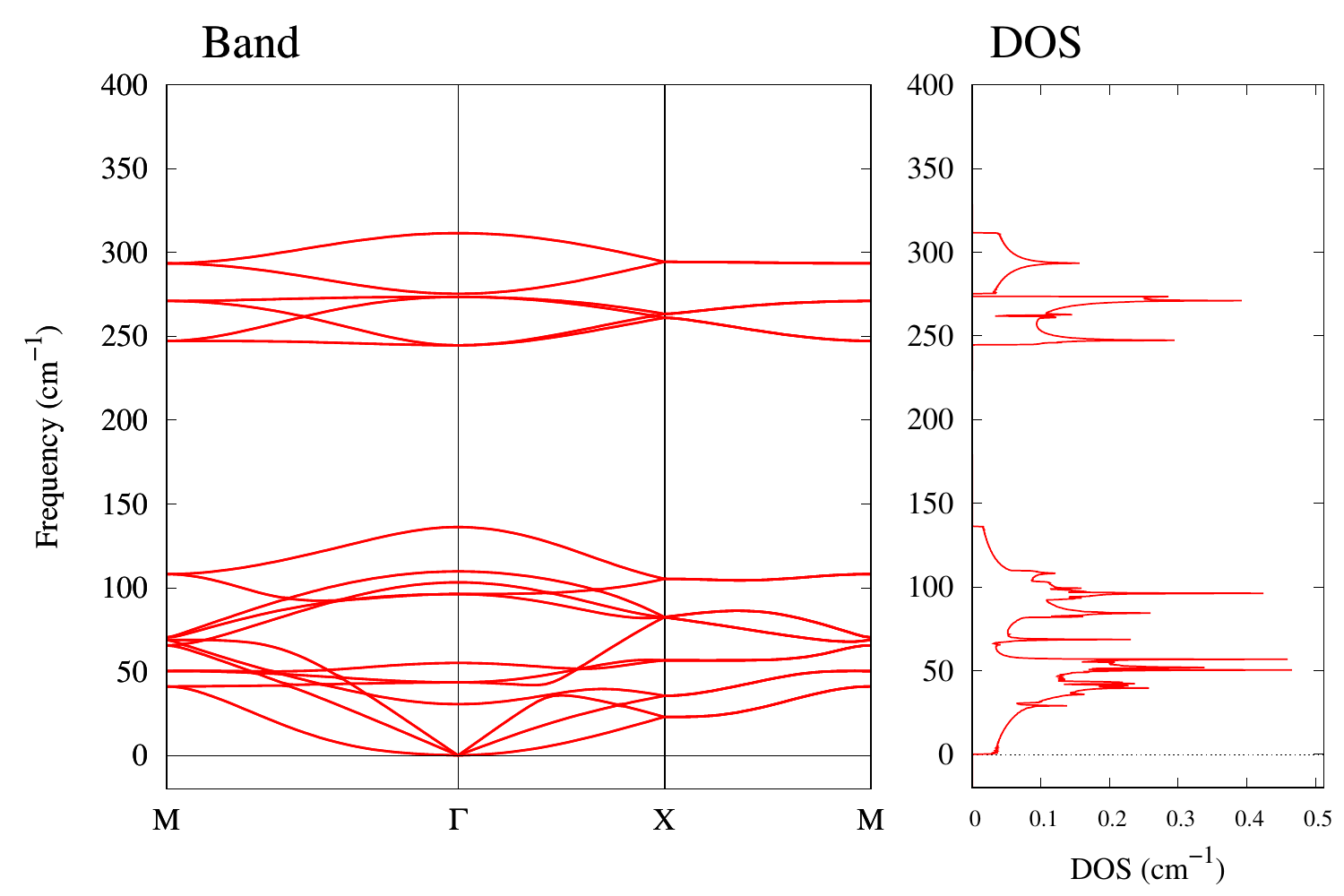}
          \subcaption{\textit{$\eta$}-Au$_2$S}
  \end{minipage}
  \begin{minipage}[b]{0.45\linewidth}
    \includegraphics[width=\linewidth]{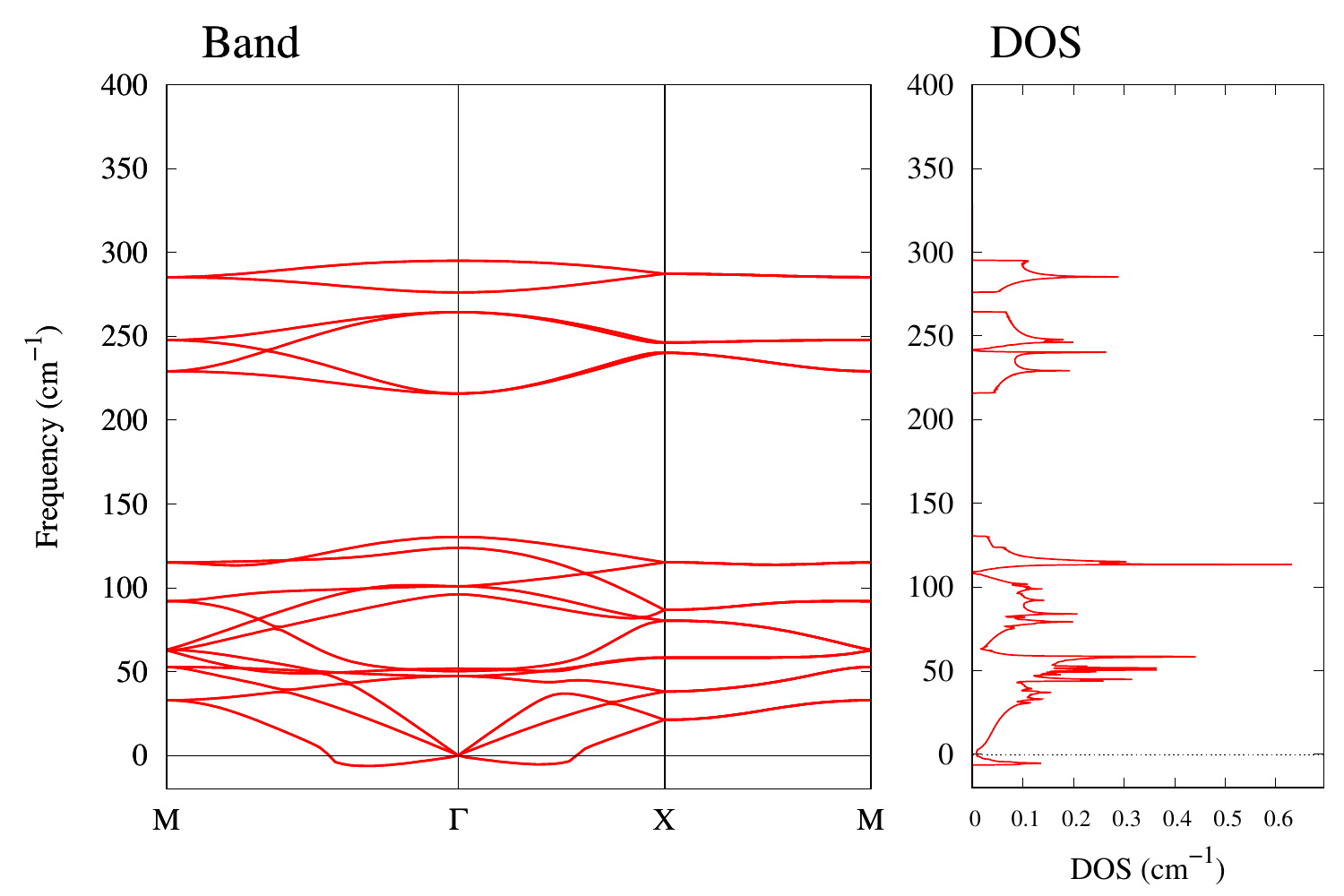}
          \subcaption{\textit{$\theta$}-Au$_2$S}
  \end{minipage}
 \caption{Phonon band and the DOS of \textit{$\eta$}- and \textit{$\theta$}-Au$_2$S monolayers.}\label{fig:Au2S_phonon}
\end{figure*}

\begin{figure}[htb]
    \centering
    \includegraphics[width=0.75\linewidth]{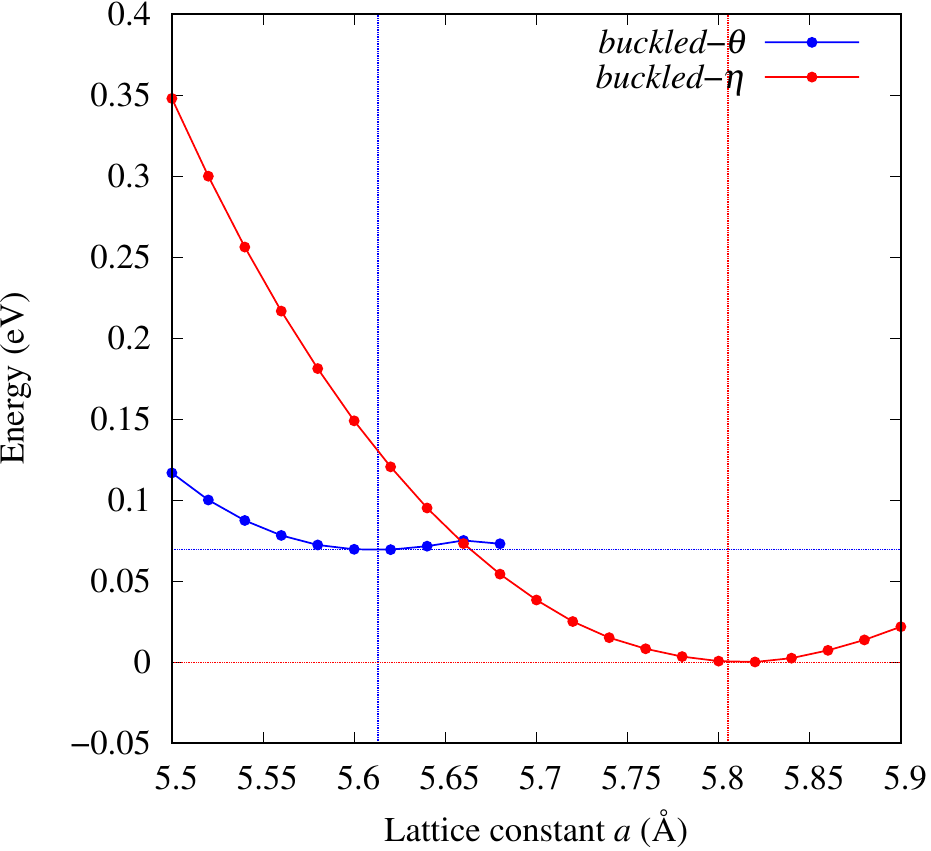}
        \caption{Energy curve of Au$_2$S monolayers. The red (blue) line represents the energy curve of $\eta$-phase ($\theta$-phase).}\label{fig:Au2S_energy_curve}
\end{figure}

The stability of these Au$_2$S monolayers can be discussed 
by phonon dispersion and the DOS in Fig.~\ref{fig:Au2S_phonon}.
Generally speaking, a negative frequency in a phonon band indicates a negative curvature of the potential energy surface, which means that the structure is unstable or including anharmonic contributions.
Therefore, the negative frequency in Fig.~\ref{fig:Au2S_phonon}(b) might indicate the instability or anharmonicity of \textit{$\theta$}-Au$_2$S.
These phonon bands are consistent with other previous research~\cite{Gao2021}.
In addition, the energy curves for lattice constants in Fig.~\ref{fig:Au2S_energy_curve} give us another insight for stability of Au$_2$S.
The potential curve of \textit{$\theta$}-Au$_2$S is so shallow that a phase transition from $\theta$-phase to $\eta$-phase is easily induced by stretching the lattice constants.
The opposite concept is also true that the $\eta$-phase can cause phase transition
to the $\theta$-phase by bi-axial compression to the lattice.
Such compression might be induced by lattice matching with a substrate or boundary matching with other monolayers.
These kinds of interactions with other stable materials can stabilize unstable monolayers.
Silicene on Ag surface~\cite{Gao2012,PhysRevB.98.195311} and silicene on ZrB$_2$~\cite{PhysRevLett.108.245501} are good examples for monolayers stabilized on the surface.
Therefore, it cannot be said that the \textit{$\theta$}-Au$_2$S is sufficiently stable as a free-standing monolayer, but it may be possible to exist on a substrate.

\begin{figure}[htb]
    \includegraphics[width=\linewidth]{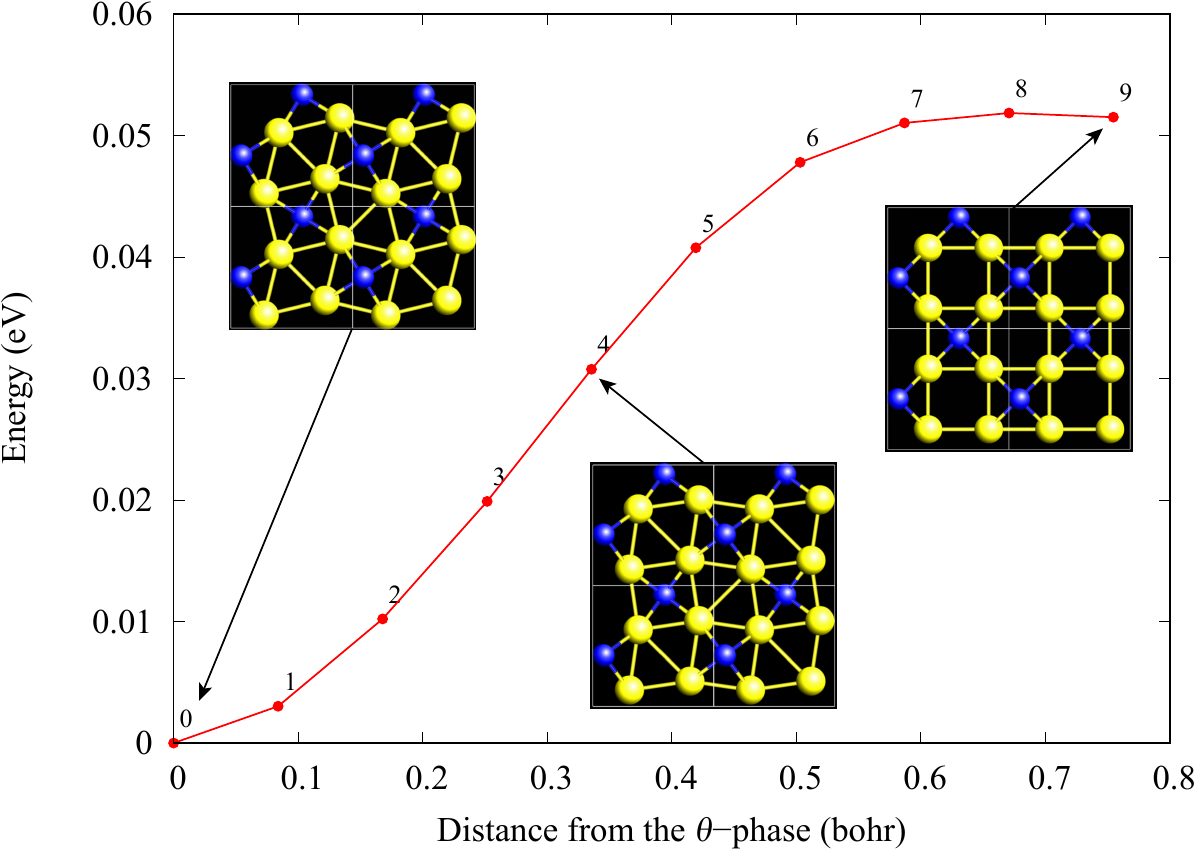}
        \caption{The change of total energy of Au$_2$S as a function of the distance (Bohr)
        from the $\theta$-Au$_2$S to $\eta$-Au$_2$S and the corresponding geometrical structures. 
        The NEB calculations were performed with the lattice constant $a=5.62$ \AA.
        Each number in the plot area represents the index of the NEB images.}\label{fig:Au2S_neb}
\end{figure}

The nudged elastic band (NEB) calculation~\cite{10.1063/1.1323224} for Au$_2$S is also performed at the lattice constant
$a=5.62$ \AA, which is close to the most stable state of \textit{$\theta$}-Au$_2$S.
Since the energy barrier from $\eta$-phase to $\theta$-phase is very small as shown in Fig.~\ref{fig:Au2S_neb},
it is concluded that the phase transition of $\eta \to \theta$ can occur easily by applying the compressive strain of about 3.5\% of the lattice constants.

\subsection{Electronic band structure} \label{sec:Au2S_eband_structure}
\begin{figure*}[htb]
   \begin{minipage}[b]{0.41\linewidth}
    \includegraphics[width=\linewidth]{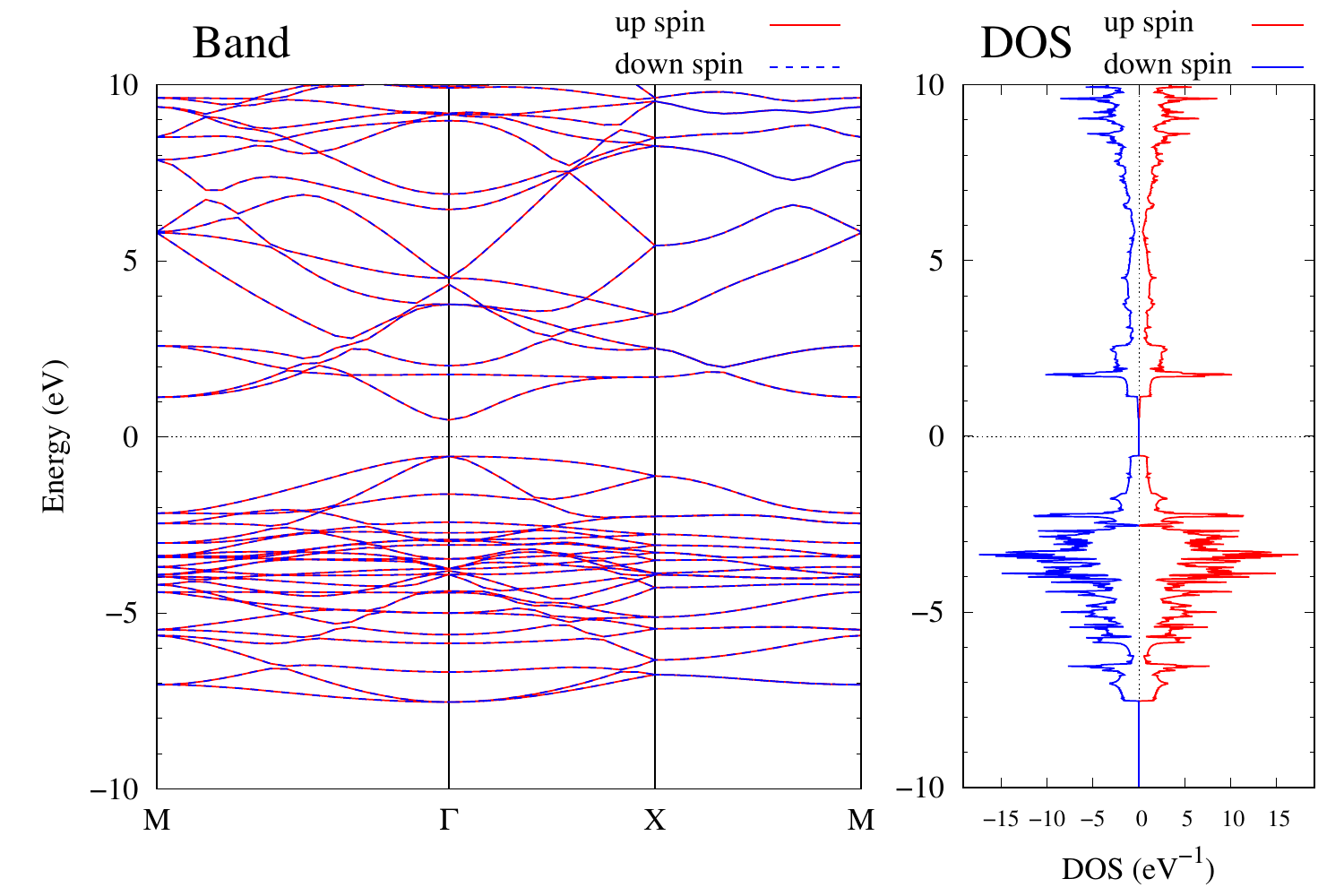}
          \subcaption{Band and the DOS of \textit{$\eta$}-Au$_2$S}
  \end{minipage}
   \begin{minipage}[b]{0.25\linewidth}
    \mbox{\raisebox{10pt}{ \includegraphics[width=\linewidth] {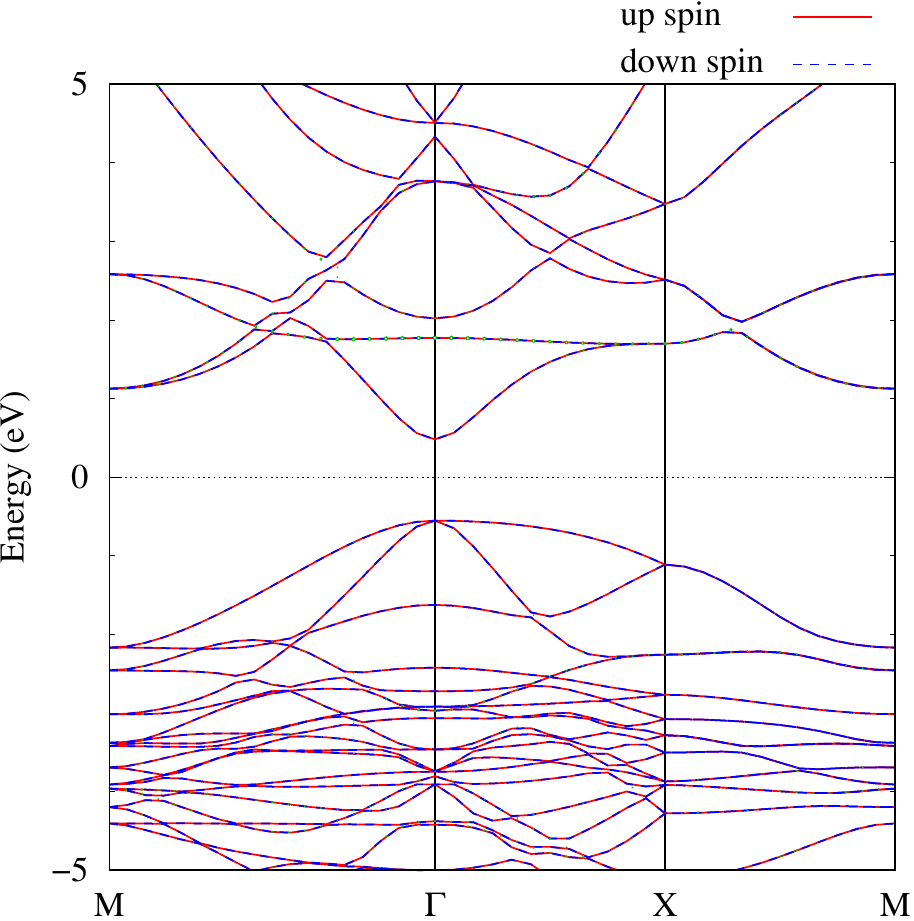}}}
          \subcaption{S-\textit{s}}
  \end{minipage}
  \begin{minipage}[b]{0.25\linewidth}
    \mbox{\raisebox{10pt}{ \includegraphics[width=\linewidth] {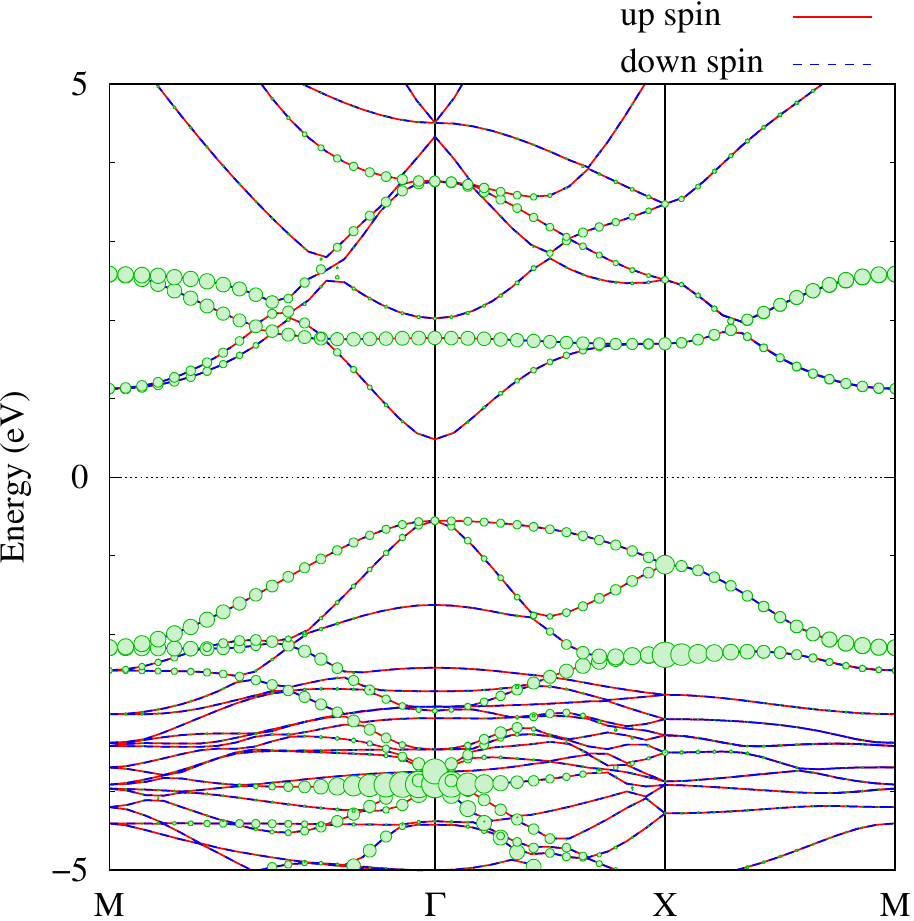}}}
          \subcaption{S-\textit{p}}
  \end{minipage}
  \\
  \begin{minipage}[b]{0.25\linewidth}
    \includegraphics[width=\linewidth]{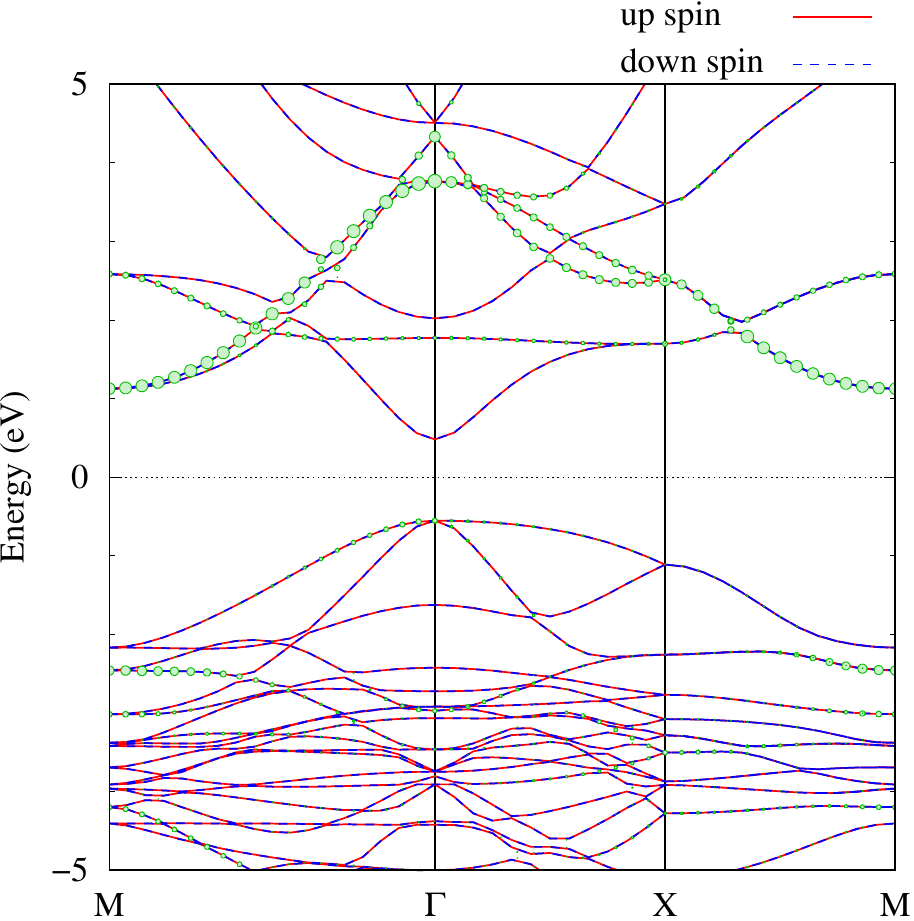}
          \subcaption{Au-\textit{s}}
  \end{minipage}
  \begin{minipage}[b]{0.25\linewidth}
    \includegraphics[width=\linewidth]{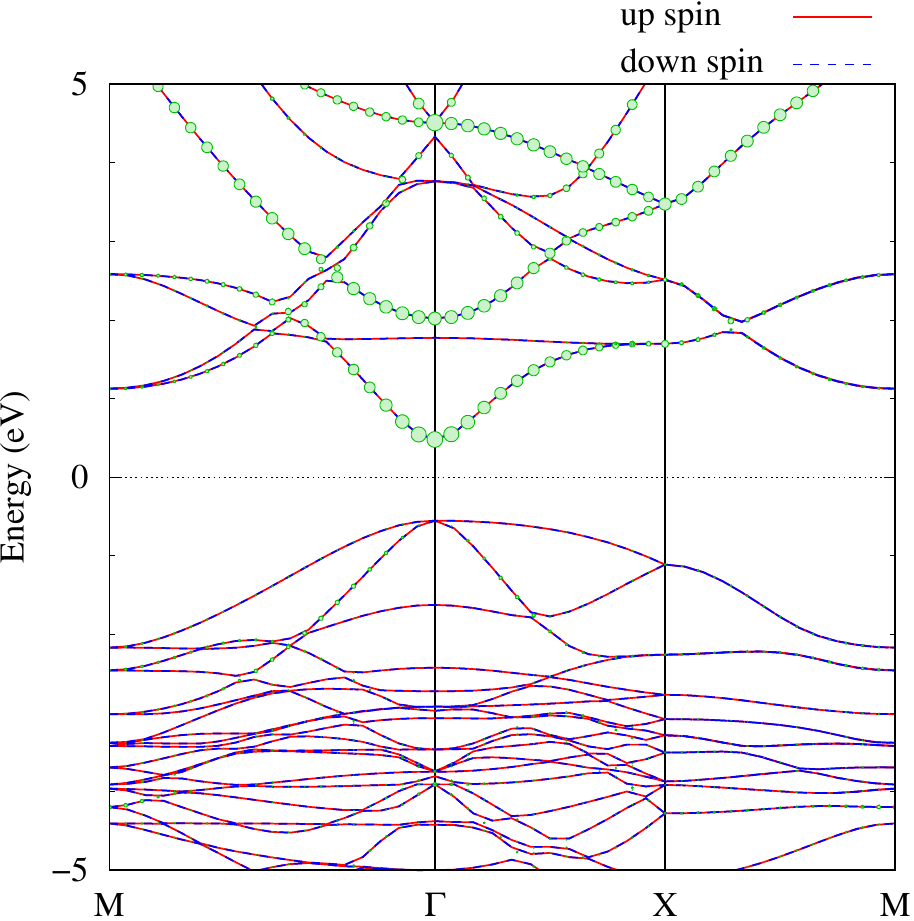}
          \subcaption{Au-\textit{p}}
  \end{minipage}
   \begin{minipage}[b]{0.25\linewidth}
    \includegraphics[width=\linewidth]{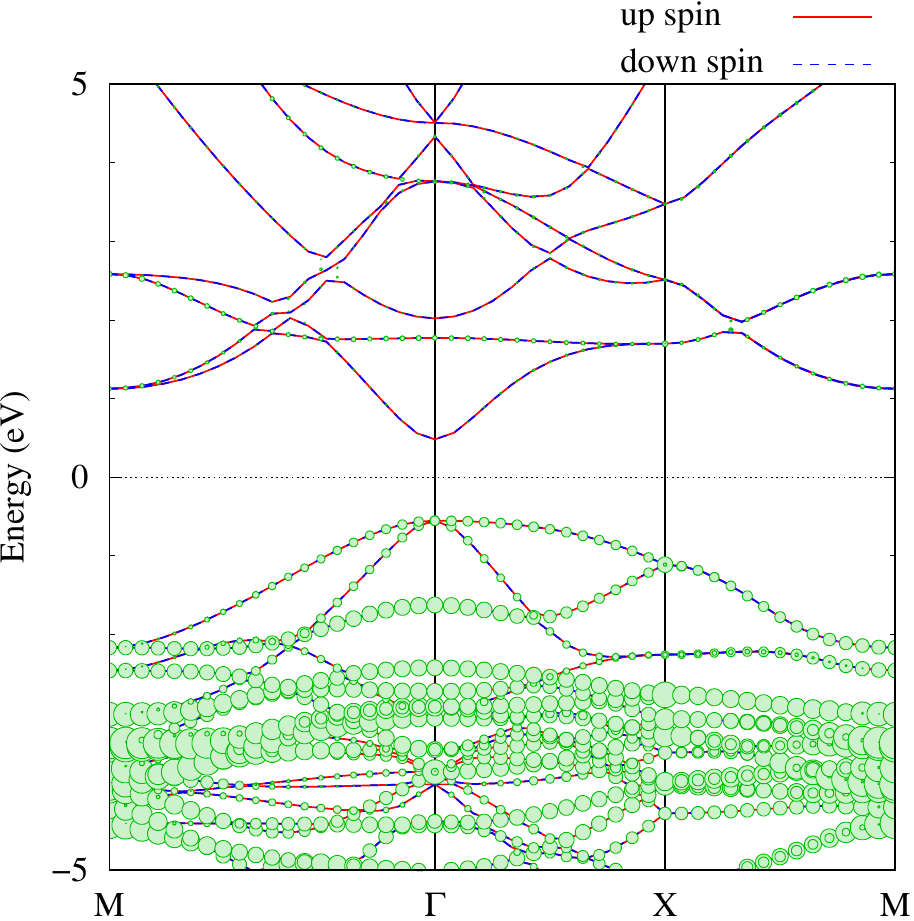}
          \subcaption{Au-\textit{d}}
  \end{minipage}
 \caption{(a) Electronic band and the DOS of \textit{$\eta$}-Au$_2$S. The $y$ axis is taken so that the Fermi energy is zero. (b-f) Weight of the electronic band of Au and S atoms of \textit{$\eta$}-Au$_2$S. The weight is projected on \textit{s}, \textit{p} and \textit{d} type functions.}\label{fig:Au2S_eband_unfolding_eta}
\end{figure*}

\begin{figure*}[htb]
   \begin{minipage}[b]{0.41\linewidth}
    \includegraphics[width=\linewidth]{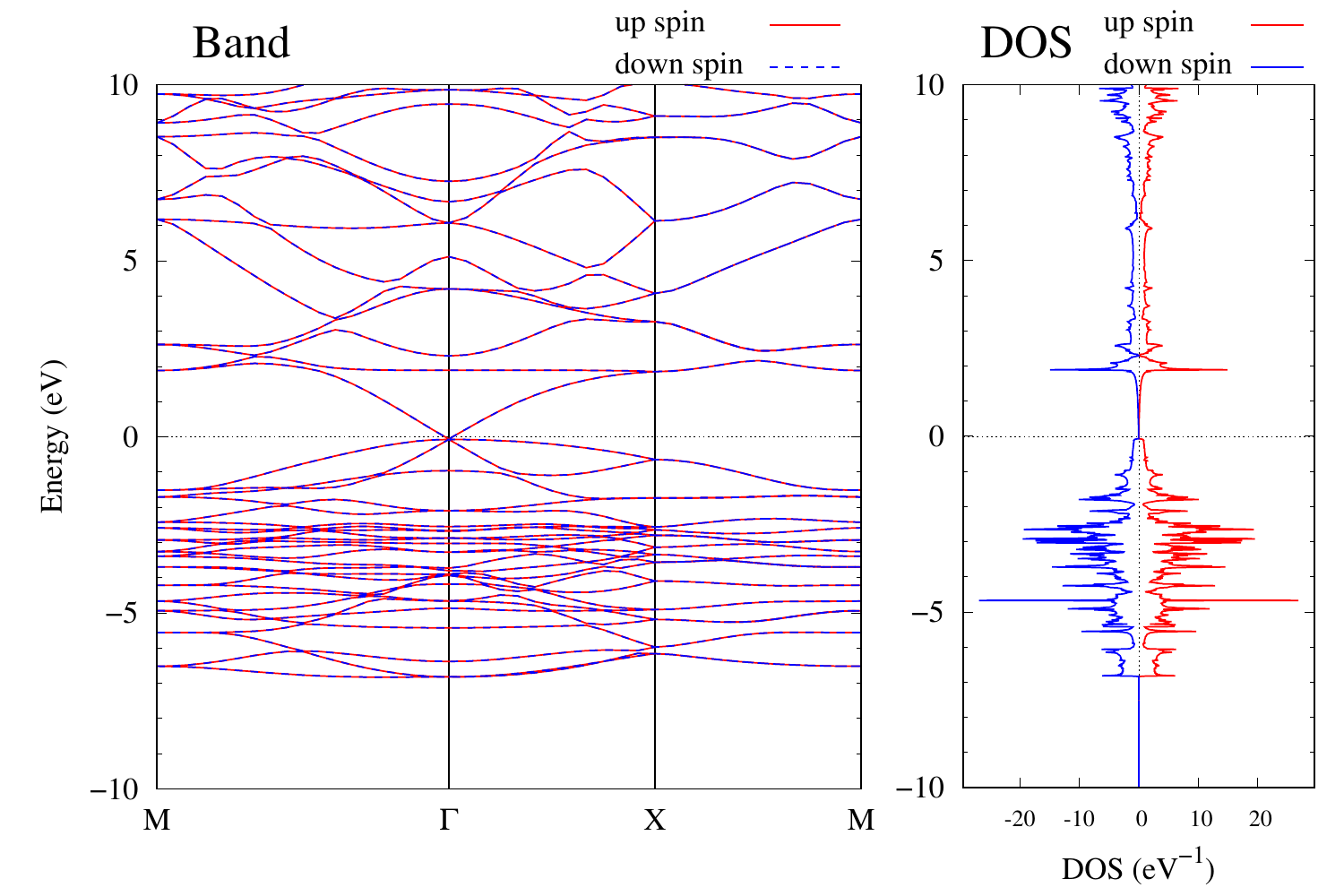}
          \subcaption{Band and the DOS of \textit{$\theta$}-Au$_2$S}
  \end{minipage}
   \begin{minipage}[b]{0.25\linewidth}
    \mbox{\raisebox{10pt}{ \includegraphics[width=\linewidth] {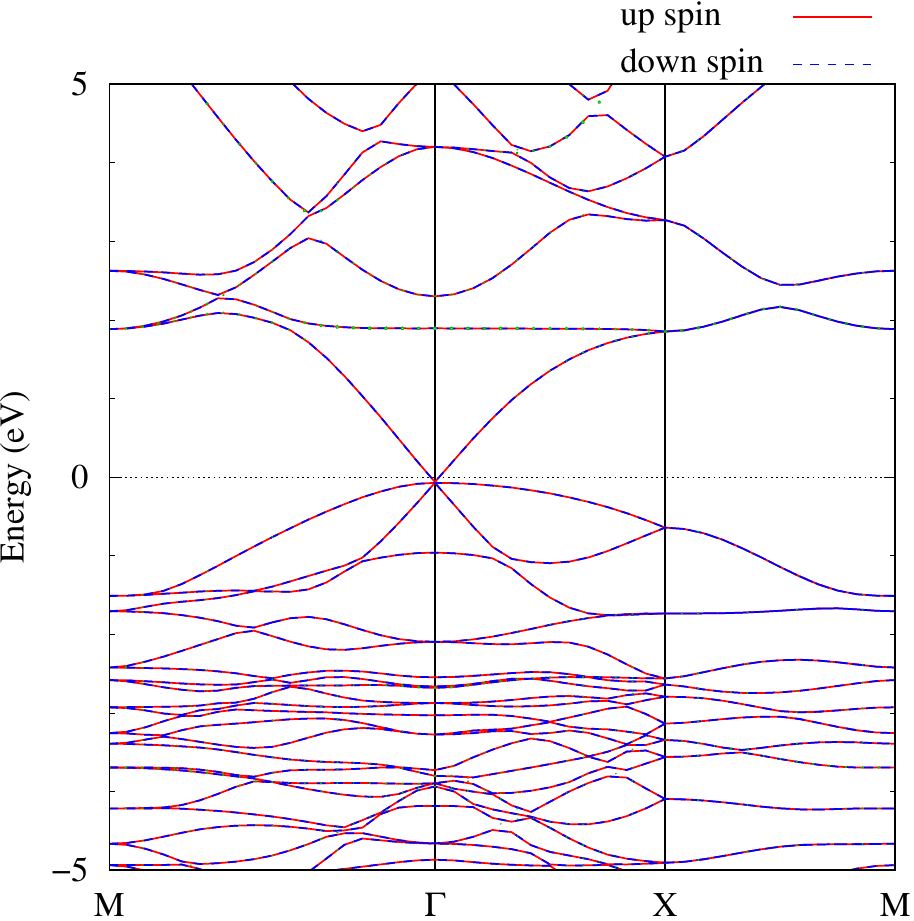}}}
          \subcaption{S-\textit{s}}
  \end{minipage}
  \begin{minipage}[b]{0.25\linewidth}
    \mbox{\raisebox{10pt}{ \includegraphics[width=\linewidth] {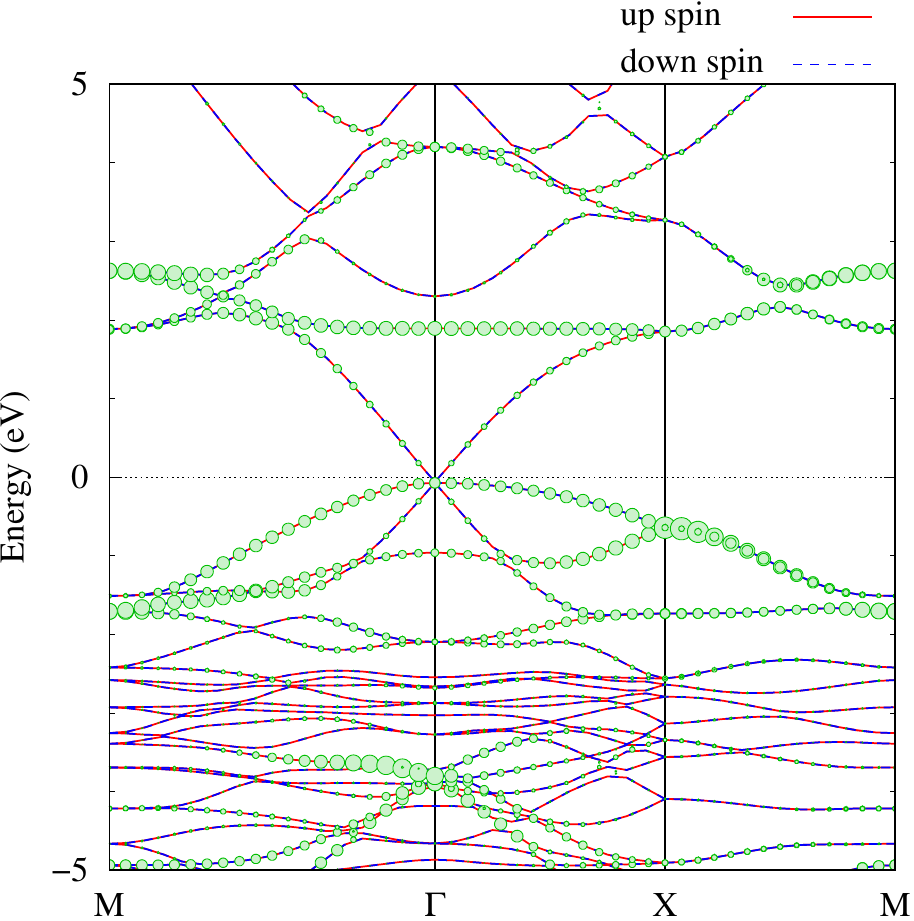}}}
          \subcaption{S-\textit{p}}
  \end{minipage}
    \\
  \begin{minipage}[b]{0.25\linewidth}
    \includegraphics[width=\linewidth]{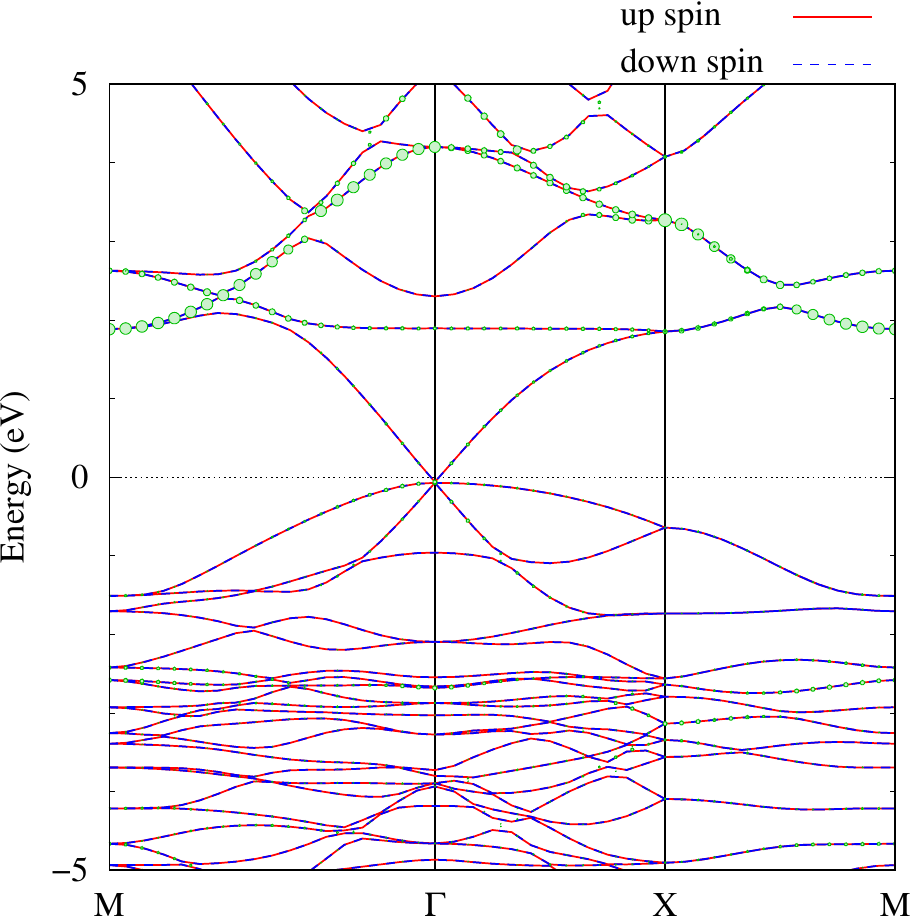}
          \subcaption{Au-\textit{s}}
  \end{minipage}
  \begin{minipage}[b]{0.25\linewidth}
    \includegraphics[width=\linewidth]{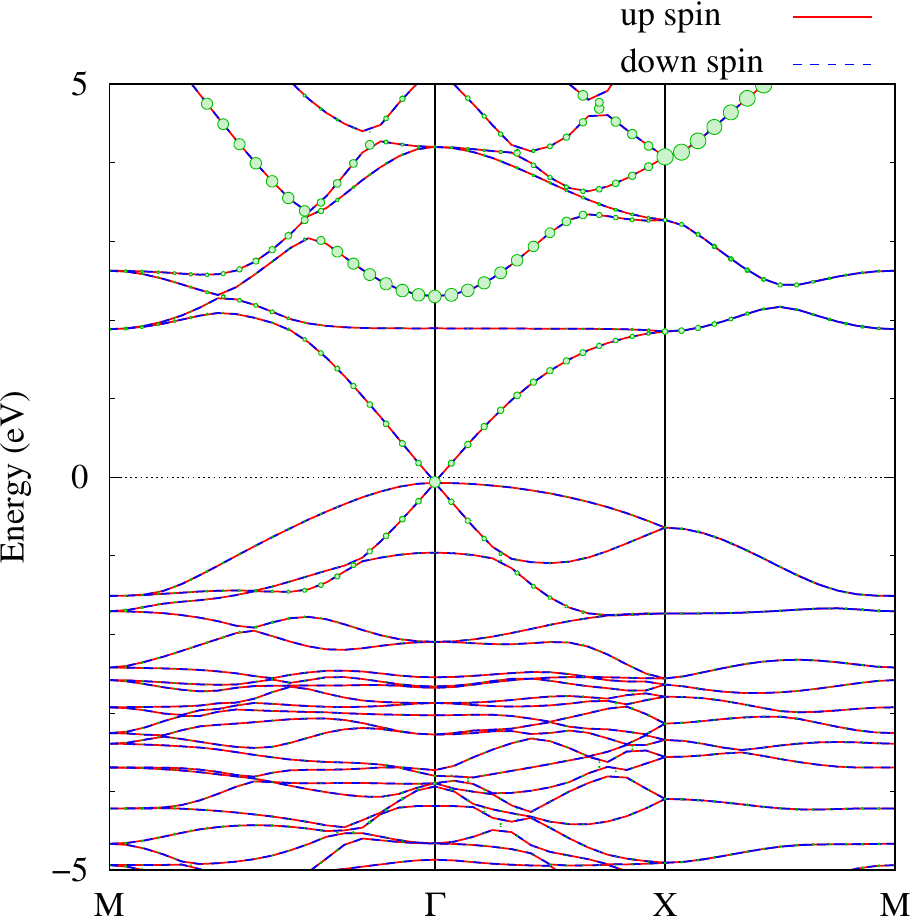}
          \subcaption{Au-\textit{p}}
  \end{minipage}
   \begin{minipage}[b]{0.25\linewidth}
    \includegraphics[width=\linewidth]{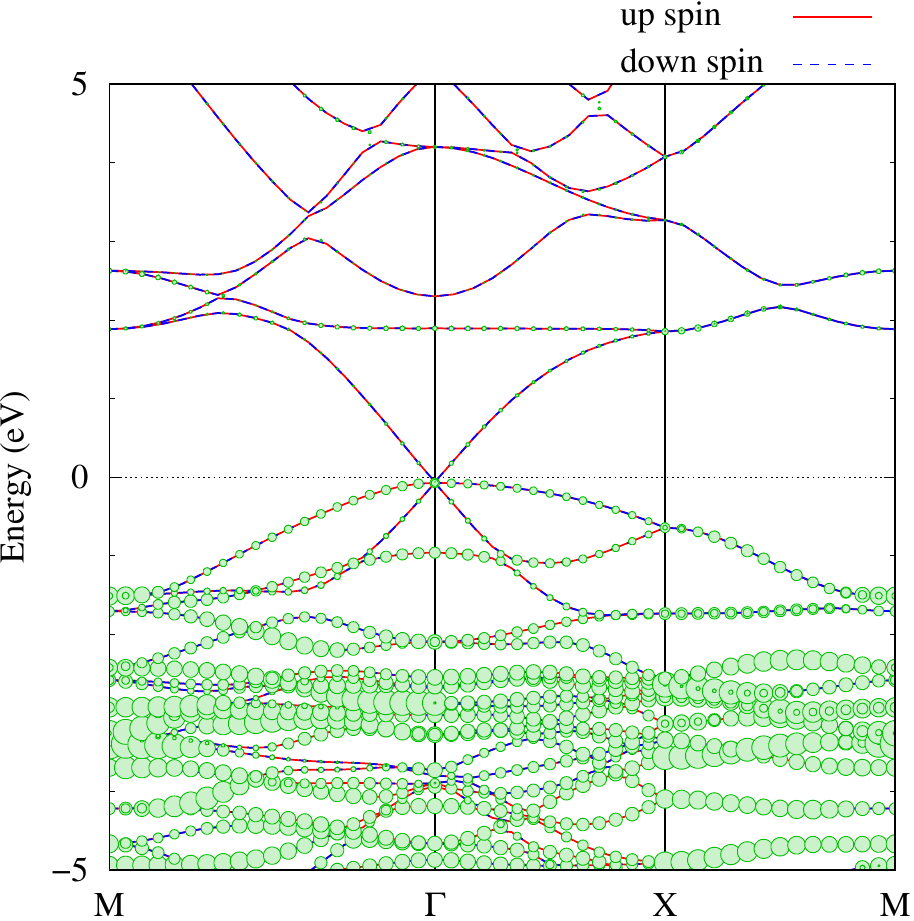}
          \subcaption{Au-\textit{d}}
  \end{minipage}
  
 \caption{(a) Electronic band and the DOS of \textit{$\theta$}-Au$_2$S. The $y$ axis is taken so that the Fermi energy is zero. (b-f) Weight of the electronic band of Au and S atoms of \textit{$\theta$}-Au$_2$S. The weight is projected on \textit{s}, \textit{p} and \textit{d} type functions.}\label{fig:Au2S_eband_unfolding_theta}

\end{figure*}

Electronic band structures and the DOS of \textit{$\eta$}- and \textit{$\theta$}-Au$_2$S monolayers are shown in Figs.~\ref{fig:Au2S_eband_unfolding_eta} (a) and \ref{fig:Au2S_eband_unfolding_theta} (a).
The \textit{$\eta$}-Au$_2$S has a direct band gap of 1.02 eV at $\Gamma$ point,
whereas the band gap is closed for the \textit{$\theta$}-Au$_2$S.
This kind of band gap modulation in the electronic bands of \textit{$\eta$}- and \textit{$\theta$}-Cu$_2$S by using uniaxial and biaxial stress has been reported previously ~\cite{C8NH00216A,Gao2021,doi:10.1021/acs.jpclett.0c00613,ZHU2019113704}.
In Ref.~\cite{doi:10.1021/acs.jpclett.0c00613},
based on the partial DOS analysis,
the researchers have revealed that the valence band maximum (VBM) of \textit{$\theta$}-Cu$_2$S is dominantly contributed by the
hybridization of $p_x$, $p_y$ atomic orbitals of S atoms and $d_{xy}$ atomic
orbitals of Cu atoms, whereas the conduction band minimum (CBM) of  \textit{$\theta$}-Cu$_2$S is
mainly from the $p_x$, $p_y$ atomic orbitals of Cu atoms.

Here, we discuss more details of the origins of the modulations of VBM and CBM around the $\Gamma$ point. 
At first, we calculated the weight of each contribution of pseudo-atomic orbital in band structures of \textit{$\eta$}- and \textit{$\theta$}-Au$_2$S as shown in Figs.~\ref{fig:Au2S_eband_unfolding_eta}(b-f) and \ref{fig:Au2S_eband_unfolding_theta}(b-f)
based on the Ref.~\cite{Lee2013-nq}.
Additionally, the partial density of states (PDOS) are shown in Figs.~S2 and S3 in Supplemental Information.
Here, we denote the contribution from $s$ type functions of Au atoms as Au-$s$.
We also denote the valence (conduction) bands as VB1 (CB1), VB2 (CB2), and so on, in order of proximity to the Fermi level.
From Fig.~\ref{fig:Au2S_eband_unfolding_eta} for \textit{$\eta$}-Au$_2$S,
it is obvious that the VB1 and VB2 have large contributions from both Au-$d$ and S-$p_x$,$p_y$.
In addition, CB1 has contributions from Au-$p_x$,$p_y$, forming bonds between Au atoms arranged in a square lattice. 
In contrast, CB2 have contributions from Au-$s$,$d$ and S-$p_z$, forming a flat band.
The shape of VB1 and CB2 of $\theta$-phase in Fig~\ref{fig:Au2S_eband_unfolding_theta} are similar to those of $\eta$-phase. 
However, two bands of VB2 and CB1 of $\theta$-phase are hybridized and forming a crossing linear dispersion
whose main contributions are Au-$p_x$,$p_y$, Au-$d$, and S-$p_x$,$p_y$.
Here, we note that the crossing linear dispersion split a little bit
when the spin-orbit interaction is included (see Fig.~S1 in Supplemental Information).

\begin{figure*}[htb]
  \begin{minipage}[b]{0.30\linewidth}
    \includegraphics[width=\linewidth]{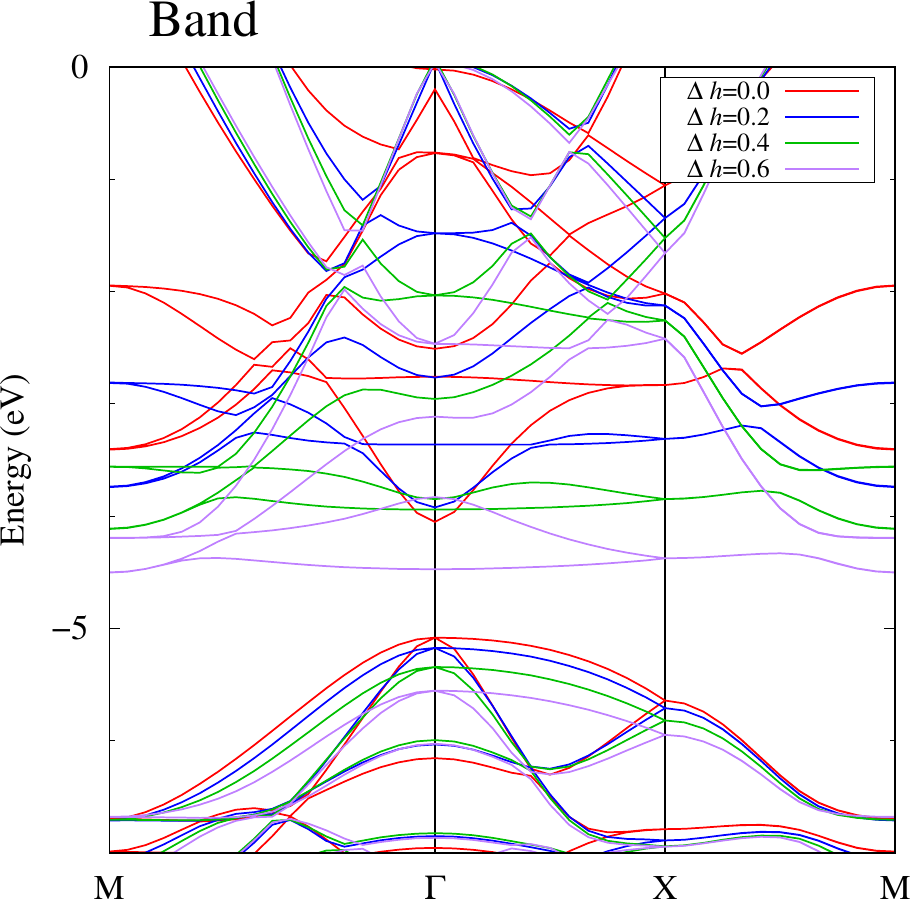}
          \subcaption{\textit{$\eta$}-Au$_2$S}
  \end{minipage}
  \begin{minipage}[b]{0.30\linewidth}
    \includegraphics[width=\linewidth]{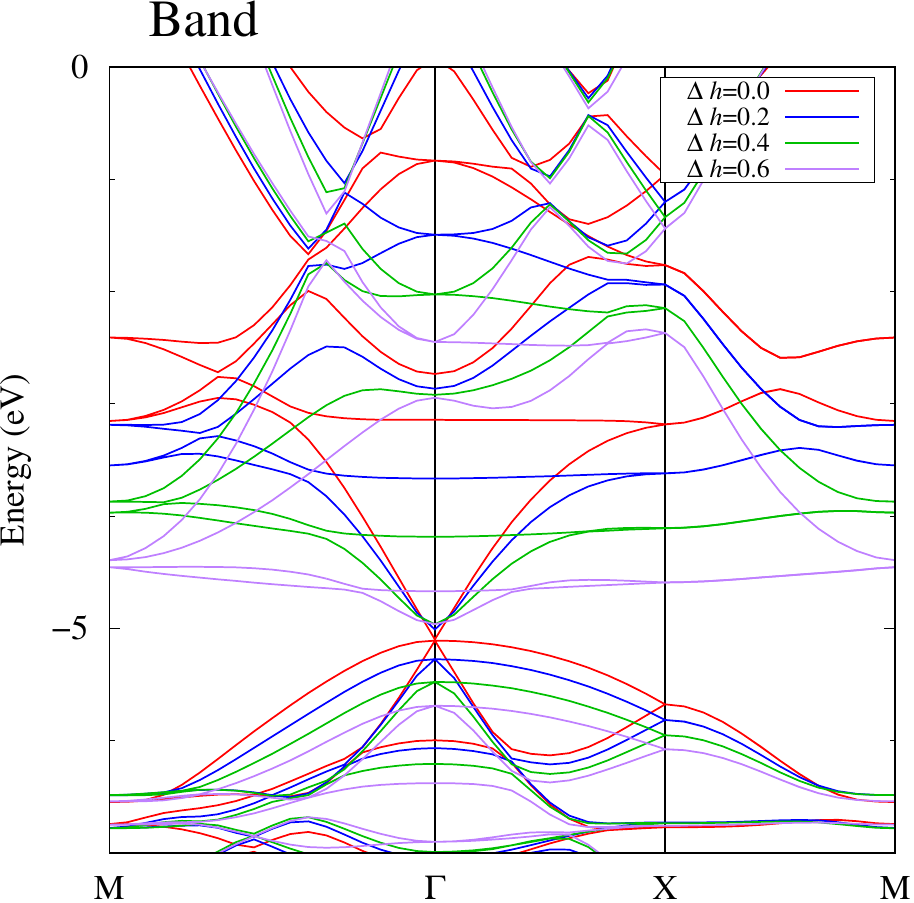}
          \subcaption{\textit{$\theta$}-Au$_2$S}
  \end{minipage}
   \begin{minipage}[b]{0.30\linewidth}
    \includegraphics[width=\linewidth]{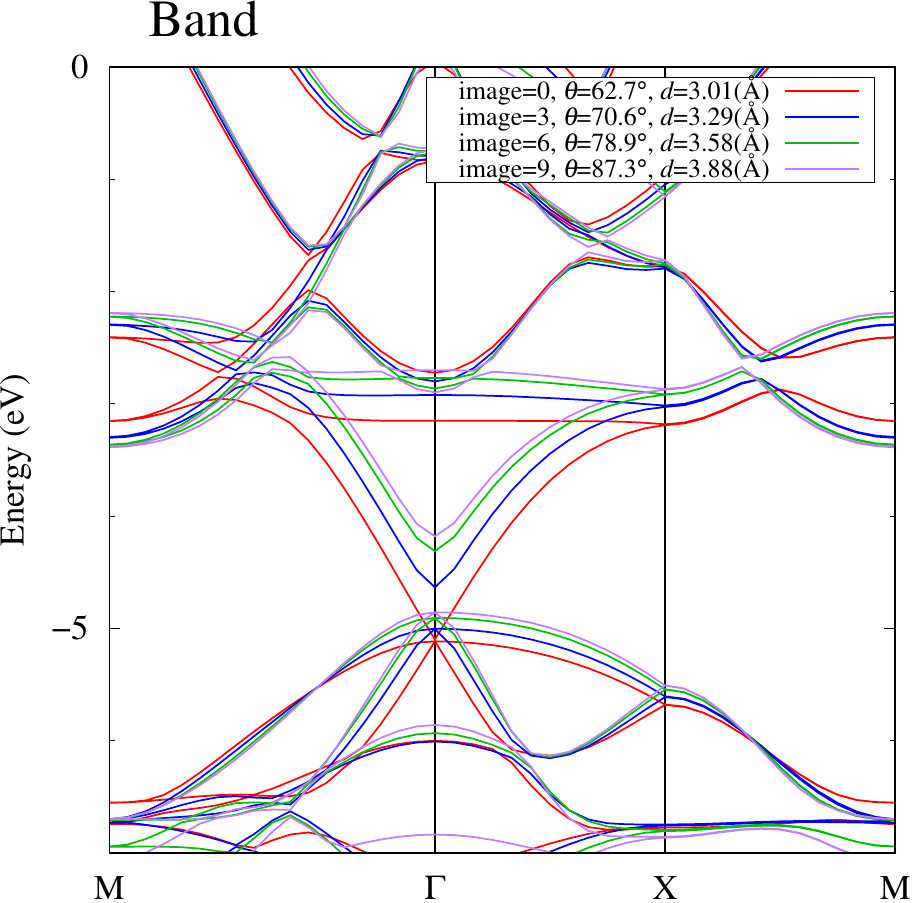}
          \subcaption{NEB images}
  \end{minipage}
 \caption{Band structures of (a) \textit{$\eta$}- and (b) \textit{$\theta$}-Au$_2$S
 for various heights ($h + \Delta h$ \AA) of S atoms with fixed positions of Au atoms.
 (c) Band structures for some NEB images in Fig.~\ref{fig:Au2S_neb}.
  The image 0 and 9 represent the \textit{$\theta$} and \textit{$\eta$}-Au$_2$S, respectively.
  The $\theta$ (deg) represents the angle of Au atoms as shown in the Fig.~\ref{fig:Au2S_structure},
  and $d$ is the Au-Au bond distance in each structure.
 }\label{fig:Au2S_eband_analysis}
 
\end{figure*}

Figures \ref{fig:Au2S_eband_analysis} (a) and (b) show band structures of \textit{$\eta$}- and \textit{$\theta$}-Au$_2$S
for various heights ($h + \Delta h$ \AA) of S atoms with fixed positions of Au atoms.
The band energy of CB1 at the $\Gamma$ point remains almost unchanged when $\Delta h$ is varied,
since the main contributions of the CB1 are only $p_x$ and $p_y$ of fixed Au atoms.
In contrast, the band energies of VB2, VB1 and CB2, including contributions from S atoms,
decrease as the S atoms move away from the Au atomic layer.
Therefore, it reveals that the CBM depends on the bond distance of Au-Au, not on the interaction between Au and S atoms.
On the other hand, the VBM largely depends on the interaction between Au and S atoms.
Note that the cases of $\Delta h= 0.4, 0.6$ of $\eta$-phase are not included in the above discussion to avoid the ambiguity,
since the order of the bands is swapped in the cases of $\Delta h= 0.4, 0.6$ of $\eta$-phase.
The flat band, whose energy decreases as $\Delta h$ is increased, is a band derived from the S-$p_z$ isolation state.
It is characteristic that the flatness remains even when $\Delta h$ is decreased.

Figure~\ref{fig:Au2S_eband_analysis}(c) shows band structures of transition states
from \textit{$\theta$}-phase to \textit{$\eta$}-phase calculated by using structures 
obtained by the NEB calculation (Fig.~\ref{fig:Au2S_neb}).
The images 0 and 9 represent the \textit{$\theta$}-phase and \textit{$\eta$}-phase, respectively.
The Au-Au bond distance $d$ (see Fig.~\ref{fig:Au2S_structure}) increases in the transition from \textit{$\theta$}-phase\ to \textit{$\eta$}-phase.
Moreover, as the Au-Au bond distance $d$ increases, the band energy of CB1, which consists of Au-$p_x$,$p_y$, increases.
Although energies of the other bands, CB2, VB1 and VB2, also increase, the energy shifts are smaller than that of CB1.

To summarize the above discussions on Fig.~\ref{fig:Au2S_eband_analysis},
the CBM can be controlled by the change of Au-Au bond length $d$ induced by strain and $\theta \longleftrightarrow \eta$ phase transition, whereas the VBM can be controlled by interaction between the Au layer and attached S atoms.


\section{A\lowercase{u}$_2$X monolayers} \label{sec:Au2X}


\subsection{Structure and stability} \label{sec:Au2X_structure}

Total energies, cohesive energies, and structural parameters defined in 
Fig.~\ref{fig:Au2S_structure} for $\eta$- and $\theta$-Au$_2$X (X=S, Se, Te, Si, Ge),
are listed in Table~\ref{tab:Au2X_properties}.
The lattice constants $a$ of all the $\theta$-phase structures in the Table~\ref{tab:Au2X_properties} are about 5.6 \AA.
The lattice constants $a$ of the $\eta$-phase structures are about 5.8 \AA~for chalcogenides
and about 5.6 \AA~for Si, Ge compounds.
This means that lattice constants of these structures are almost independent of the species of atoms
attached to the above and below the Au atomic layer.
On the other hand, the heights $h$ of X atoms are consistent with the trend in the length of the covalent bond radius~\cite{https://doi.org/10.1002/chem.200800987}.
Since the angle $\theta$ and the Au-Au bond length $d$ defined in Fig.~\ref{fig:Au2S_structure} can be
measures of the strength of the interaction between Au atoms,
Table~\ref{tab:Au2X_properties} indicates that $\theta$-Au$_2$S, $\theta$-Au$_2$Se and $\theta$-Au$_2$Te might have stronger Au-Au interactions in that order.

The cohesive energies of monolayers and bulks in Table~\ref{tab:Au2X_properties} are defined as 
\begin{align}
\nonumber
&E_{\rm coh, monolayer} = ( 4\times E_{\rm Au-atom} + 2\times E_{\rm X-atom}\\
\nonumber
& \hspace{75pt} - E_{\rm Au_2X-monolayer} )/6,
\end{align}
\begin{align}
\nonumber
&E_{\rm coh, bulk} = \{ 4( E_{\rm Au-atom} - E_{\rm Au-bulk} / N_{\rm Au-bulk} )\\
\nonumber
& \hspace{60pt}  + 2 ( E_{\rm X-atom} -  E_{\rm coh, X-bulk}/N_{\rm X-bulk} ) \}/6.
\end{align}
In our results, only $\eta$- and $\theta$-Au$_2$S satisfy the condition $ E_{\rm coh, bulk} < E_{\rm coh, system}$,
which is one of the measures of stability.
Therefore, $\eta$- and $\theta$-Au$_2$S might be good candidates of stable monolayers.
However, it cannot be said that the other monolayers do not exist as a stabilized monolayer, since these kinds of monolayers might be synthesized on a substrate.

\begin{figure*}
  \begin{minipage}[b]{0.32\linewidth}
    \centering
    \includegraphics[width=\linewidth]{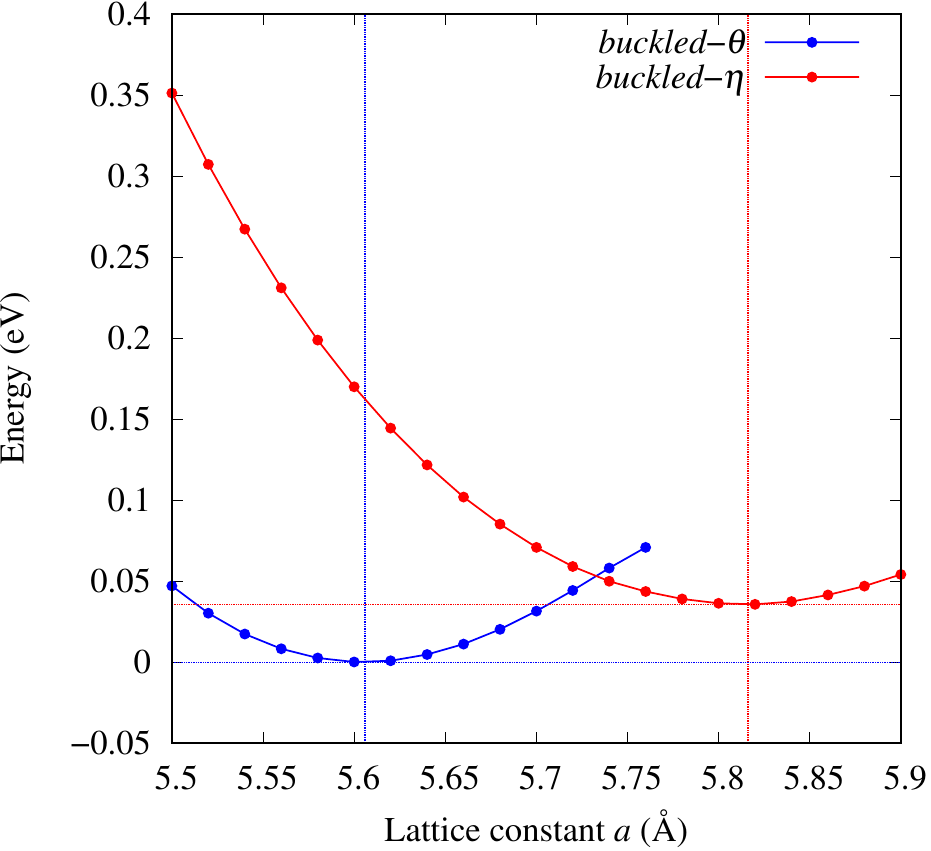}
          \subcaption{Au$_2$Se}
  \end{minipage}
  \begin{minipage}[b]{0.32\linewidth}
    \includegraphics[width=\linewidth]{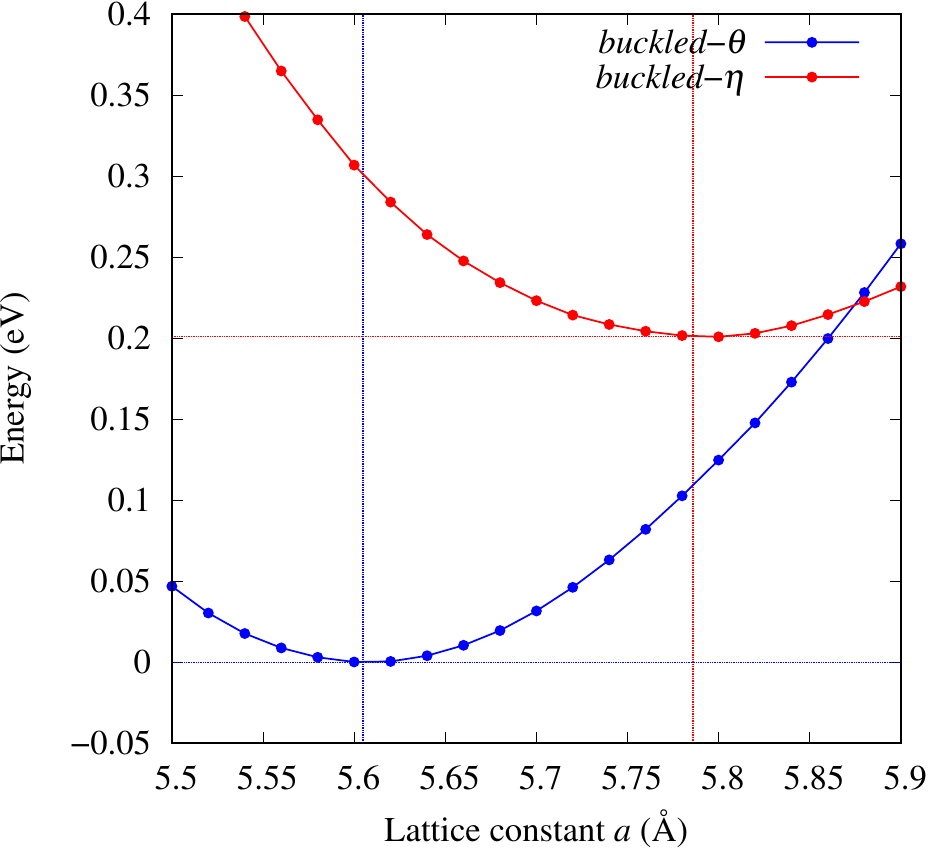}
          \subcaption{Au$_2$Te}
  \end{minipage}
  \begin{minipage}[b]{0.32\linewidth}
    \includegraphics[width=\linewidth]{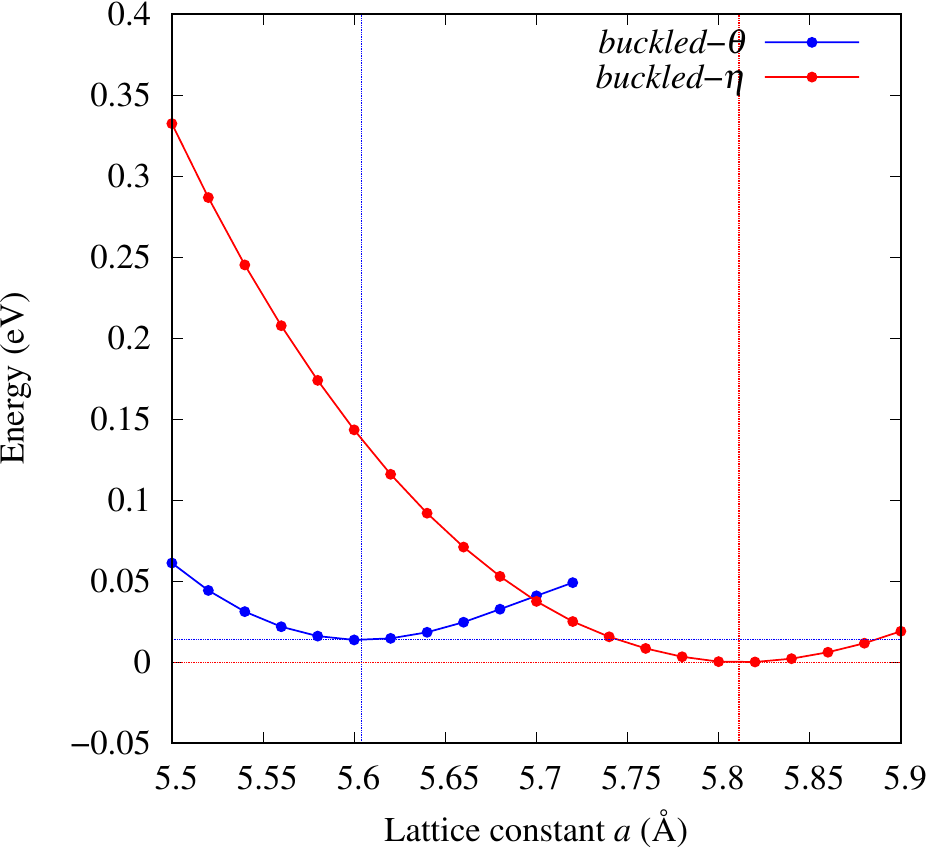}
          \subcaption{Au$_4$SSe}
  \end{minipage}
  
  \vspace{5pt}
  \begin{minipage}[b]{0.32\linewidth}
    \centering
    \includegraphics[width=\linewidth]{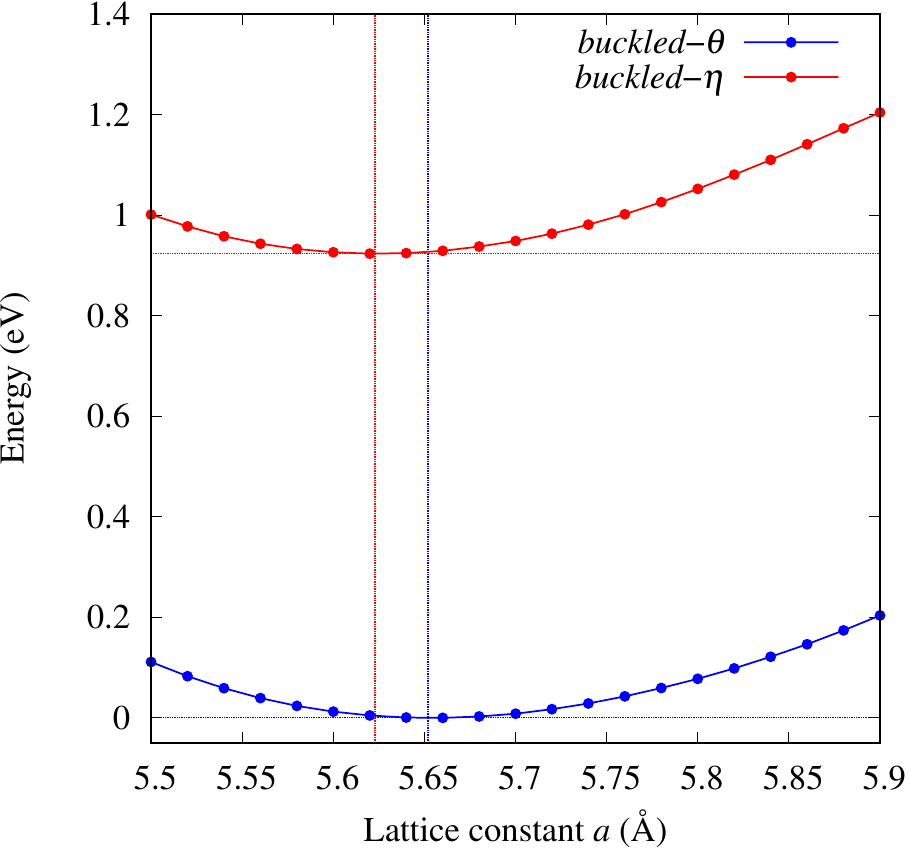}
          \subcaption{Au$_2$Si}
  \end{minipage}
  \begin{minipage}[b]{0.32\linewidth}
    \includegraphics[width=\linewidth]{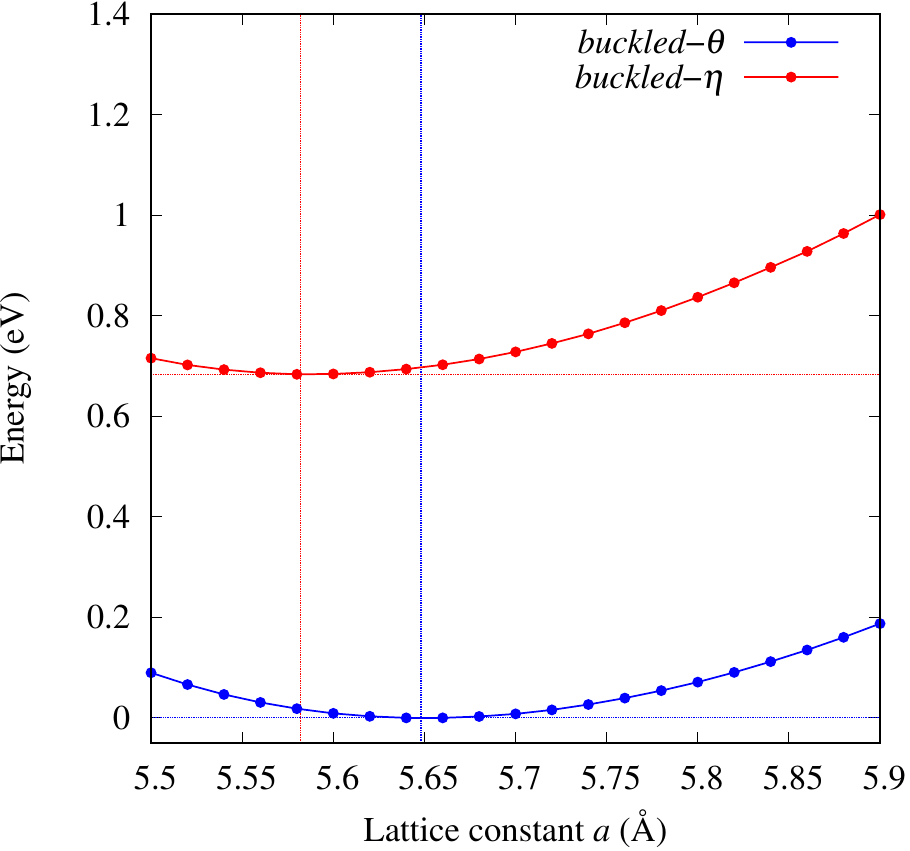}
          \subcaption{Au$_2$Ge}
  \end{minipage} 
  \begin{minipage}[b]{0.32\linewidth}
    \includegraphics[width=\linewidth]{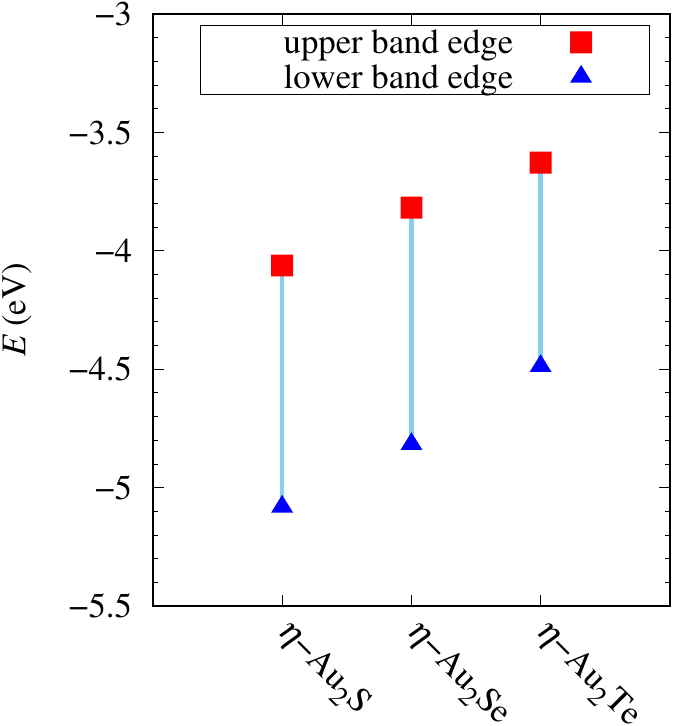}
          \subcaption{Band gaps}
  \end{minipage} 
  
        \caption{Energy curves of $\eta$- and $\theta$-Au$_2$X (X=Se, Te, Si, Ge) and Au$_4$SSe monolayers (a-e).
        The red (blue) line represents the energy curve of $\eta$-phase ($\theta$-phase).
        Band gaps at the $\Gamma$ point of Au$_2$X (X=S, Se, Te) (f).
        The red squares and the blue triangles represent CBM and VBM, respectively.
        }\label{fig:Au2X_energy_curve}
\end{figure*}

From Figs.~S6 and S7 in Supplemental Information, the phonon bands of Au$_2$X monolayers around zero energy have a similar shape to Au$_2$S except for $\eta$-Au$_2$Si and $\eta$-Au$_2$Ge, which have large negative frequencies indicating instability.
It is also clear from the energy curve in Fig.~\ref{fig:Au2X_energy_curve} that 
Au$_2$Si, Au$_2$Ge, Au$_2$Se, and Au$_2$Te prefer to be $\theta$-phase,
whereas Au$_2$S prefers to be $\eta$-phase.
In other words, 
the type of more stable phase can be changed by controlling the species of atoms adsorbed on the Au lattice.

As an example of controlling the stability of the phase,
we calculated energy curves of $\eta$- and $\theta$-Au$_4$SSe, 
whose structure is generated by replacing a half of S atoms in Au$_2$S to Se atoms,
as shown in Fig.~\ref{fig:Au2X_energy_curve} (c).
The $\theta$-phase, which is unstable in Au$_2$S, stabilizes and the energy of the most stable state approaches that of the $\eta$-phase.
It means that the $\eta \leftrightarrow \theta$ phase transition of Au$_4$SSe can be easily induced by tensile and compressive strain of about 3.5\% of the lattice constants.
Therefore, by partially substituting S atoms,
it might be possible to create devices with partially different mechanical and physical properties in a single monolayer.

\begin{table*}[htp]
    \centering
\caption{Properties of Au$_2$X.
        Lattice constant $a$, X atom's height $h$, angle $\theta$, and Au-Au bond length $d$ are defined in Fig~\ref{fig:Au2S_structure}.
        The relative energy $E_{\rm rel}$ represents the energy difference from the energy of $\eta$-phase.}
        \label{tab:Au2X_properties}
\begin{tabular}{cccccccc}
\hline
Name &  $E_{\rm rel}$ (eV)  & $E_{\rm coh,system} (eV)$ & $E_{\rm coh,bulk} (eV)$ & $a$ (\AA) & Height $h$ (\AA) & Angle $\theta$ (deg) & $d$ (\AA)\\
\hline
\textit{$\eta  $}-Au$_2$S  & $\ \ 0.0000$  & 3.062 & 2.978 & 5.805 & 1.35 & 89.97 & 4.10\\
\textit{$\theta$}-Au$_2$S  & $\ \ 0.0697$  & 3.051 & 2.978 & 5.613 & 1.41 & 62.73 & 3.01\\
\cline{1-8}
\textit{$\eta  $}-Au$_2$Se &  $\ \ 0.0000$  & 2.966 & 3.033 & 5.816 & 1.56 & 90.01 & 4.11 \\
\textit{$\theta$}-Au$_2$Se & $-0.0354$  & 2.972 & 3.033 & 5.606 & 1.61 & 60.72 & 2.93\\
\cline{1-8}
\textit{$\eta  $}-Au$_2$Te &  $\ \ 0.0000$  & 2.866 & 2.969 & 5.786 & 1.87 & 90.02 & 4.09 \\
\textit{$\theta$}-Au$_2$Te & $-0.2005$  & 2.900 & 2.969 & 5.605 & 1.88 & 58.70 & 2.85\\
\cline{1-8}
\textit{$\eta  $}-Au$_2$Si &  $\ \ 0.0000$  & 3.111 & 3.723 & 5.623 & 1.54 & 90.03 & 3.98\\
\textit{$\theta$}-Au$_2$Si & $-0.9233$  & 3.265 & 3.723 & 5.652 & 1.32 & 54.30 & 2.71\\
\cline{1-8}
\textit{$\eta  $}-Au$_2$Ge &  $\ \ 0.0000$  & 2.940 & 3.418 & 5.582 & 1.76 & 89.97 & 3.95\\
\textit{$\theta$}-Au$_2$Ge & $-0.6835$  & 3.054 & 3.418 & 5.648 & 1.52 & 54.32 & 2.71\\
\hline
\end{tabular}
\end{table*}

\subsection{Electronic band structure} \label{sec:Au2X_eband_structure}
\begin{figure*}[htb]
  \begin{minipage}[b]{0.43\linewidth}
    \centering
    \includegraphics[width=\linewidth]{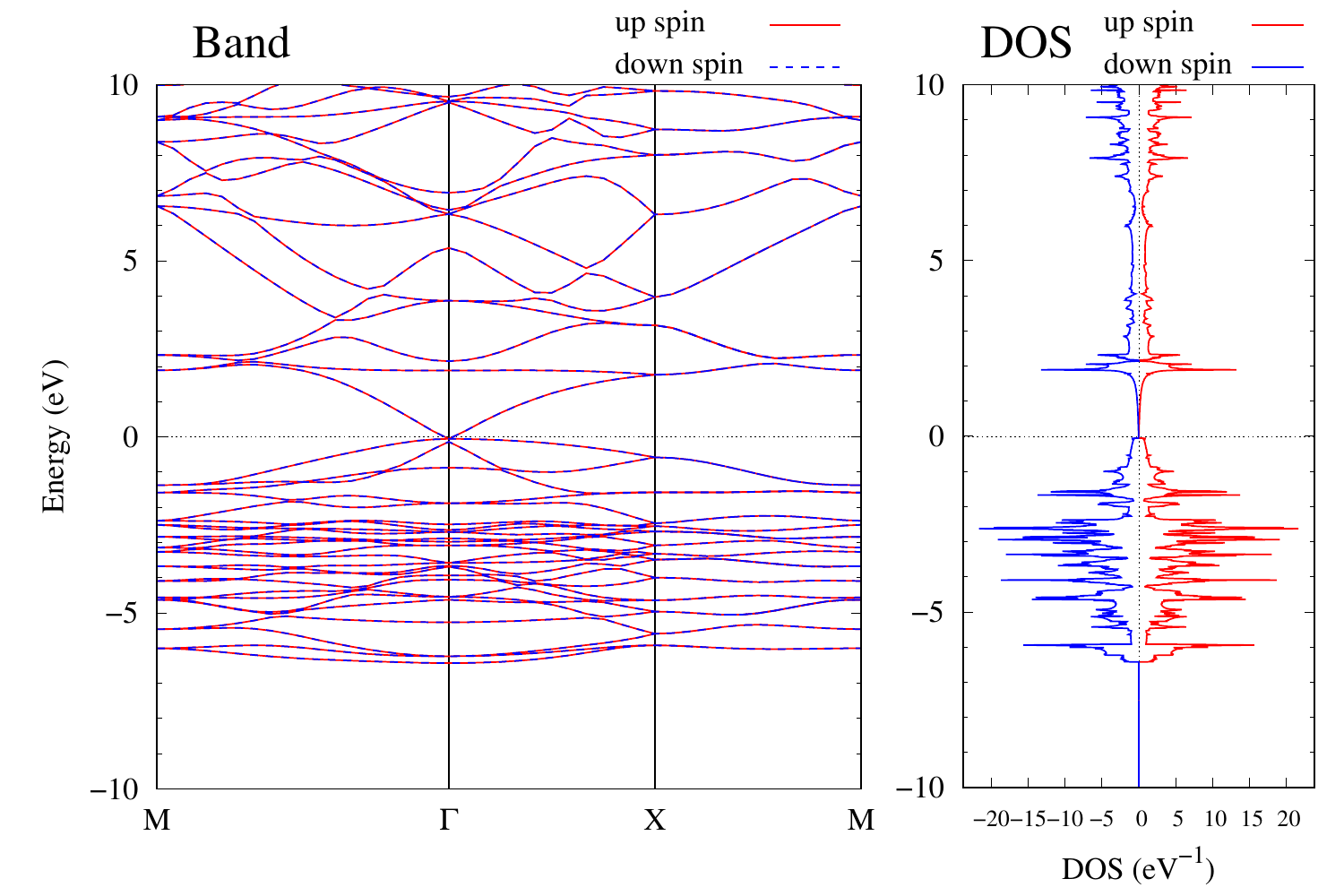}
          \subcaption{\textit{$\theta$}-Au$_2$Se}
  \end{minipage}
  \begin{minipage}[b]{0.43\linewidth}
    \centering
    \includegraphics[width=\linewidth]{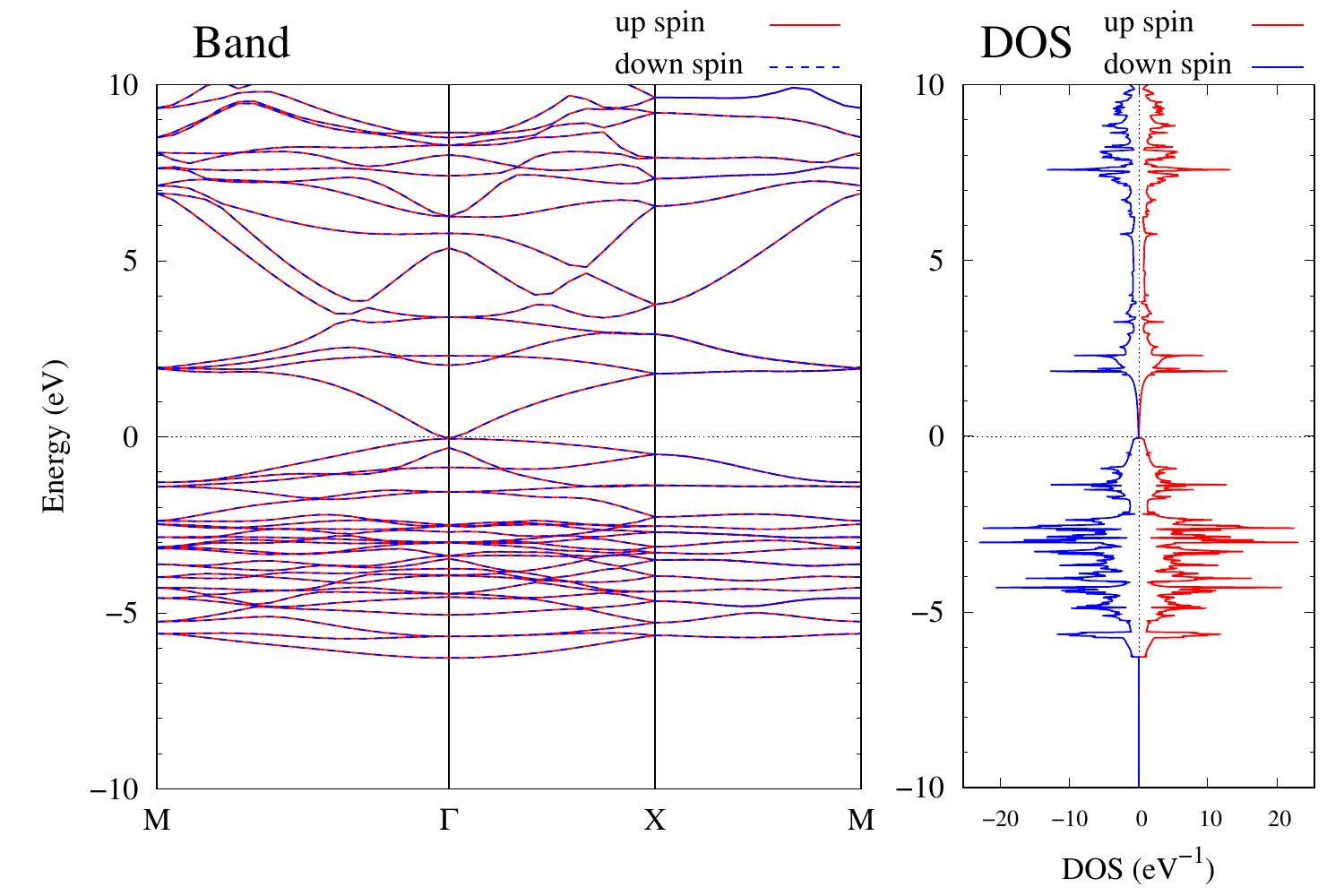}
          \subcaption{\textit{$\theta$}-Au$_2$Te}
  \end{minipage}
        \caption{Electronic band structures and the DOSs of \textit{$\theta$}-Au$_2$X (X=Se, Te) monolayers.}\label{fig:Au2Se_Au2Te_eband}
\end{figure*}

Electronic band structures and the DOSs of \textit{$\theta$}-Au$_2$X (X=Se, Te) and \textit{$\eta$}- and \textit{$\theta$}-Au$_2$Si are shown in Figs.~\ref{fig:Au2Se_Au2Te_eband} and S10, respectively.
The electronic band structures of the other Au$_2$X are shown in the Supplemental Information.
The shape of CB2, CB1, VB1, and VB2 of \textit{$\theta$}-Au$_2$Se is similar to \textit{$\theta$}-Au$_2$S.
However, VB2 of \textit{$\theta$}-Au$_2$Te is a little bit away from the Fermi level.
Therefore, though the band gap of \textit{$\theta$}-Au$_2$Te is closing,
it does not exist a linear dispersion, which is seen in \textit{$\theta$}-Au$_2$S and \textit{$\theta$}-Au$_2$Se.
The shape of the band structure of \textit{$\eta$}-Au$_2$X (X=S, Se, Te) is almost identical,
and their band gaps are shown in Fig.~\ref{fig:Au2X_energy_curve}(f).
In contrast, there is no band gap in \textit{$\eta$}-Au$_2$Si and \textit{$\eta$}-Au$_2$Ge. 
The shapes of the band structures of \textit{$\eta$}- and \textit{$\theta$}-Au$_2$Si are totally different from those of Au$_2$X (X=S, Se, Te),
since the Fermi level is lower in Au$_2$Si than in Au$_2$S.
The linear dispersion and the flat band of $\eta$-Au$_2$Si on about 0.5 eV at the $\Gamma$ point have the similar origin to the VB1, VB2 and CB1 of $\eta$-Au$_2$S.
This means that the replacement of X atoms affects the change of the Fermi level, resulting in a drastic change of the band structure.

\begin{figure*}[htb]
  \begin{minipage}[b]{0.43\linewidth}
    \centering
    \includegraphics[width=\linewidth]{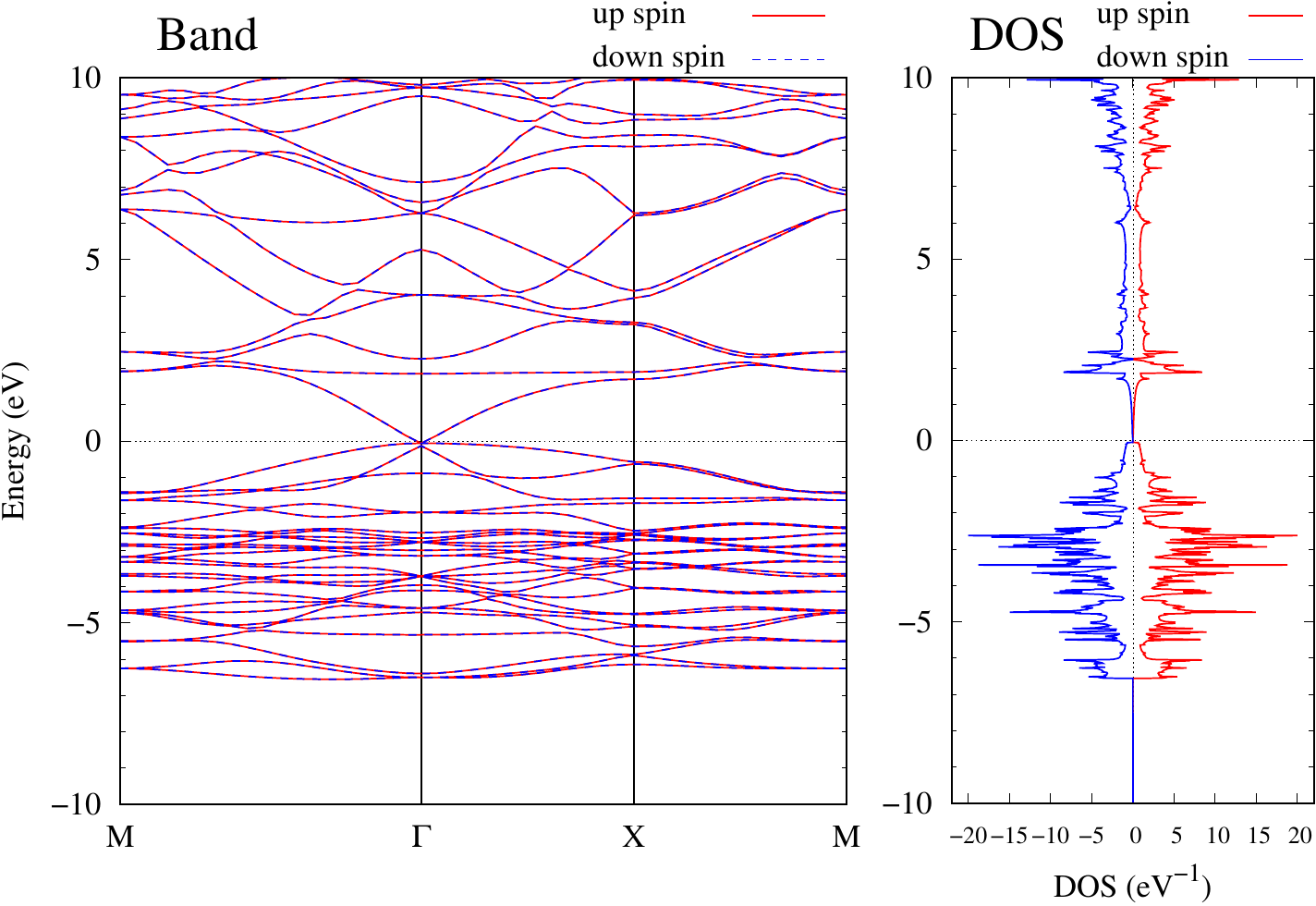}
          \subcaption{\textit{$\theta$}-Au$_4$SSe}
  \end{minipage}
  \begin{minipage}[b]{0.43\linewidth}
    \includegraphics[width=\linewidth]{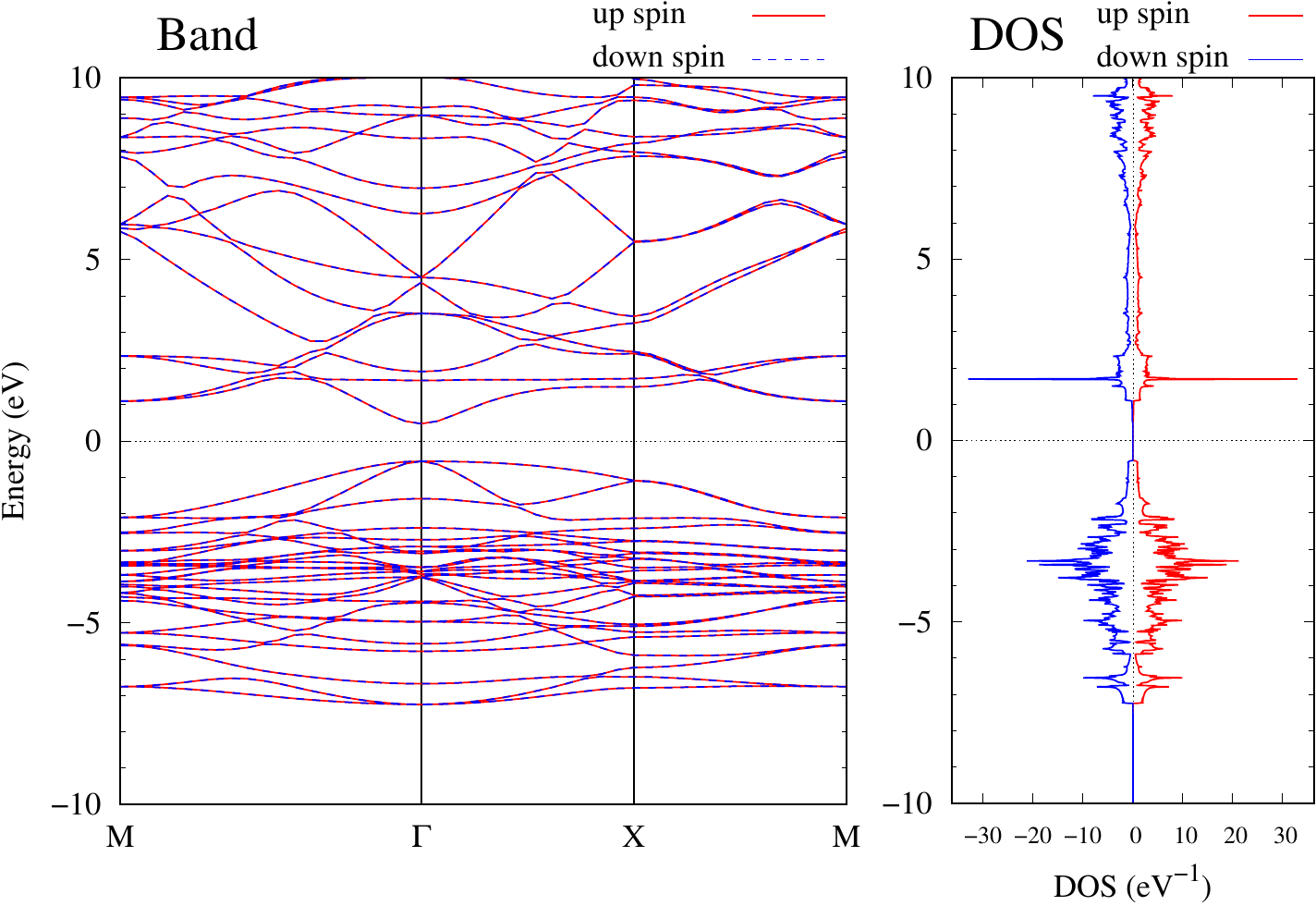}
          \subcaption{\textit{$\eta$}-Au$_4$SSe}
  \end{minipage}
        \caption{Electronic band structures and the DOSs of Au$_4$SSe monolayers.}\label{fig:Au4SSe_eband}
\end{figure*}

Finally, electronic band structures and the DOSs of \textit{$\eta$}- and \textit{$\theta$}-Au$_4$SSe, are shown in Fig.~\ref{fig:Au4SSe_eband}.
The band structure of Au$_4$SSe is almost identical to that of Au$_2$S and Au$_2$Se,
since the number of valence electron in Au$_4$SSe is the same as that of Au$_2$S and Au$_2$Se.
Therefore, Au$_4$SSe has a band structure similar to Au$_2$S and Au$_2$Se, and is equally stable in the \textit{$\eta$}- and \textit{$\theta$}-phases.


\section{Conclusions and perspectives} \label{sec:Conclusions}
In this paper, we have analyzed the $\eta \leftrightarrow \theta$ phase transition
and the change of electronic band structures of Au$_2$S induced by strain.
We have pointed out that about 3.5\% compressive biaxial strain on $\eta$-Au$_2$S,
which is induced, e.g., by lattice matching with a substrate,
can cause a transition to $\theta$-Au$_2$S.
It has also shown that the direct band gap of 1.02 eV in $\eta$-Au$_2$S disappears by this $\eta \leftrightarrow \theta$ phase transition.
Therefore, it has revealed that the band gap and the phase of Au$_2$S can be controlled by strain.
In addition, we have investigated the origin of band modulations around the $\Gamma$ point.
We have concluded that the band energy of CB1, which corresponds to CBM, is independent of the interaction between Au and S atoms,
while the increase of the Au-Au bond distance $d$ contributes significantly to the increase of the band energy of CB1.
We also have found that the band energy of VB1, which corresponds to VBM, depends on both Au-S and Au-Au interactions.
In particular, the band energy of VB1 decreases as the interaction between Au atoms and S atoms decreases.
In the $\theta$-Au$_2$S, CB1 and VB2 intersect on the $\Gamma$ point, creating linear dispersion.
These perspectives are significant for band gap engineering.

In the latter section, we have investigated \textit{$\eta$}- and \textit{$\theta$}-Au$_2$X (X=Se, Te, Si, Ge) and Au$_4$SSe.
Au$_2$Se and Au$_2$Te differ from Au$_2$S in that the $\theta$-phase is more stable than the $\eta$-phase,
although the electronic band structures are similar to those of Au$_2$S.
The energy curve of Au$_4$SSe has revealed that
we can modify the electronic state to be equally stable in the $\theta$ and $\eta$ phases by replacing some of the S atoms to Se atoms.
The electronic band structure of Au$_2$Si have revealed that
the Fermi level can be shifted by replacing S atoms with Si atoms, which have different valence electron numbers, resulting in changing from semiconductor to metal.
This means that, by replacing S atoms in a part of $\eta$-Au$_2$S that we want to give conductivity with other atoms, electronic circuits can be constructed on a single monolayer.

Thus, this family of Au$_2$X type monolayers has a potential to acquire a diversity of physical quantities
by replacing the X element without breaking the basic Au lattice network.
The family of Au$_2$X type monolayers may be a candidate of materials of atomic scale network devices.

\section{Acknowledgement}
The computation in this work has been done using the facilities of
the Supercomputer Center, the Institute for Solid State Physics, the
University of Tokyo.
The authors acknowledge financial support from JSPS KAKENHI Grant Number 20H00328.

\bibliography{ref}



\clearpage
\newpage

\newcommand{\beginsupplement}{%
    \setcounter{equation}{0}
    \setcounter{figure}{0}
    \setcounter{table}{0}
    \setcounter{page}{1}
    \makeatletter
    \renewcommand{\theequation}{S\arabic{equation}}
    \renewcommand{\thefigure}{S\arabic{figure}}
    \renewcommand{\thetable}{S\arabic{table}}
    \renewcommand{\thesection}{S\arabic{section}}
    \renewcommand{\thepage}{S\arabic{page}}
}

\section*{Supplemental Information}
\beginsupplement

\begin{figure*}[htb]
  \begin{minipage}[b]{0.45\linewidth}
    \includegraphics[width=\linewidth]{\PATHFIG/Au2S_buckled_theta_band_dos.pdf}
          \subcaption{\textit{$\theta$}-Au$_2$S}
  \end{minipage}
  \begin{minipage}[b]{0.45\linewidth}
    \includegraphics[width=\linewidth]{\PATHFIG/Au2S_buckled_eta_band_dos.pdf}
          \subcaption{\textit{$\eta$}-Au$_2$S}
  \end{minipage}
  \\
  \begin{minipage}[b]{0.45\linewidth}
    \includegraphics[width=\linewidth]{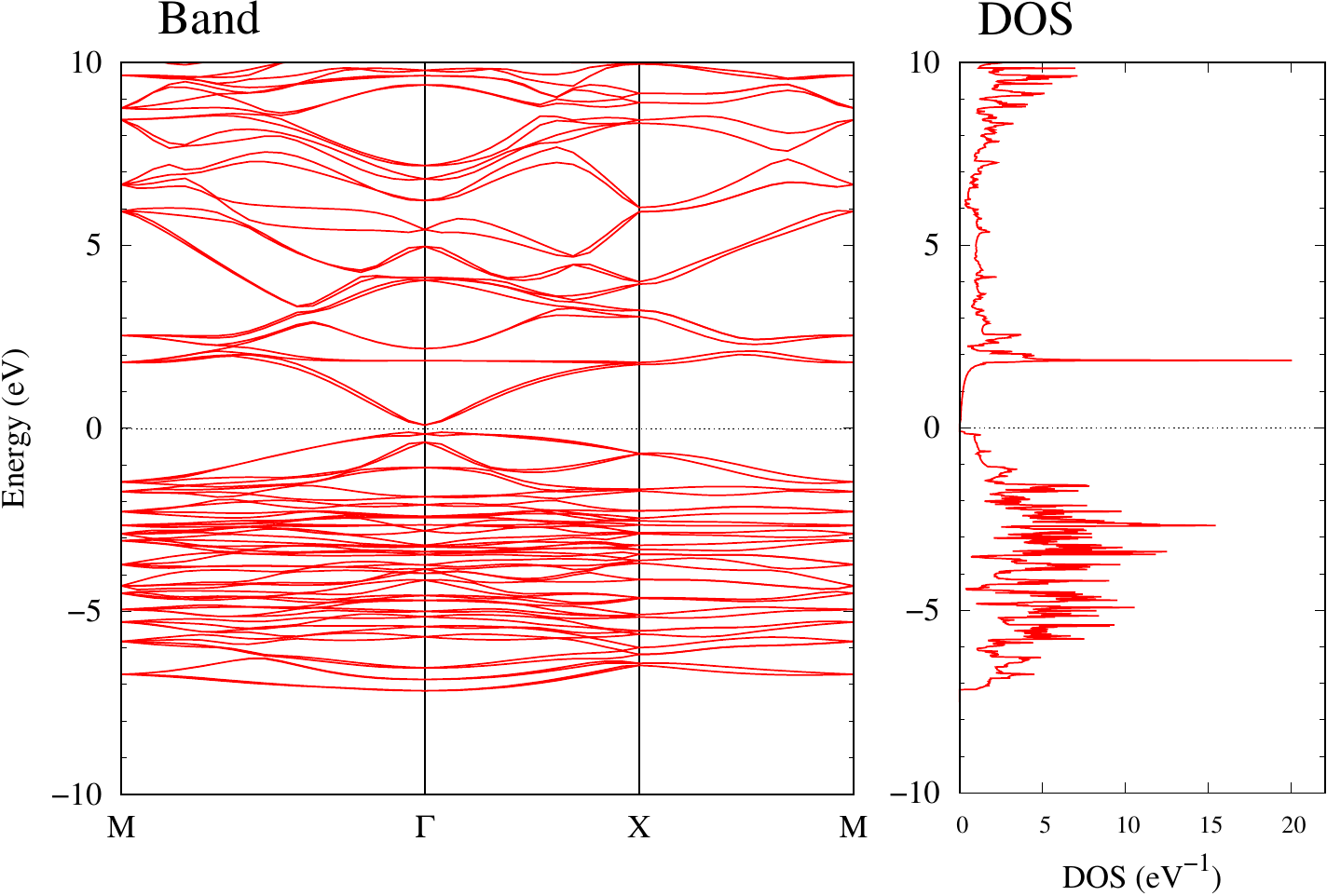}
          \subcaption{\textit{$\theta$}-Au$_2$S (rel)}
  \end{minipage}
  \begin{minipage}[b]{0.45\linewidth}
    \includegraphics[width=\linewidth]{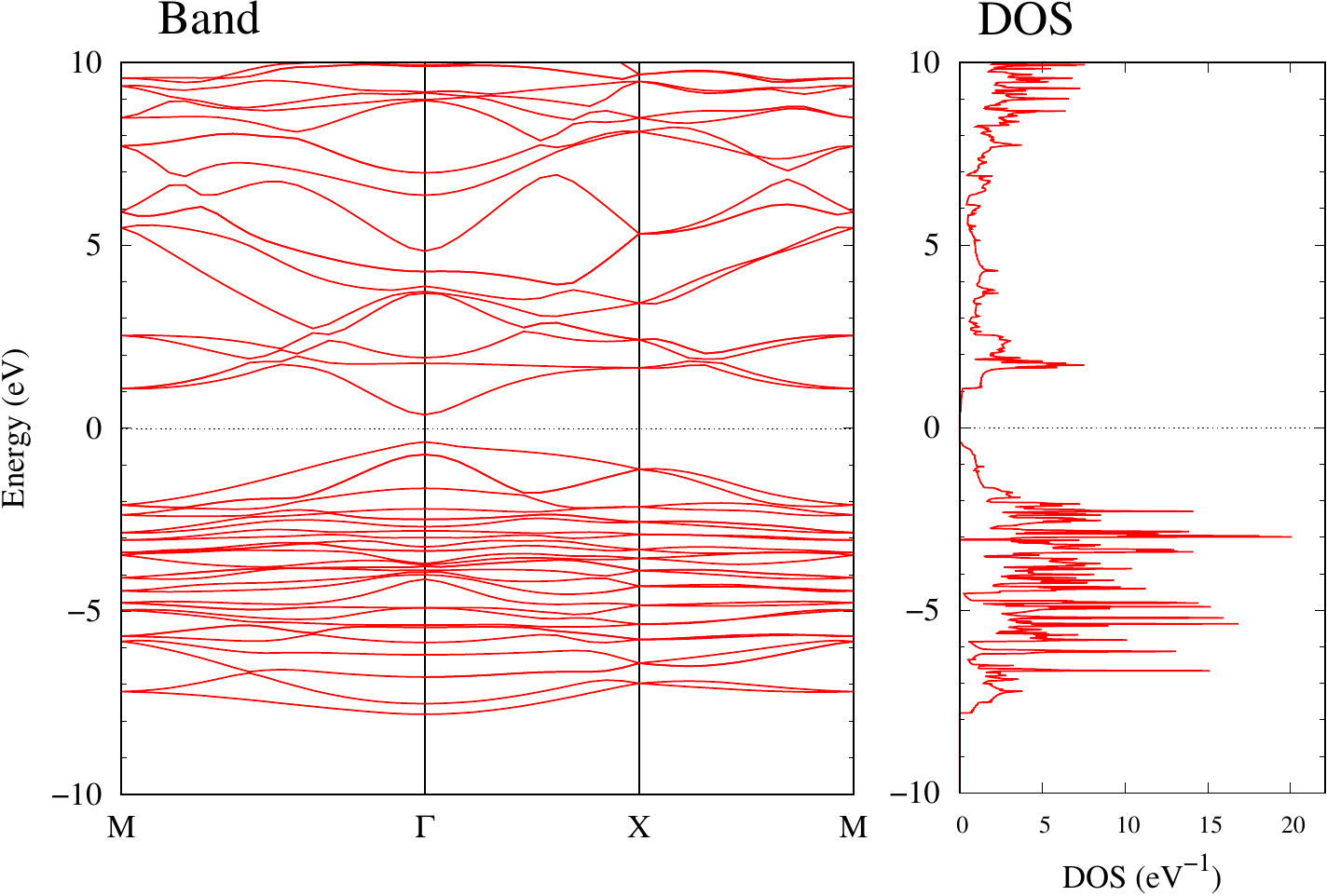}
          \subcaption{\textit{$\eta$}-Au$_2$S (rel)}
  \end{minipage}
  \caption{ Electronic band structures and the DOSs of Au$_2$S monolayers
  based on non-relativistic DFT (a,b) and that with spin-orbit coupling term (c,d).
  Non-relativistic DFT calculations also include relativistic effect in the pseudopotentials.
  The $y$ axis is taken so that the Fermi energy is zero.}\label{SI_SI_fig:Au2S_eband}
\end{figure*}

\begin{figure*}[htb]
  \begin{minipage}[b]{0.30\linewidth}
    \includegraphics[width=\linewidth]{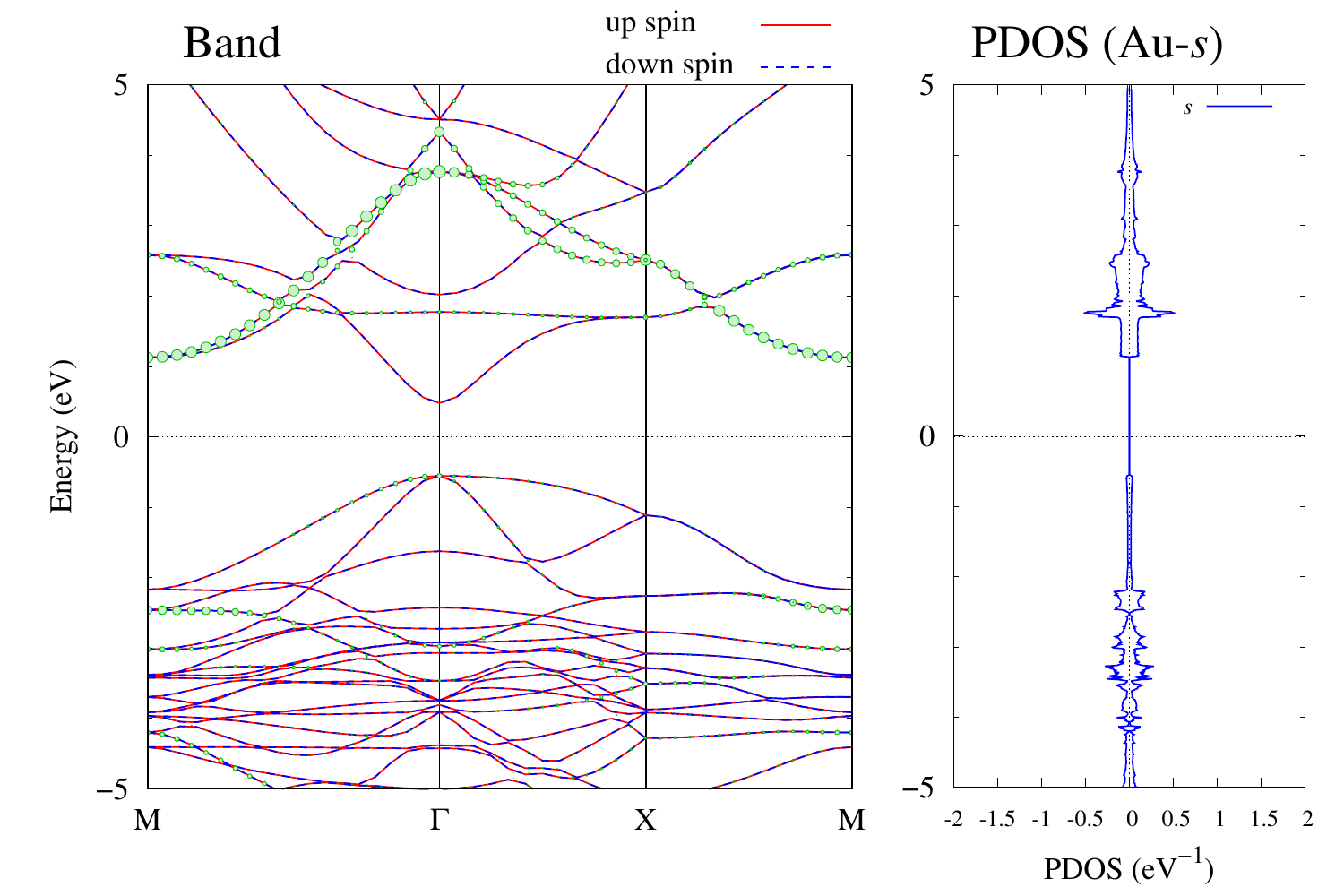}
          \subcaption{Au-\textit{s}}
  \end{minipage}
  \begin{minipage}[b]{0.30\linewidth}
    \includegraphics[width=\linewidth]{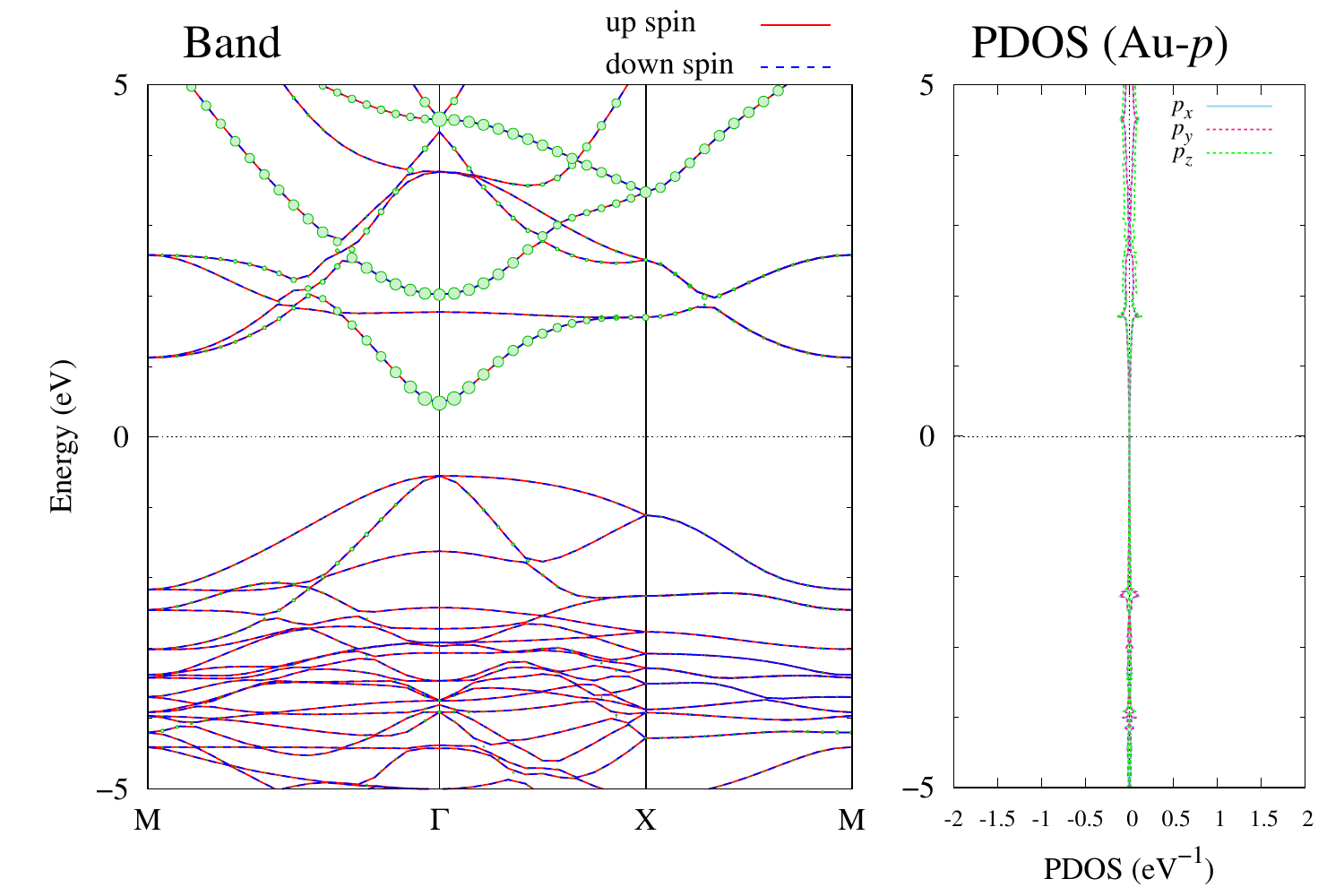}
          \subcaption{Au-\textit{p}}
  \end{minipage}
   \begin{minipage}[b]{0.30\linewidth}
    \includegraphics[width=\linewidth]{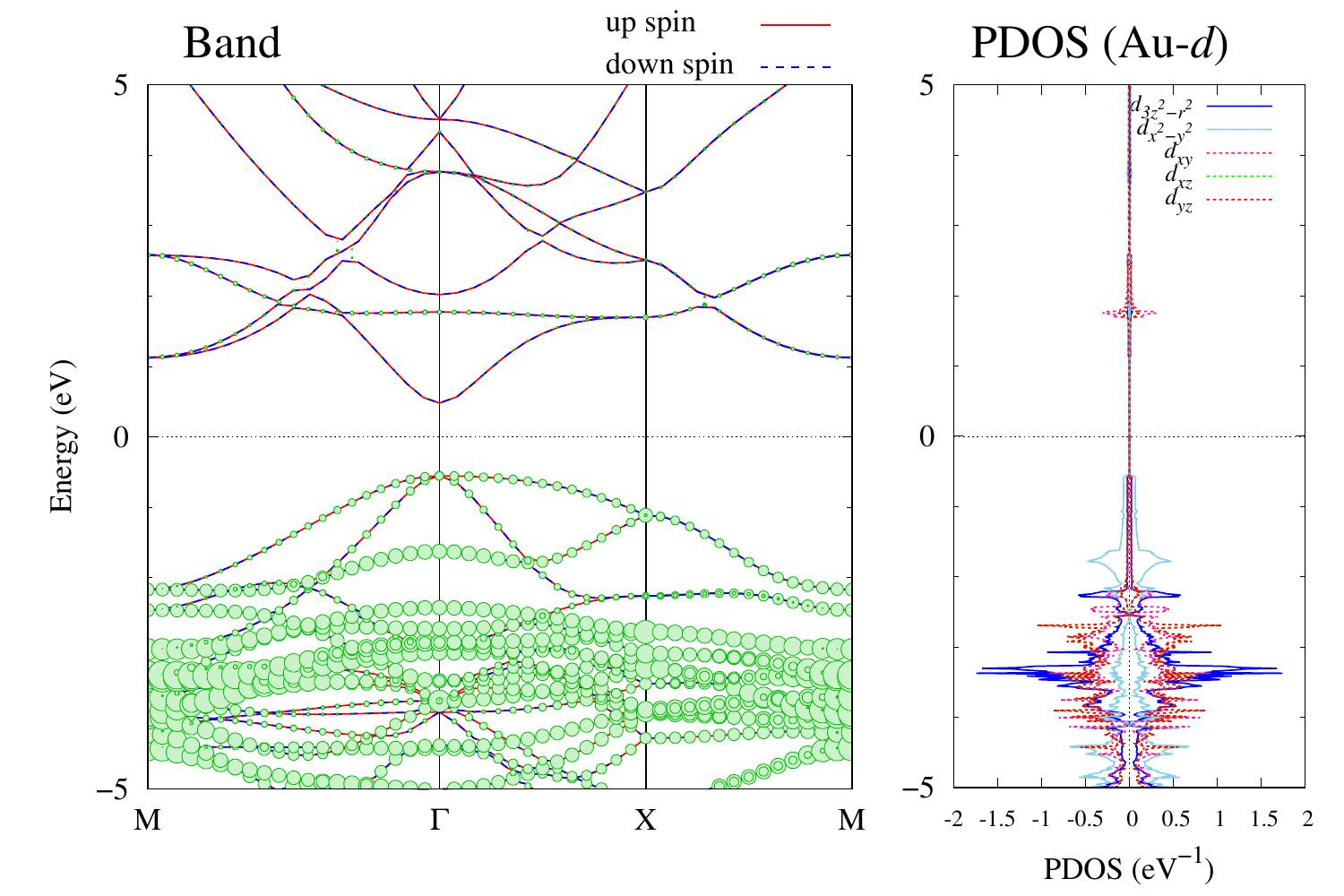}
          \subcaption{Au-\textit{d}}
  \end{minipage}
   \\
   \begin{minipage}[b]{0.30\linewidth}
    \includegraphics[width=\linewidth]{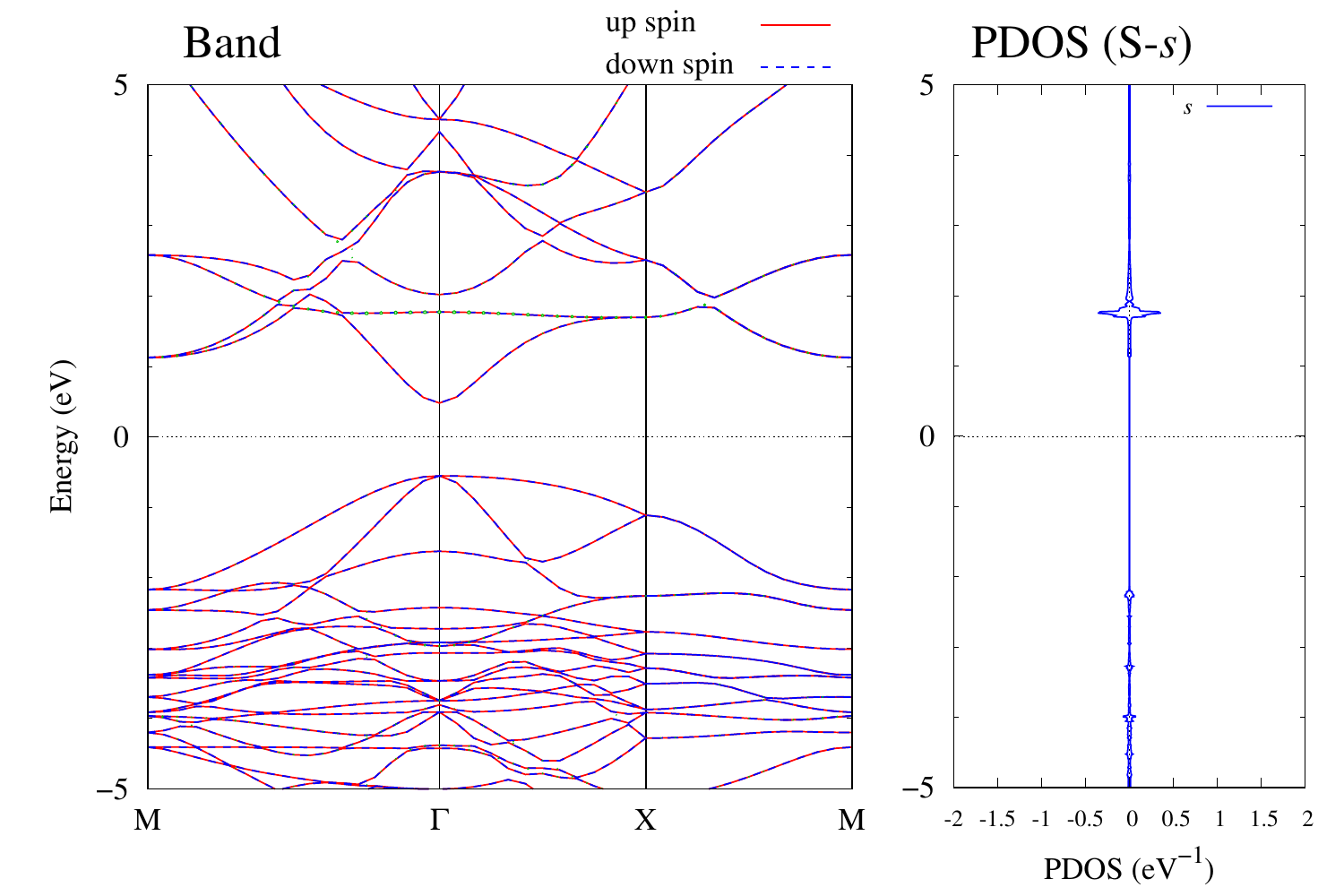}
          \subcaption{S-\textit{s}}
  \end{minipage}
  \begin{minipage}[b]{0.30\linewidth}
    \includegraphics[width=\linewidth]{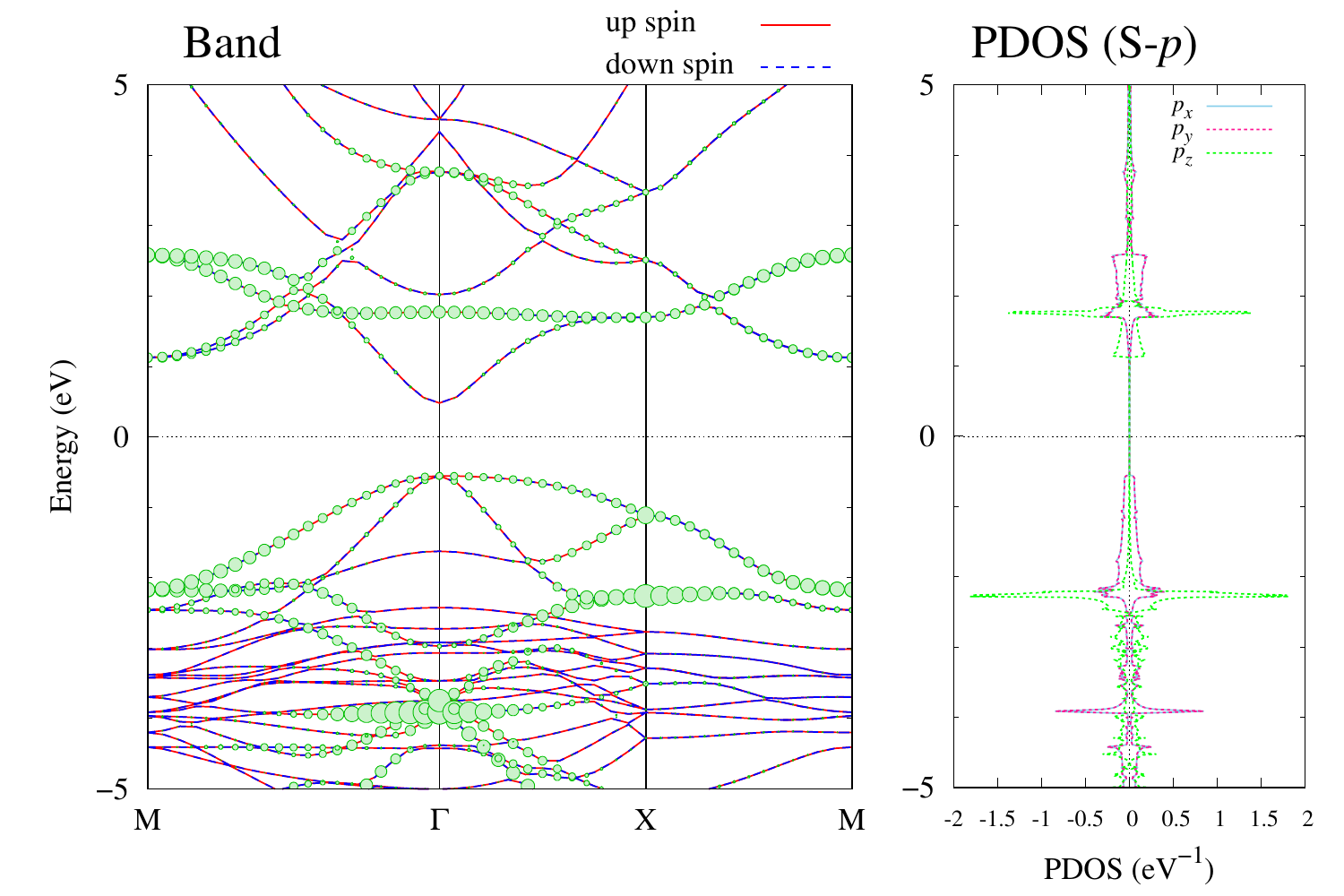}
          \subcaption{S-\textit{p}}
  \end{minipage}
 
 \caption{Weight of the electronic band and the partial DOS of Au and S atoms of \textit{buckled-$\eta$}-Au$_2$S monolayers. The weight is projected on \textit{s}, \textit{p} and \textit{d} type functions.}\label{SI_fig:Au2S_eband_unfolding_eta}
\end{figure*}

\begin{figure*}[htb]
  \begin{minipage}[b]{0.30\linewidth}
    \includegraphics[width=\linewidth]{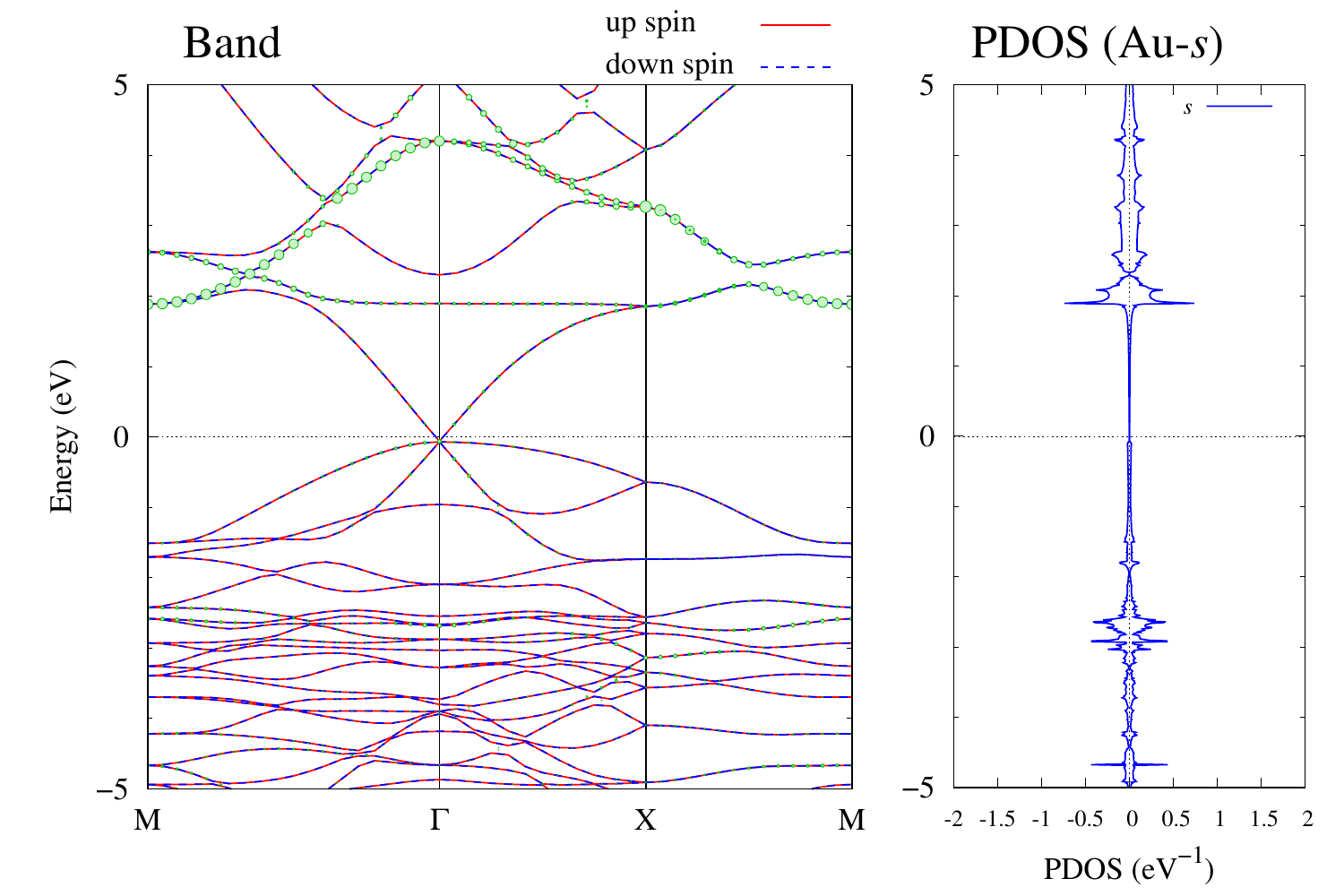}
          \subcaption{Au-\textit{s}}
  \end{minipage}
  \begin{minipage}[b]{0.30\linewidth}
    \includegraphics[width=\linewidth]{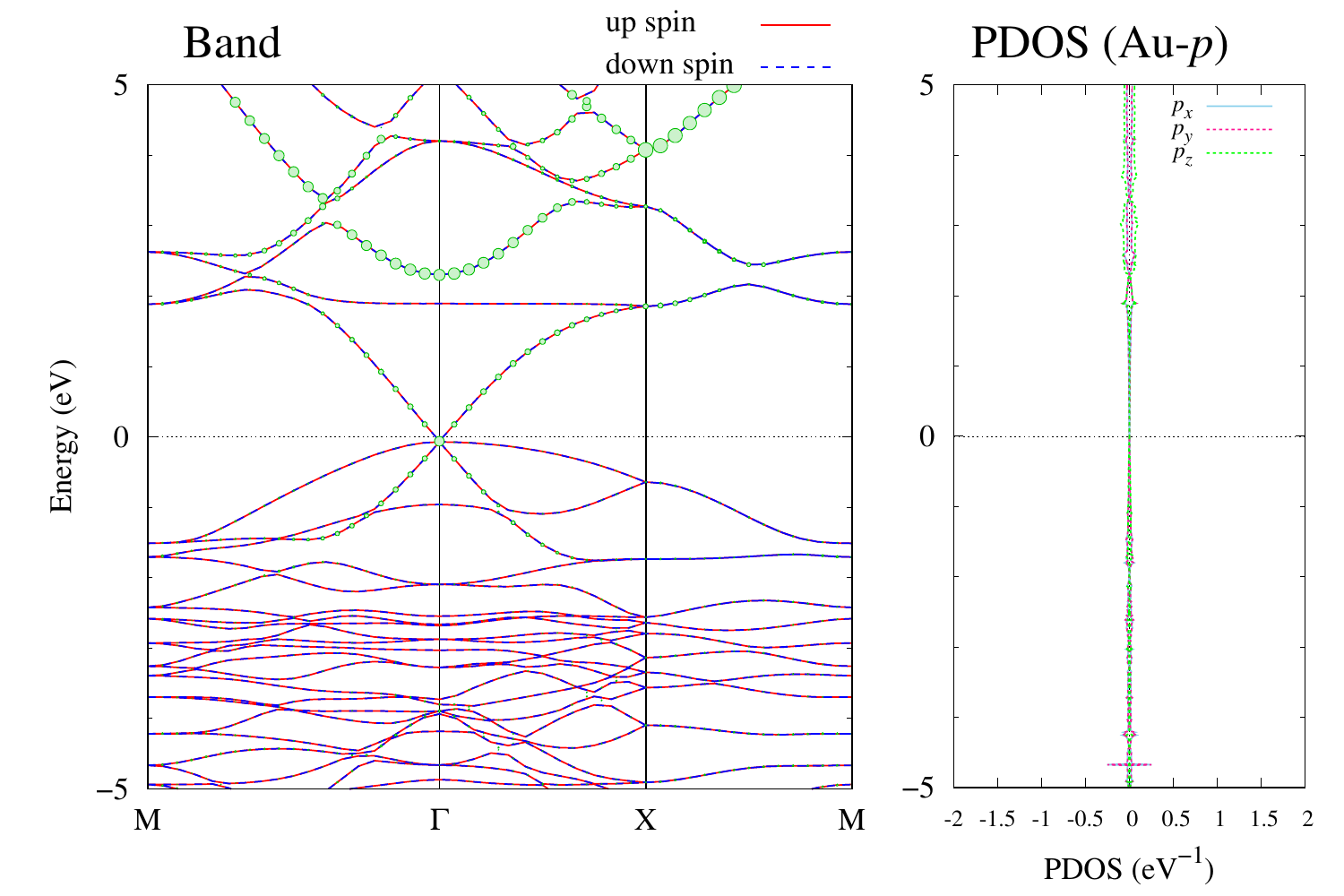}
          \subcaption{Au-\textit{p}}
  \end{minipage}
   \begin{minipage}[b]{0.30\linewidth}
    \includegraphics[width=\linewidth]{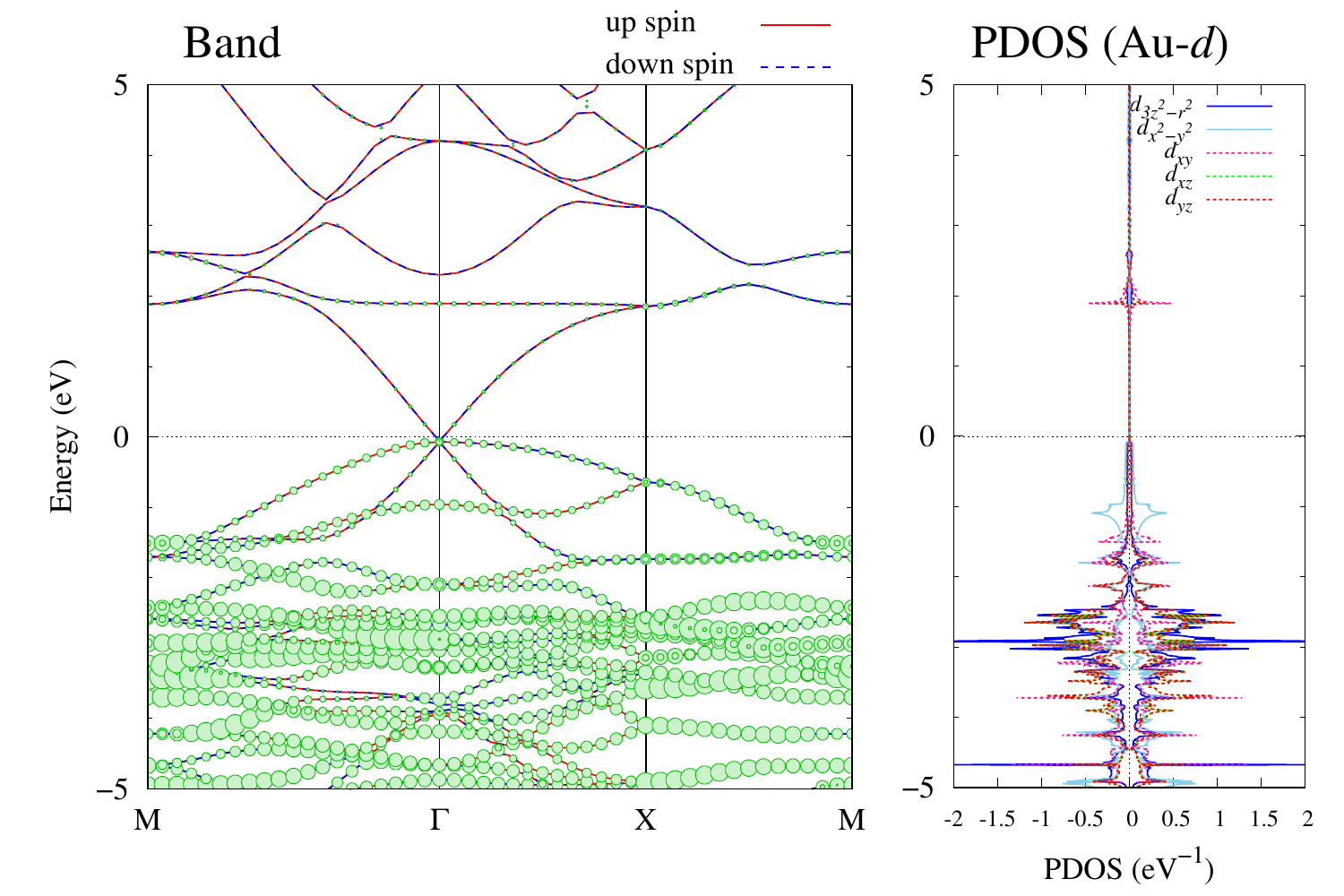}
          \subcaption{Au-\textit{d}}
  \end{minipage}
  \\
   \begin{minipage}[b]{0.30\linewidth}
    \includegraphics[width=\linewidth]{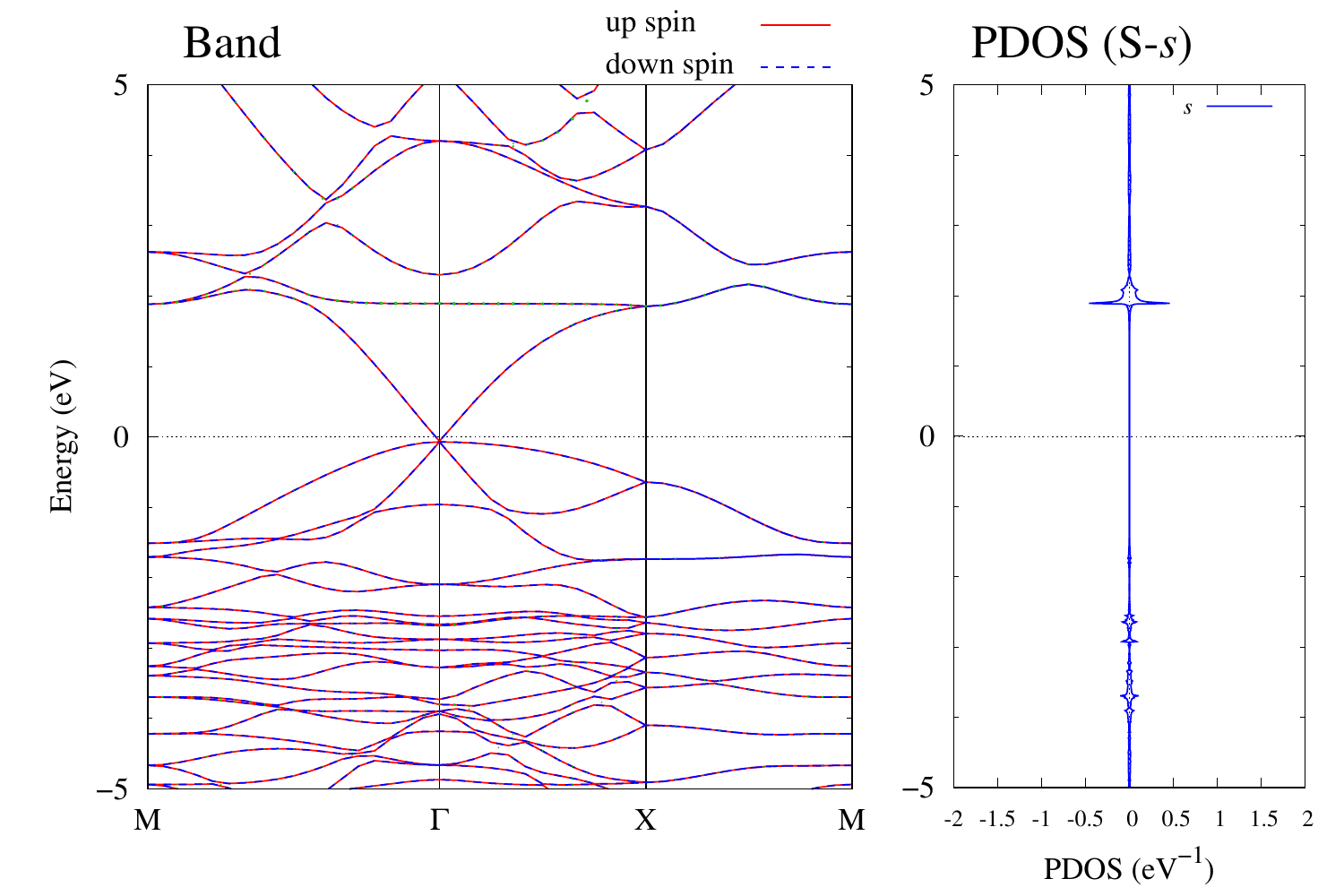}
          \subcaption{S-\textit{s}}
  \end{minipage}
  \begin{minipage}[b]{0.30\linewidth}
    \includegraphics[width=\linewidth]{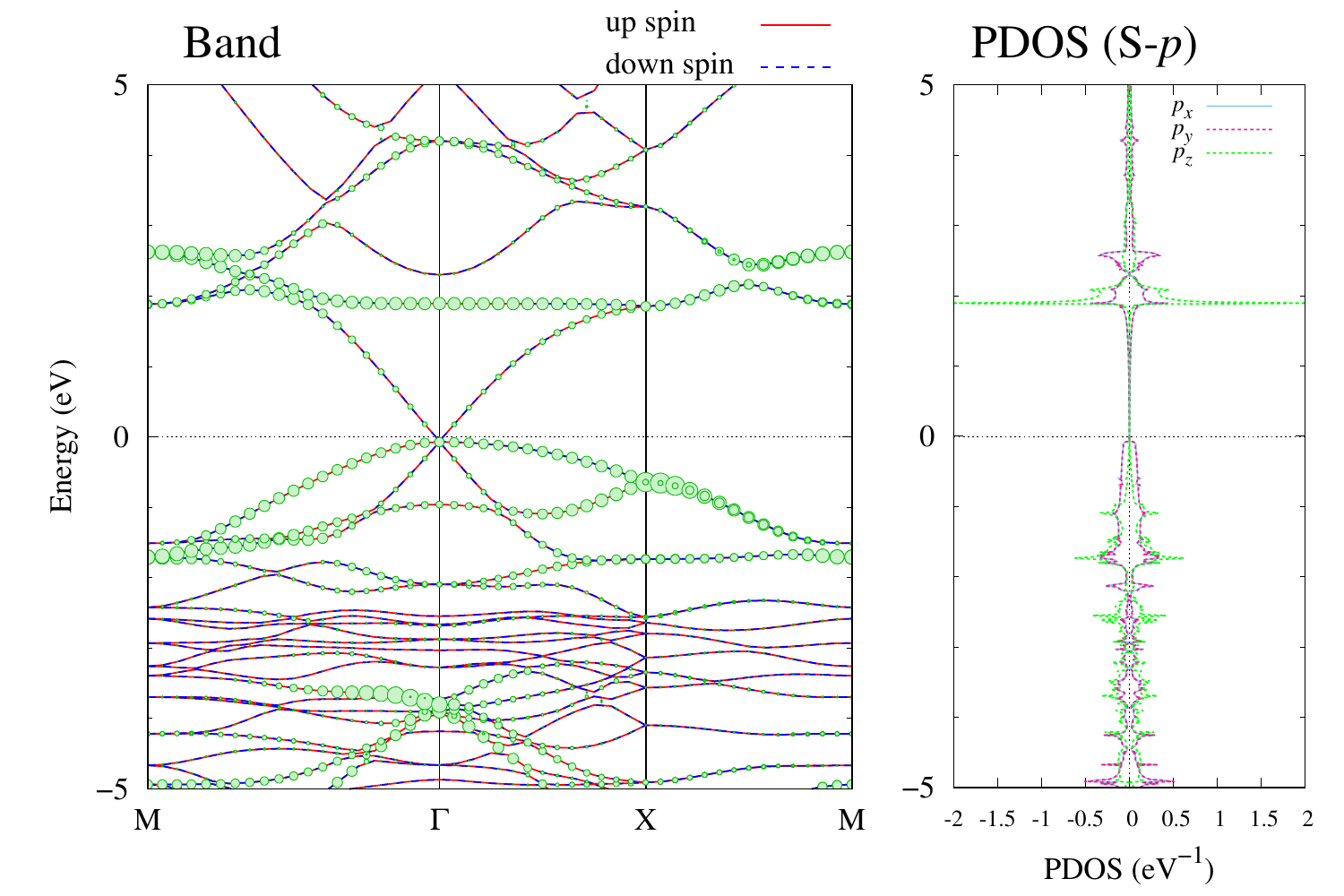}
          \subcaption{S-\textit{p}}
  \end{minipage}
 
 \caption{Weight of the electronic band and the partial DOS of Au and S atoms of \textit{buckled-$\theta$}-Au$_2$S monolayers. The weight is projected on \textit{s}, \textit{p} and \textit{d} type functions.}\label{SI_fig:Au2S_eband_unfolding_theta}
\end{figure*}

\begin{table*}[h]
\centering
\caption{Function types for each band.}
\label{tab:band-function-types}
\begin{tabular}{cc}
\hline
Band & Function Types \\ \hline
CB2           & Au-$s$,$d$, S-$p_z$        \\
CB1           & Au-$p_x$,$p_y$               \\
VB1           & Au-$d$, S-$p_x$,$p_y$       \\
VB2           & Au-$d$, S-$p_x$,$p_y$       \\ \hline
\end{tabular}
\end{table*}

\begin{figure*}[htb]
    \includegraphics[width=\linewidth]{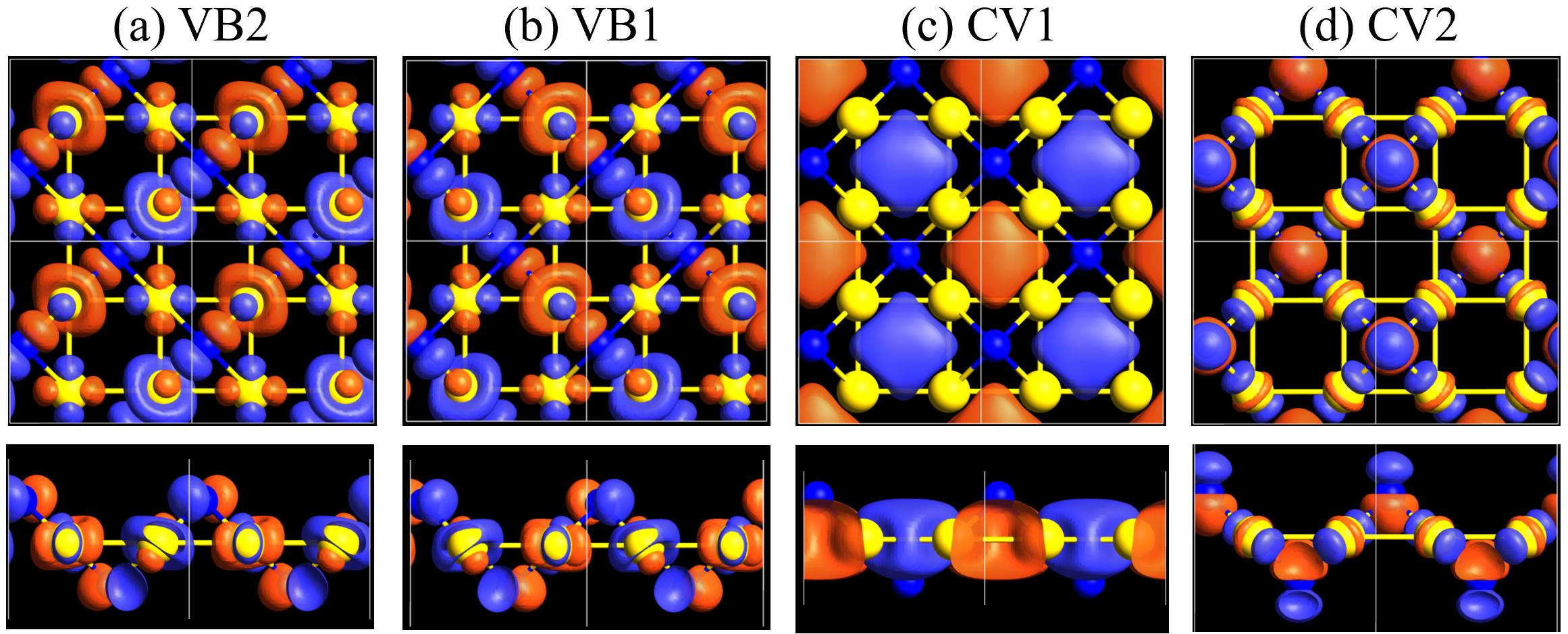}
 \caption{Molecular orbitals (MO) of VB2, VB1, CV1, and CV2 of \textit{buckled-$\eta$}-Au$_2$S monolayers at $\Gamma$ point. The threshold value of isosurfaces of the MOs is taken as 0.04 a.u. .}\label{SI_fig:eta_au2s_MO}
\end{figure*}

\begin{figure*}[htb]
    \includegraphics[width=\linewidth]{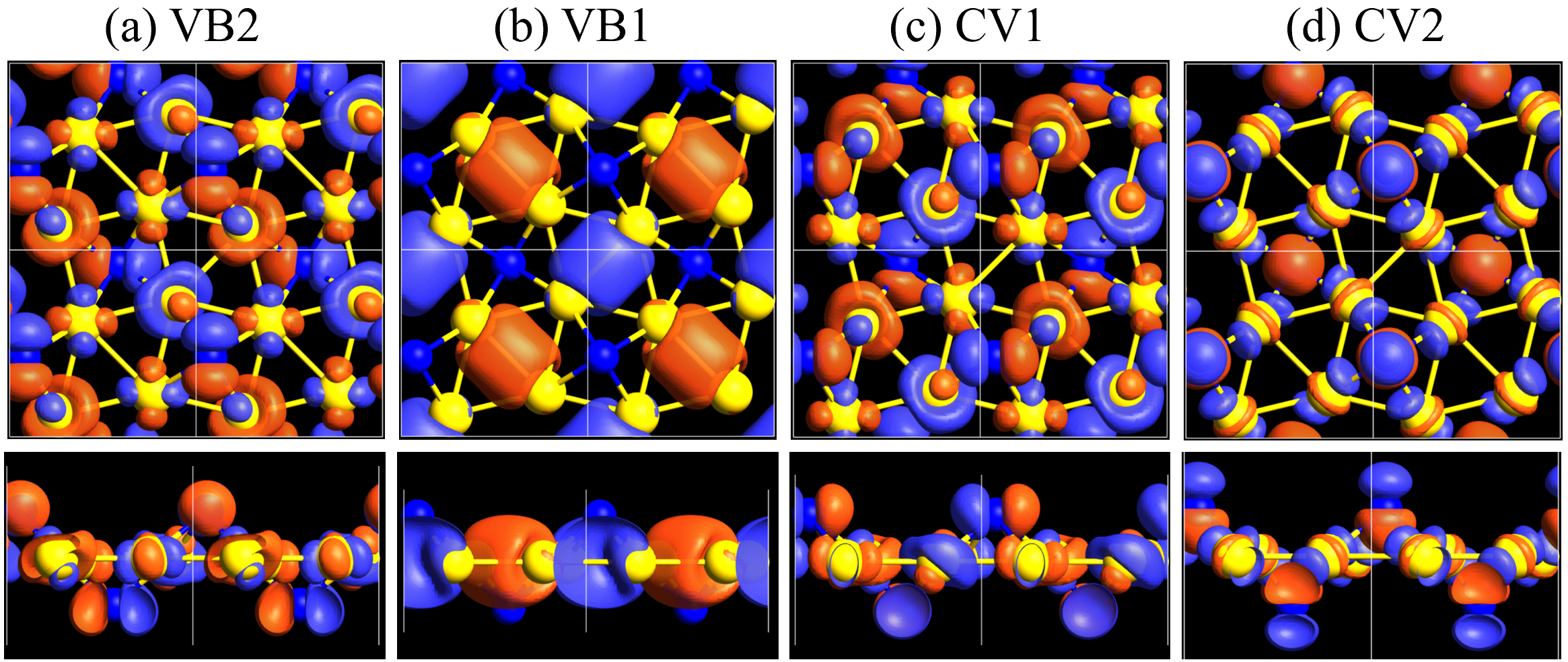}
 \caption{Molecular orbitals (MO) of VB2, VB1, CV1, and CV2 of \textit{buckled-$\theta$}-Au$_2$S monolayers at $\Gamma$ point. The threshold value of isosurfaces of the MOs is taken as 0.04 a.u. .
 The CB1, VB1 and VB2 are almost degenerate at $\Gamma$ point.}\label{SI_fig:theta_au2s_MO}
\end{figure*}

\clearpage

\begin{figure*}[htb]
  \begin{minipage}[b]{0.30\linewidth}
    \includegraphics[width=\linewidth]{\PATHFIG/phonon_buckled_eta_Au2S.pdf}
          \subcaption{Au$_2$S}
  \end{minipage}
  \begin{minipage}[b]{0.30\linewidth}
    \includegraphics[width=\linewidth]{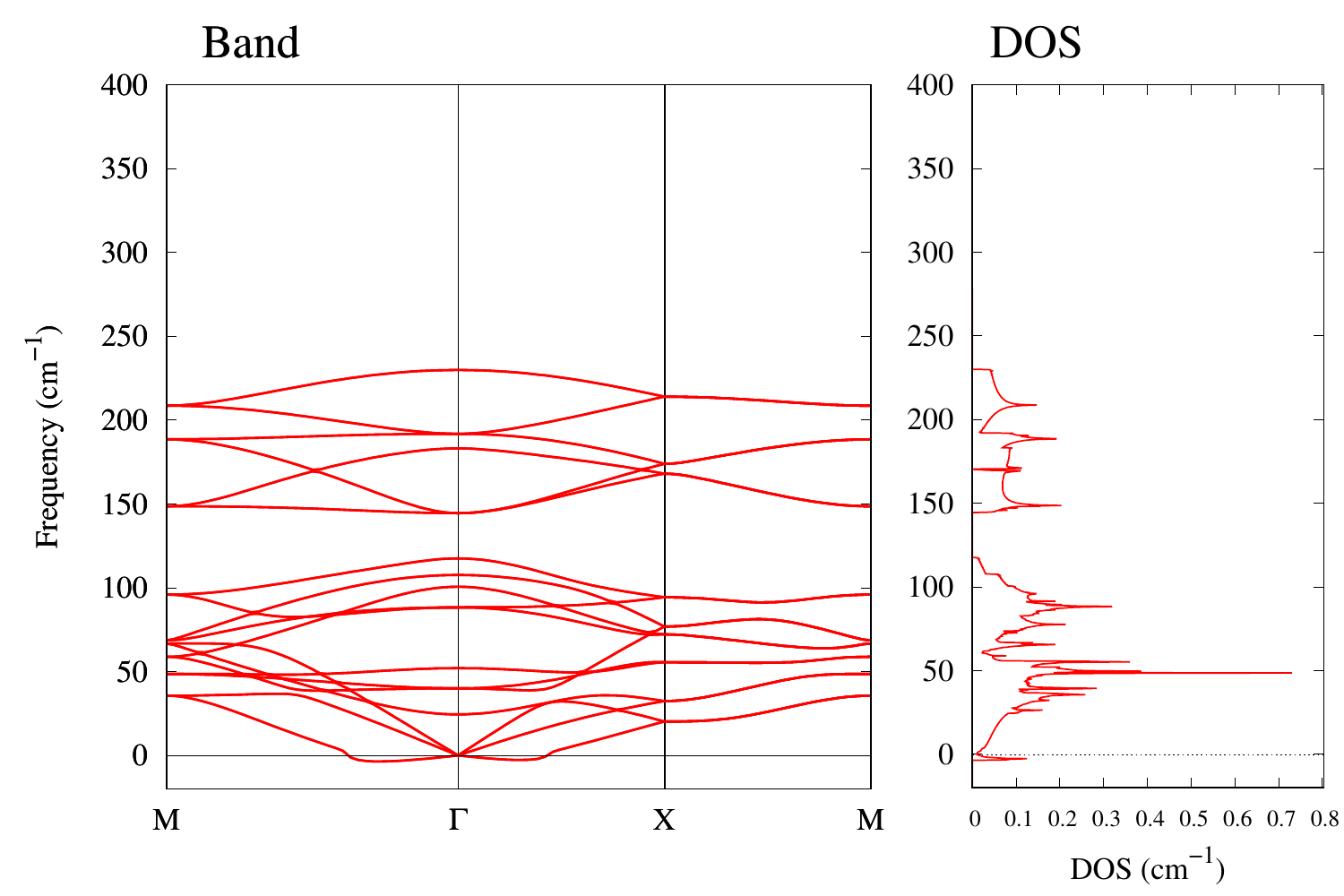}
          \subcaption{Au$_2$Se}
  \end{minipage}
   \begin{minipage}[b]{0.30\linewidth}
    \includegraphics[width=\linewidth]{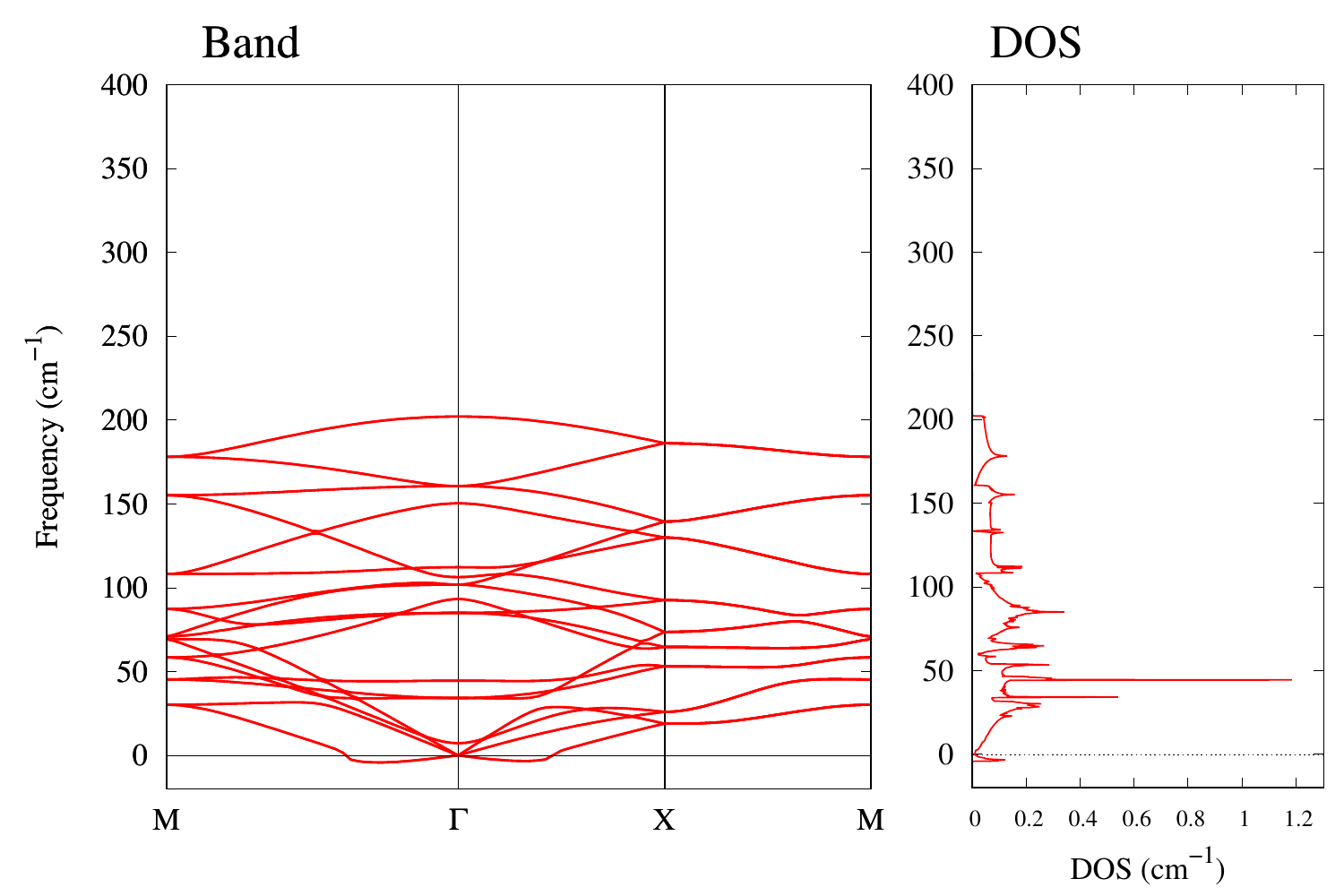}
          \subcaption{Au$_2$Te}
  \end{minipage}
  \\
   \begin{minipage}[b]{0.30\linewidth}
    \includegraphics[width=\linewidth]{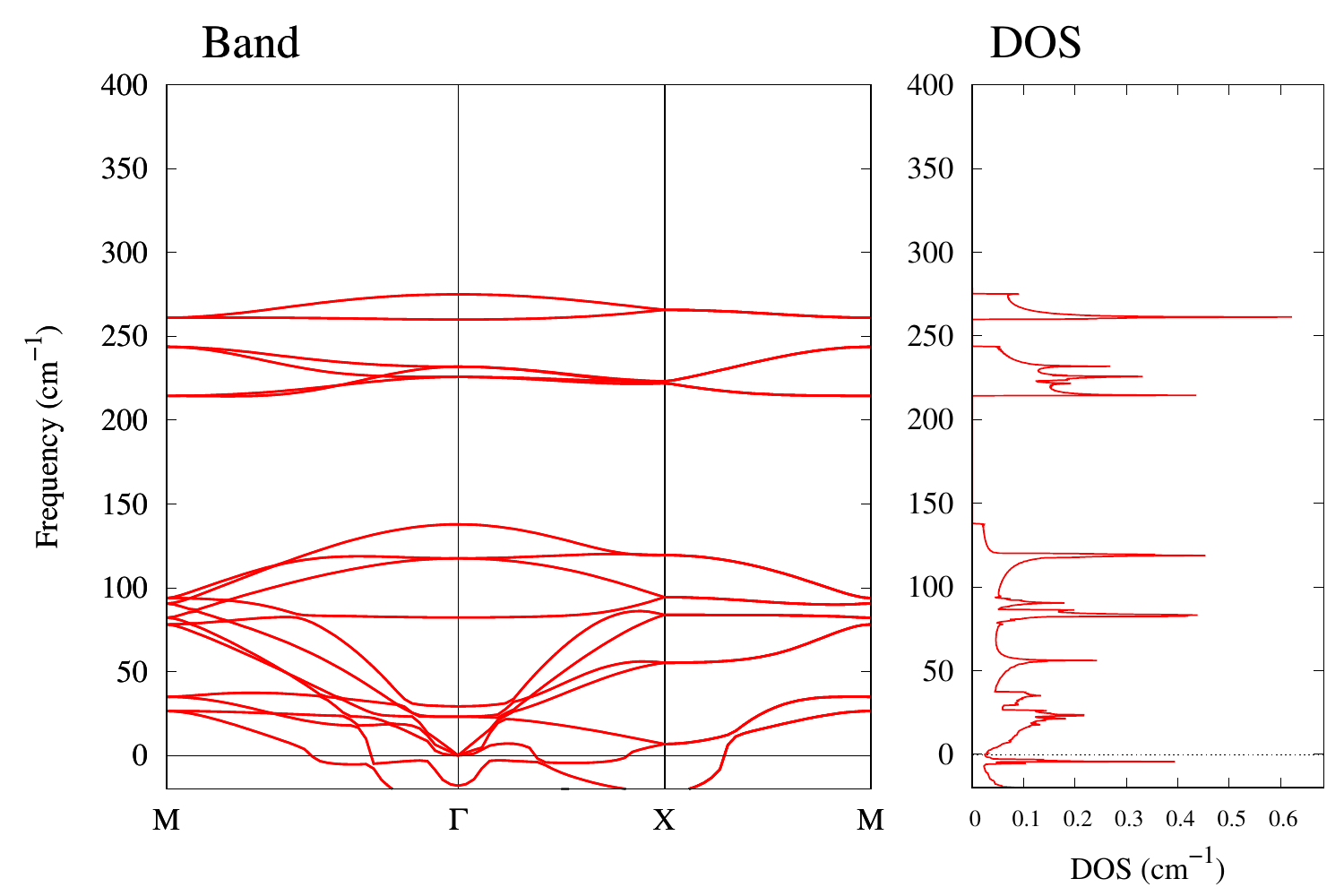}
          \subcaption{Au$_2$Si}
  \end{minipage}
  \begin{minipage}[b]{0.30\linewidth}
    \includegraphics[width=\linewidth]{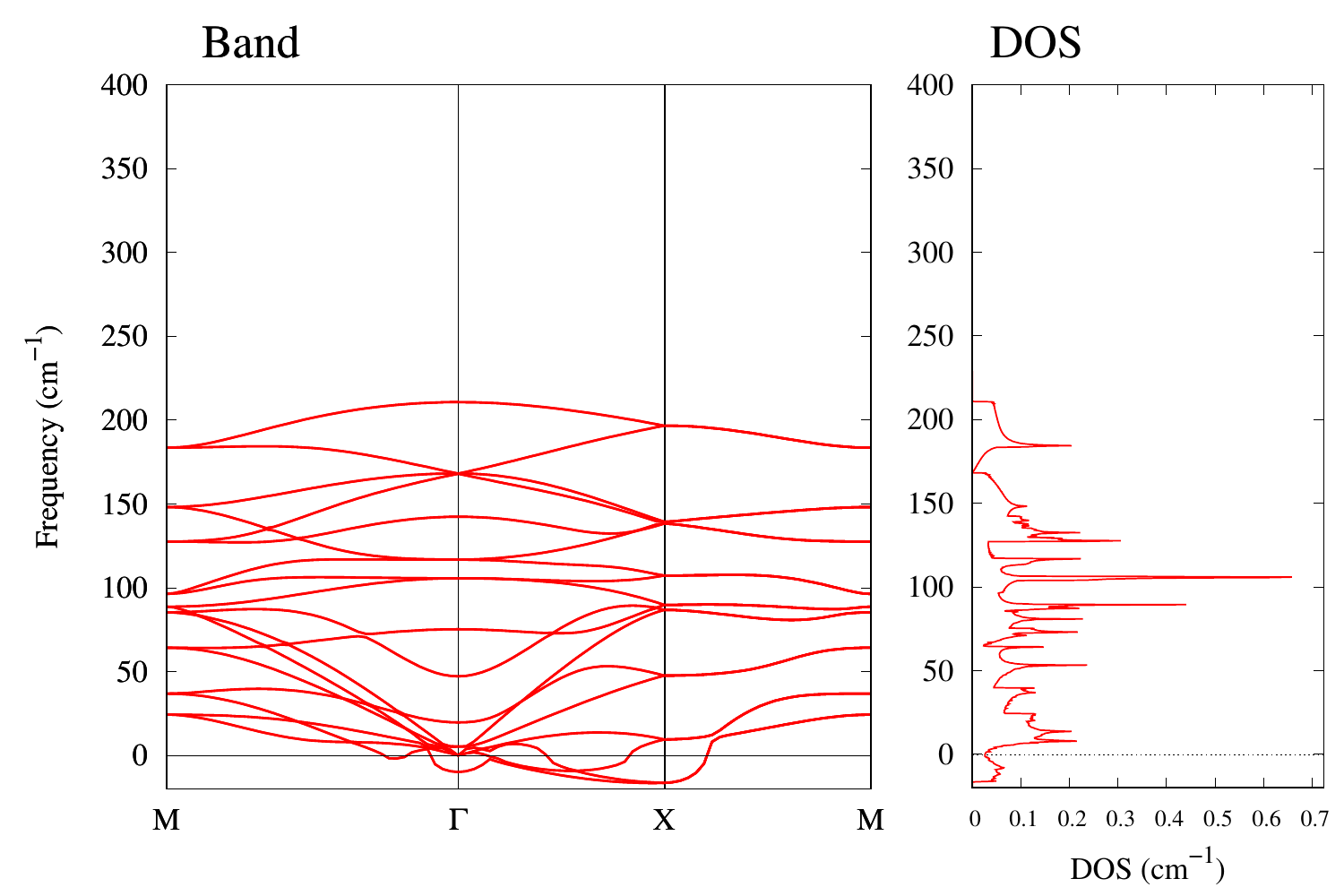}
          \subcaption{Au$_2$Ge}
  \end{minipage}
 
 \caption{Phonon band and the DOS of \textit{buckled-$\eta$}-Au$_2$X (X=S, Se, Te, Si, Ge).}\label{SI_fig:buckled_etaAu2X_phonon}
\end{figure*}

\begin{figure*}[htb]
  \begin{minipage}[b]{0.30\linewidth}
    \includegraphics[width=\linewidth]{\PATHFIG/phonon_buckled_theta_Au2S.pdf}
          \subcaption{Au$_2$S}
  \end{minipage}
  \begin{minipage}[b]{0.30\linewidth}
    \includegraphics[width=\linewidth]{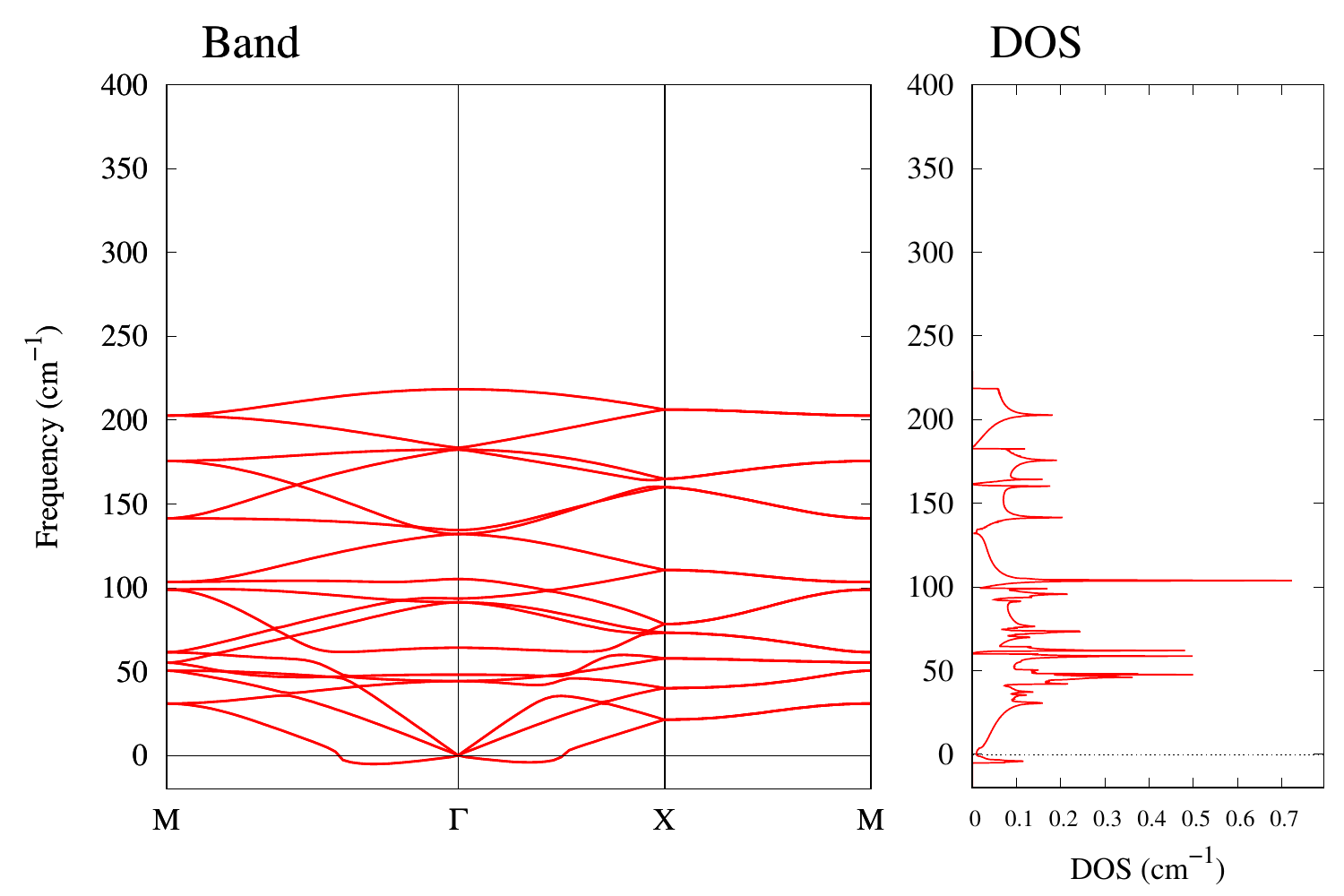}
          \subcaption{Au$_2$Se}
  \end{minipage}
   \begin{minipage}[b]{0.30\linewidth}
    \includegraphics[width=\linewidth]{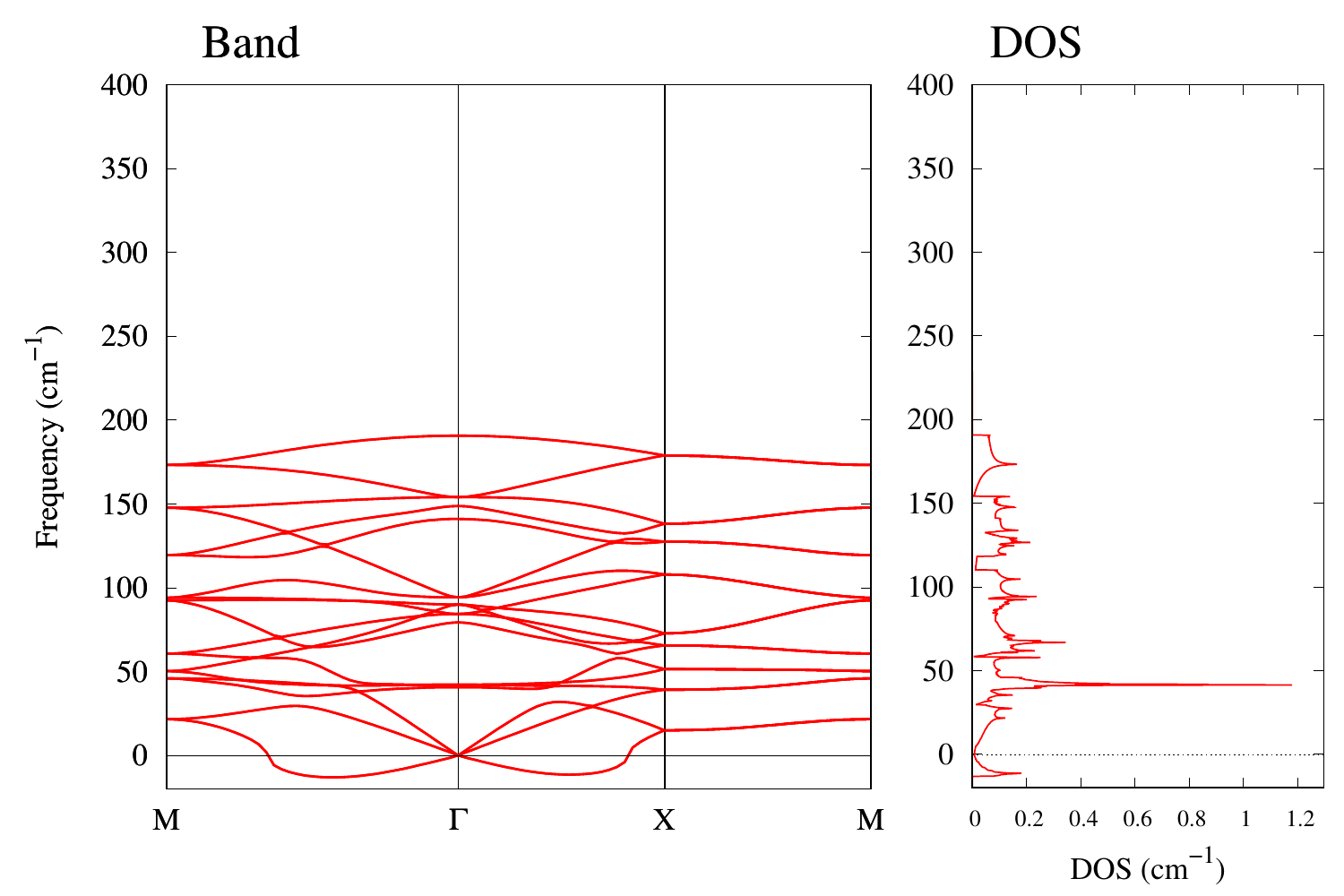}
          \subcaption{Au$_2$Te}
  \end{minipage}
  \\
   \begin{minipage}[b]{0.30\linewidth}
    \includegraphics[width=\linewidth]{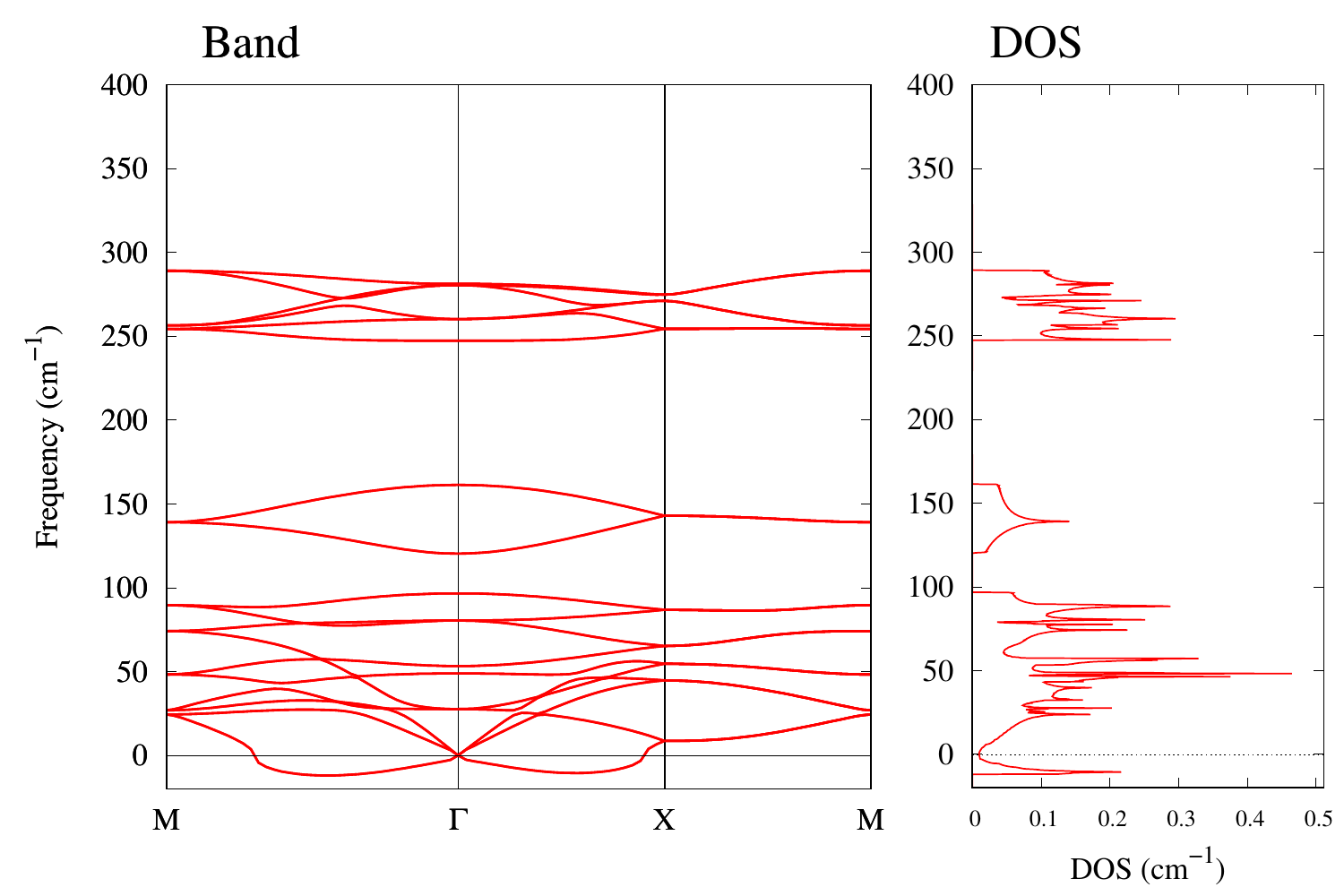}
          \subcaption{Au$_2$Si}
  \end{minipage}
  \begin{minipage}[b]{0.30\linewidth}
    \includegraphics[width=\linewidth]{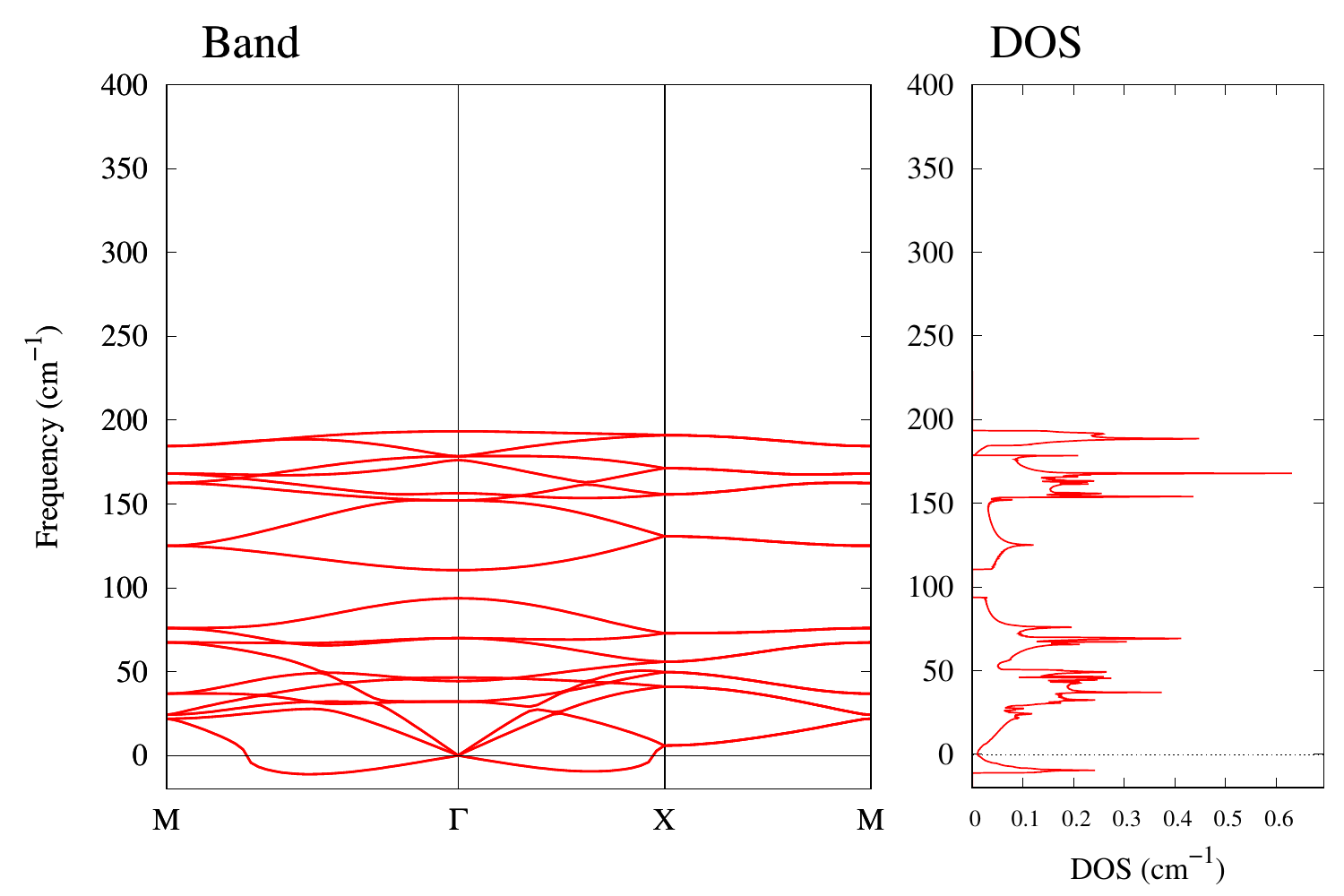}
          \subcaption{Au$_2$Ge}
  \end{minipage}
 
 \caption{Phonon band and the DOS of \textit{buckled-$\theta$}-Au$_2$X (X=S, Se, Te, Si, Ge).}\label{SI_fig:buckled_theta_Au2X_phonon}
\end{figure*}

\begin{figure*}[htb]
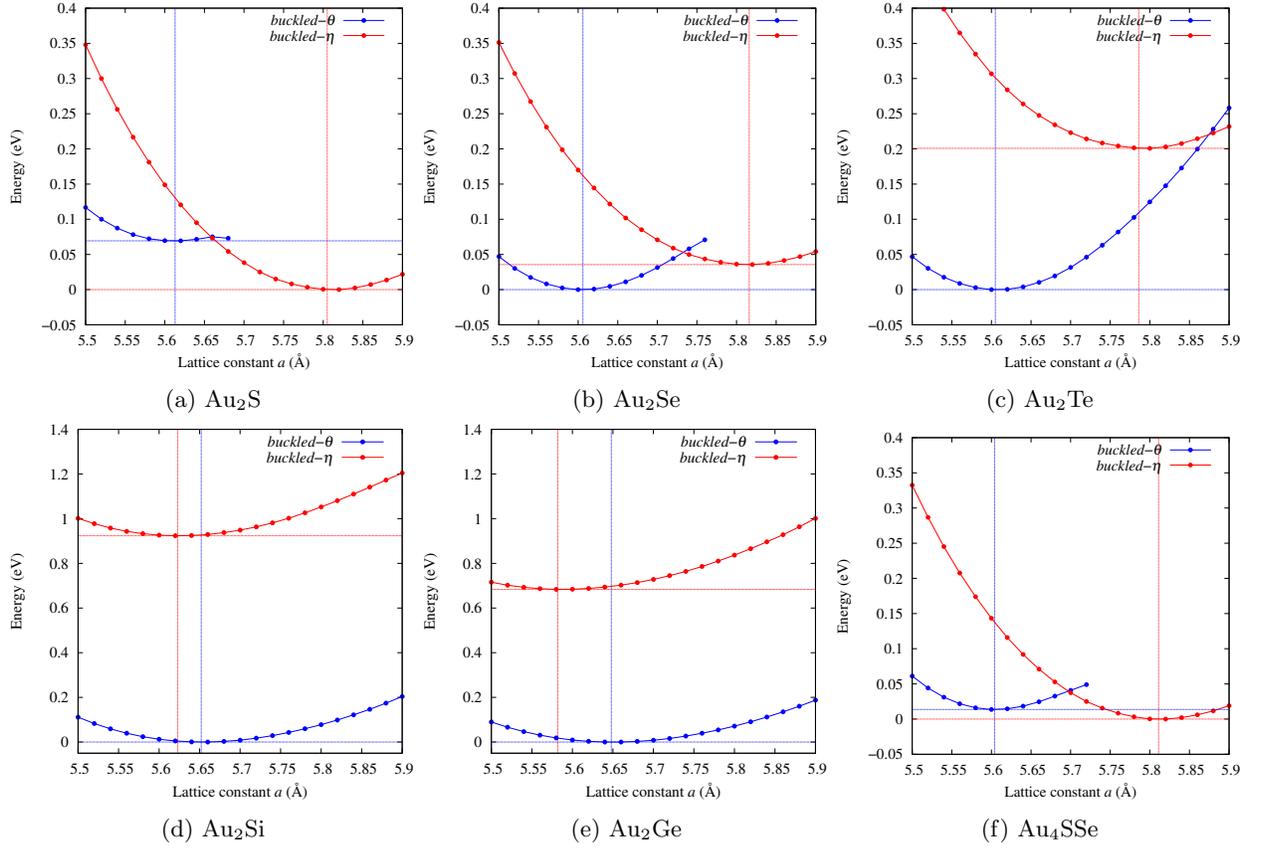

  \begin{minipage}[b]{0.30\linewidth}
    \includegraphics[width=\linewidth]{\PATHFIG/Au2S_energy.pdf}
          \subcaption{Au$_2$S}
  \end{minipage}
  \begin{minipage}[b]{0.30\linewidth}
    \includegraphics[width=\linewidth]{\PATHFIG/Au2Se_energy.pdf}
          \subcaption{Au$_2$Se}
  \end{minipage}
   \begin{minipage}[b]{0.30\linewidth}
    \includegraphics[width=\linewidth]{\PATHFIG/Au2Te_energy.pdf}
          \subcaption{Au$_2$Te}
  \end{minipage}
  \\
   \begin{minipage}[b]{0.30\linewidth}
    \includegraphics[width=\linewidth]{\PATHFIG/Au2Si_energy.pdf}
          \subcaption{Au$_2$Si}
  \end{minipage}
  \begin{minipage}[b]{0.30\linewidth}
    \includegraphics[width=\linewidth]{\PATHFIG/Au2Ge_energy.pdf}
          \subcaption{Au$_2$Ge}
  \end{minipage}
  \begin{minipage}[b]{0.30\linewidth}
    \includegraphics[width=\linewidth]{\PATHFIG/Au4SSe_energy.pdf}
          \subcaption{Au$_4$SSe}
  \end{minipage}
 
 \caption{Energy curve of \textit{buckled-$\eta$}- and \textit{buckled-$\theta$}- Au$_2$X (X=S, Se, Te, Si, Ge) and Au$_4$SSe monolayers.
 The red (blue) line represents the energy curve of $\eta$-phase ($\theta$-phase).
 }\label{SI_fig:Au2X_energy_curve}
\end{figure*}

\begin{figure*}[htb]
  \begin{minipage}[b]{0.48\linewidth}
    \centering
    \includegraphics[width=\linewidth]{\PATHFIG/Au2Se_buckled_theta_band_dos.pdf}
          \subcaption{\textit{buckled-$\theta$}-Au$_2$Se}
  \end{minipage}
  \begin{minipage}[b]{0.48\linewidth}
    \includegraphics[width=\linewidth]{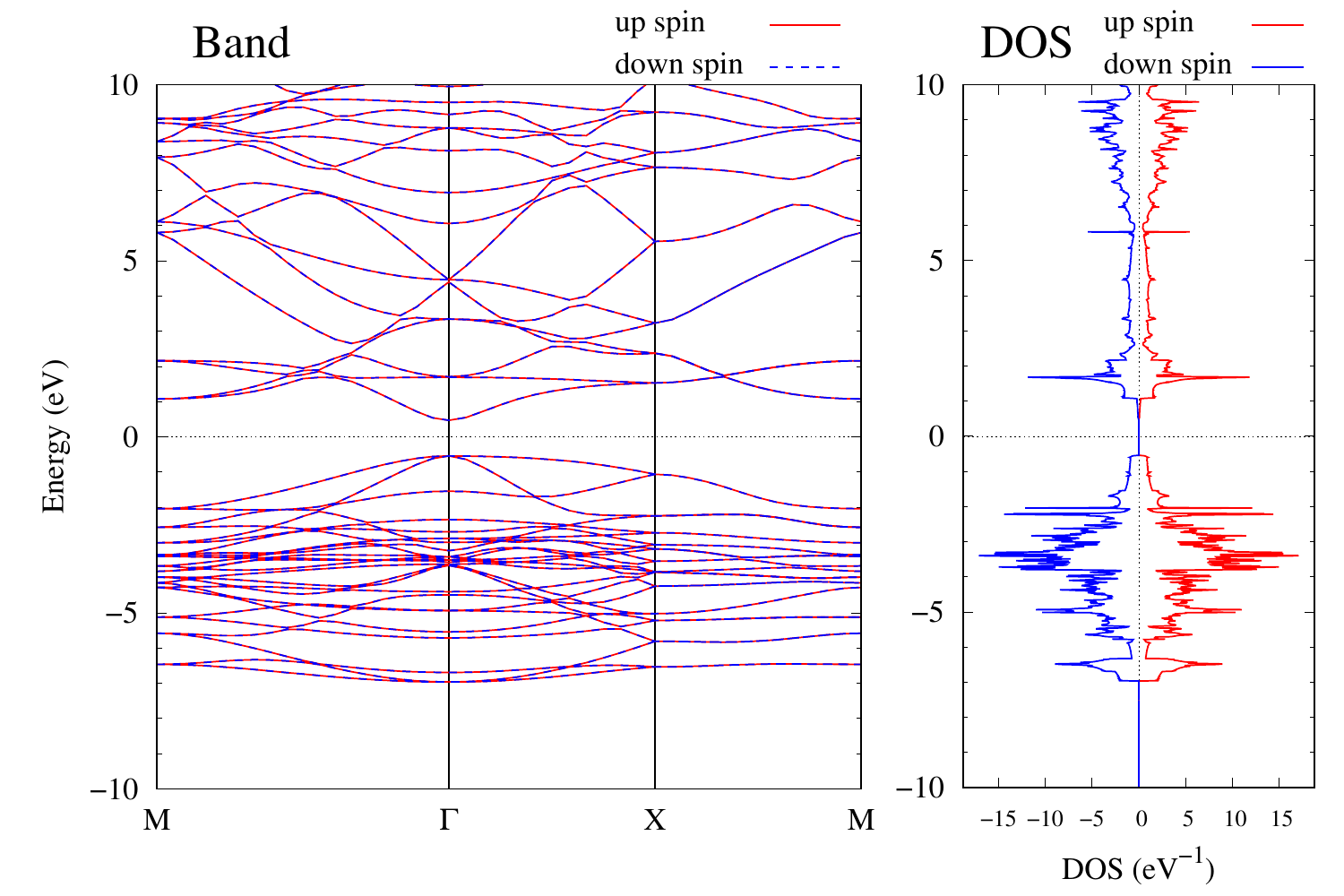}
          \subcaption{\textit{buckled-$\eta$}-Au$_2$Se}
  \end{minipage}

  \vspace{5pt}
  \begin{minipage}[b]{0.48\linewidth}
    \centering
    \includegraphics[width=\linewidth]{\PATHFIG/Au2Te_buckled_theta_band_dos.pdf}
          \subcaption{\textit{buckled-$\theta$}-Au$_2$Te}
  \end{minipage}
  \begin{minipage}[b]{0.48\linewidth}
    \includegraphics[width=\linewidth]{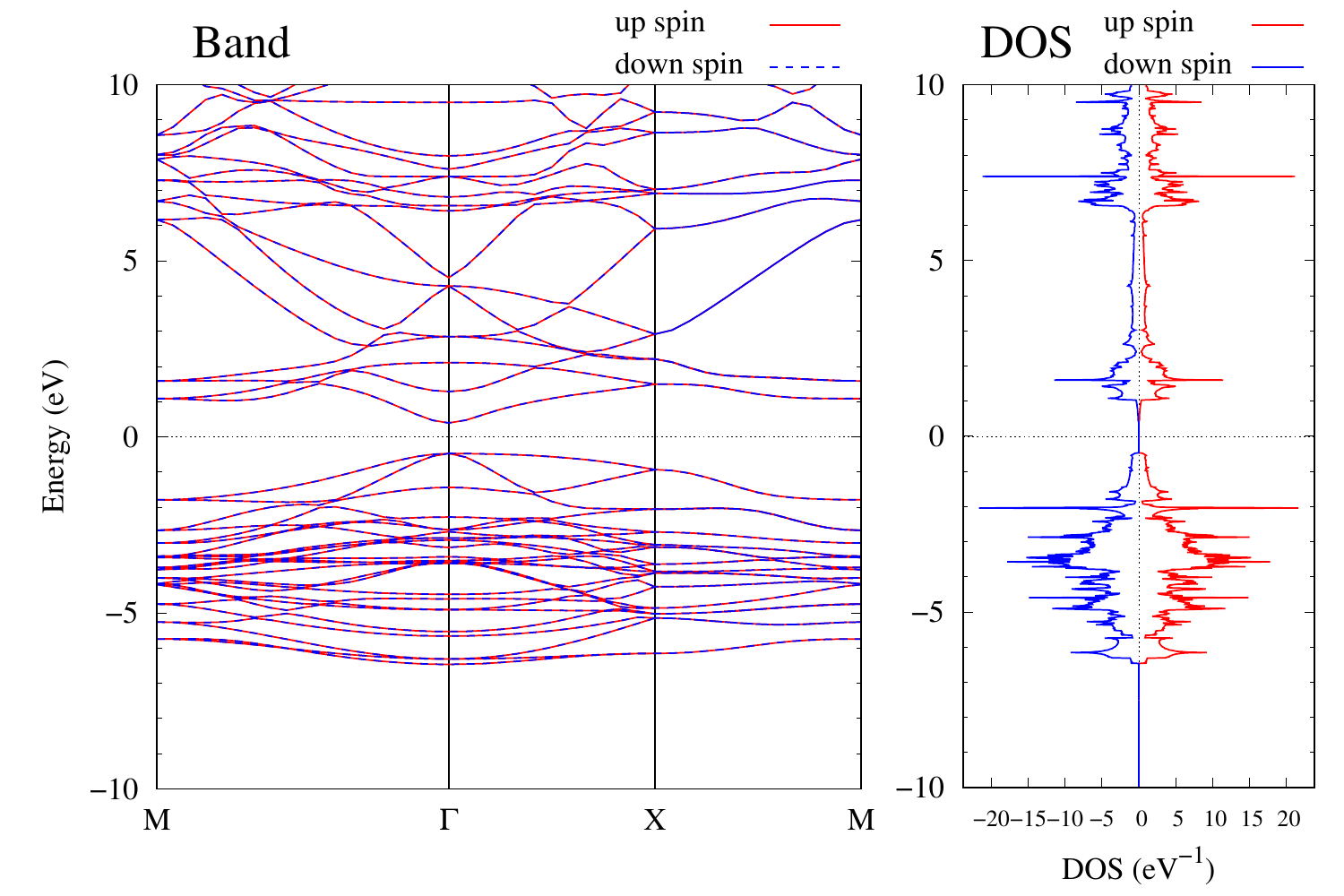}
          \subcaption{\textit{buckled-$\eta$}-Au$_2$Te}
  \end{minipage}
  
  \vspace{5pt}
  \begin{minipage}[b]{0.48\linewidth}
    \centering
    \includegraphics[width=\linewidth]{\PATHFIG/Au4SSe_buckled_theta_band_dos.pdf}
          \subcaption{\textit{buckled-$\theta$}-Au$_4$SSe}
  \end{minipage}
  \begin{minipage}[b]{0.48\linewidth}
    \includegraphics[width=\linewidth]{\PATHFIG/Au4SSe_buckled_eta_band_dos.pdf}
          \subcaption{\textit{buckled-$\eta$}-Au$_4$SSe}
  \end{minipage}
        \caption{Electronic band and the DOS of Au$_2$X (X=Se, Te) and Au$_4$SSe monolayers.}\label{SI_fig:Au2Se_Au2Te_eband}
\end{figure*}

\begin{figure*}[htb]
  \begin{minipage}[b]{0.48\linewidth}
    \centering
    \includegraphics[width=\linewidth]{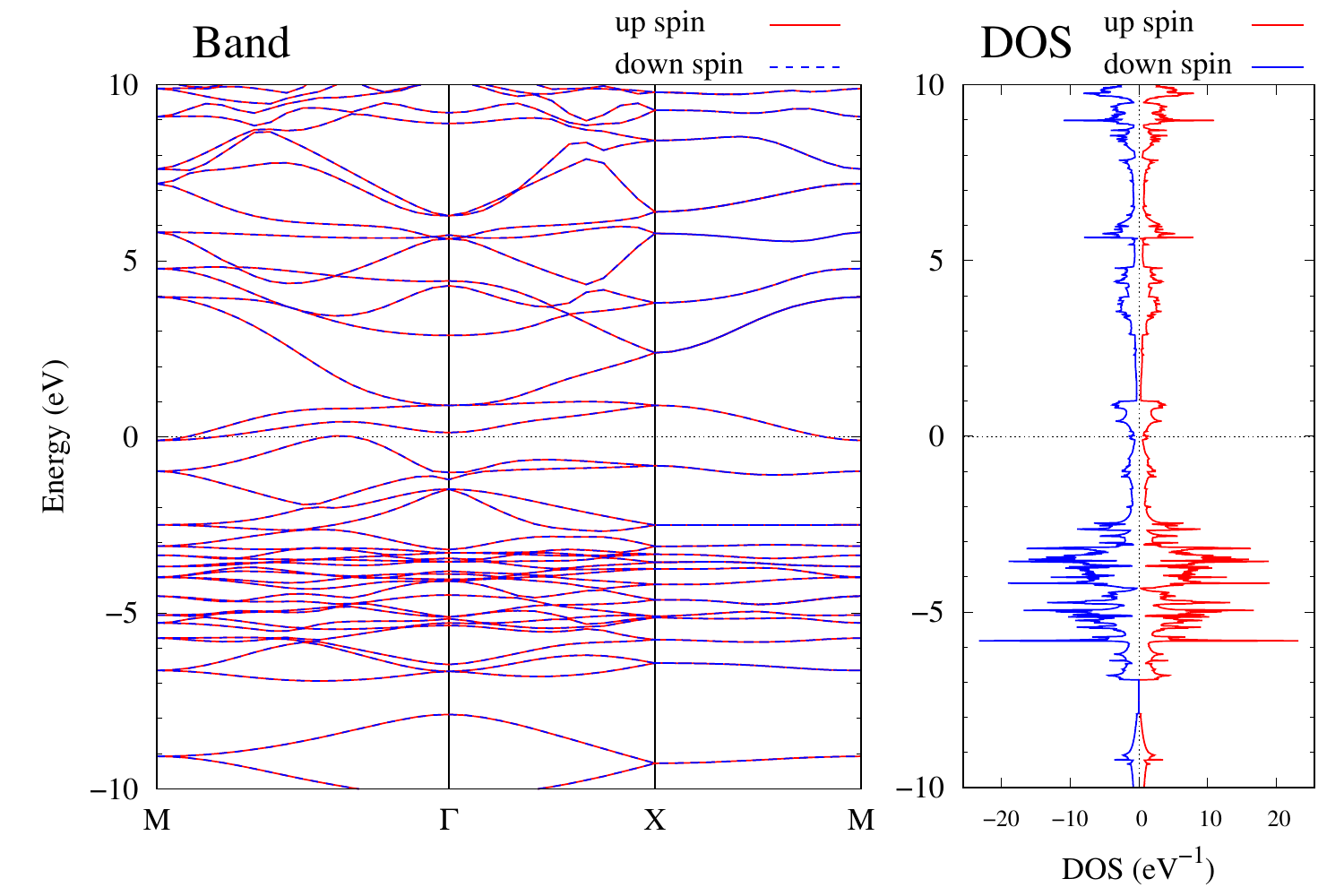}
          \subcaption{\textit{buckled-$\theta$}-Au$_2$Si}
  \end{minipage}
  \begin{minipage}[b]{0.48\linewidth}
    \includegraphics[width=\linewidth]{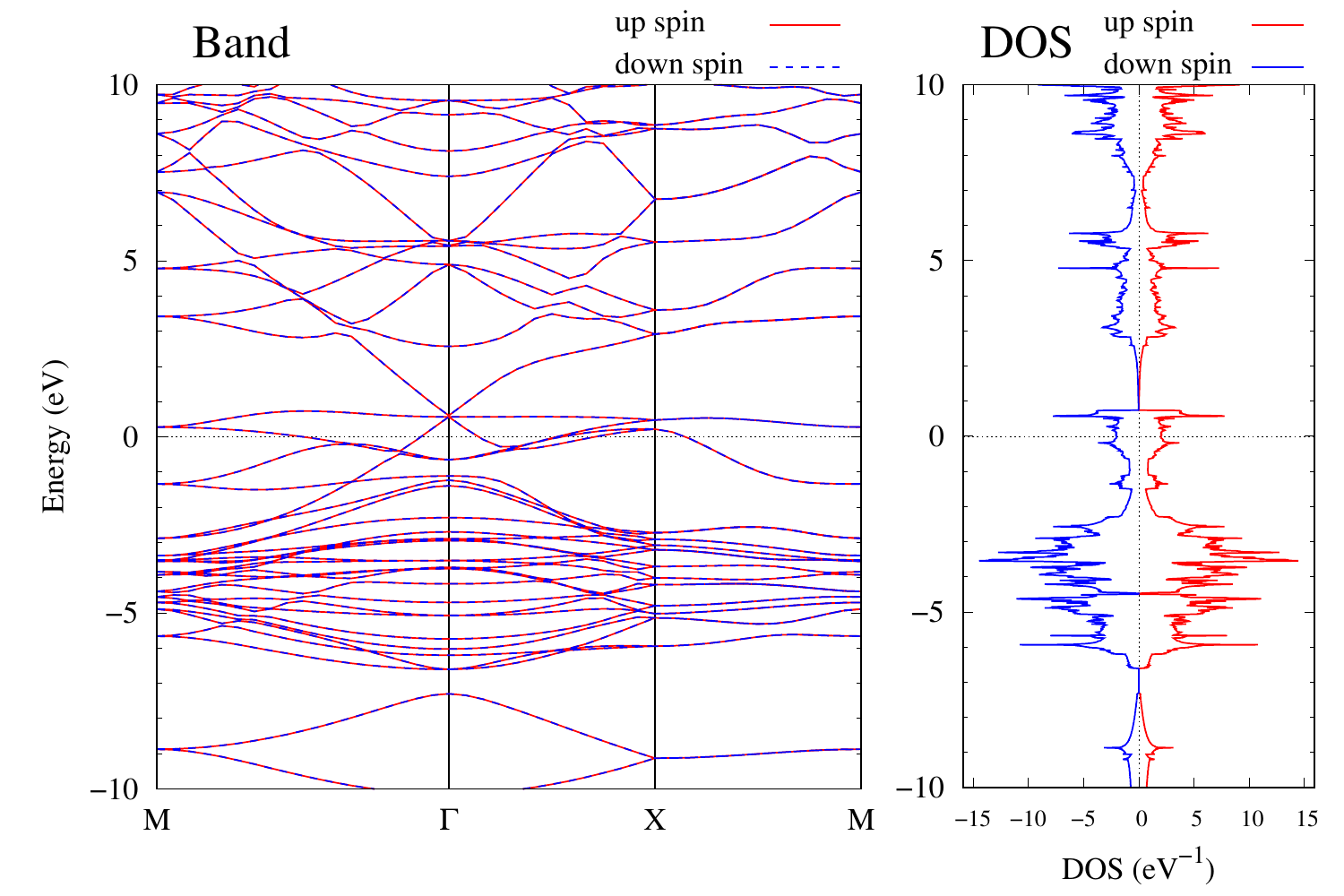}
          \subcaption{\textit{buckled-$\eta$}-Au$_2$Si}
  \end{minipage}

  \vspace{5pt}
  \begin{minipage}[b]{0.48\linewidth}
    \centering
    \includegraphics[width=\linewidth]{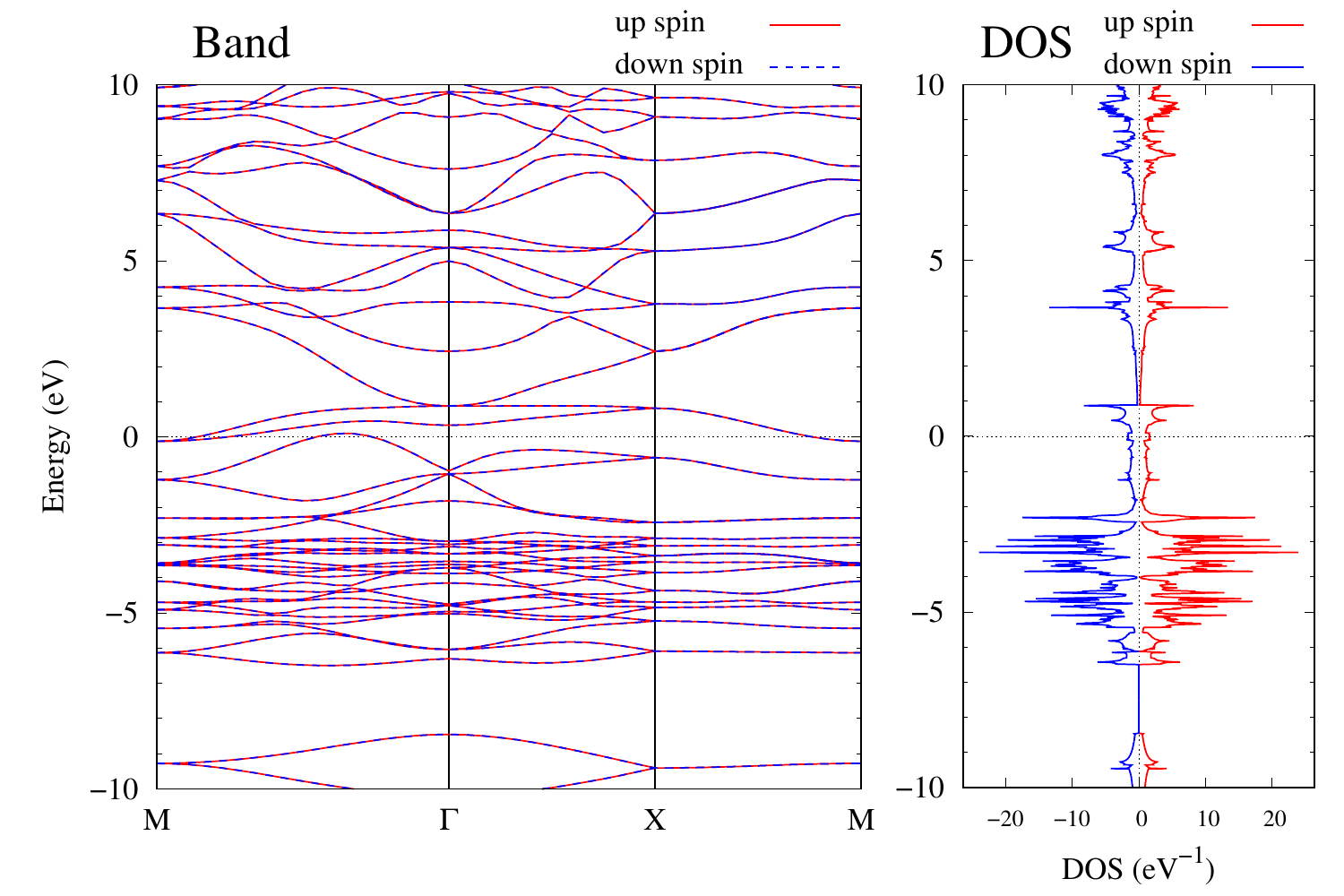}
          \subcaption{\textit{buckled-$\theta$}-Au$_2$Ge}
  \end{minipage}
  \begin{minipage}[b]{0.48\linewidth}
    \includegraphics[width=\linewidth]{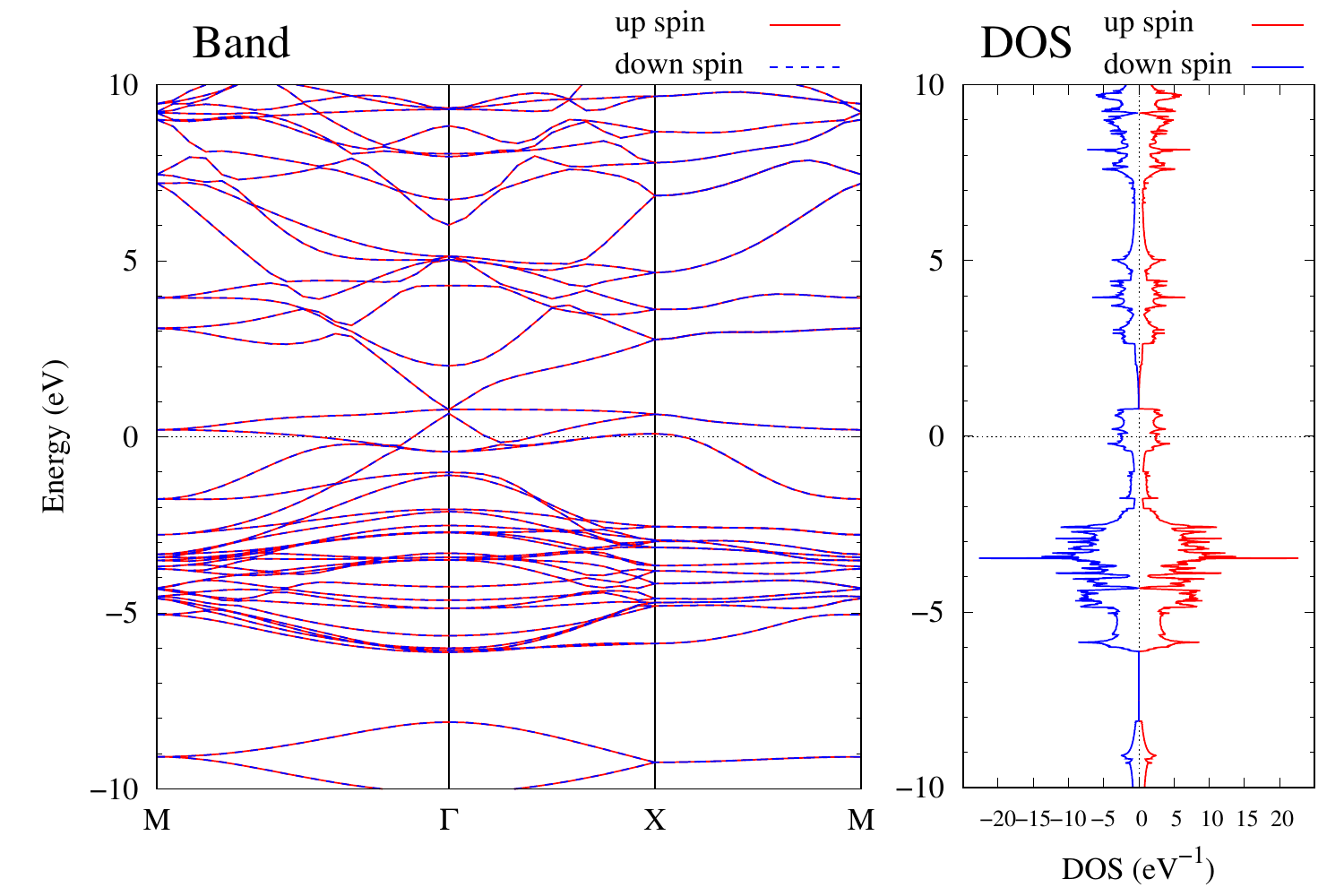}
          \subcaption{\textit{buckled-$\eta$}-Au$_2$Ge}
  \end{minipage}
        \caption{Electronic band and the DOS of Au$_2$X (X=Si, Ge) monolayers.}\label{SI_fig:Au2Si_Au2Ge_eband}
\end{figure*}

\begin{figure*}[htb]
  \begin{minipage}[b]{0.20\linewidth}
    \centering
    \includegraphics[width=\linewidth]{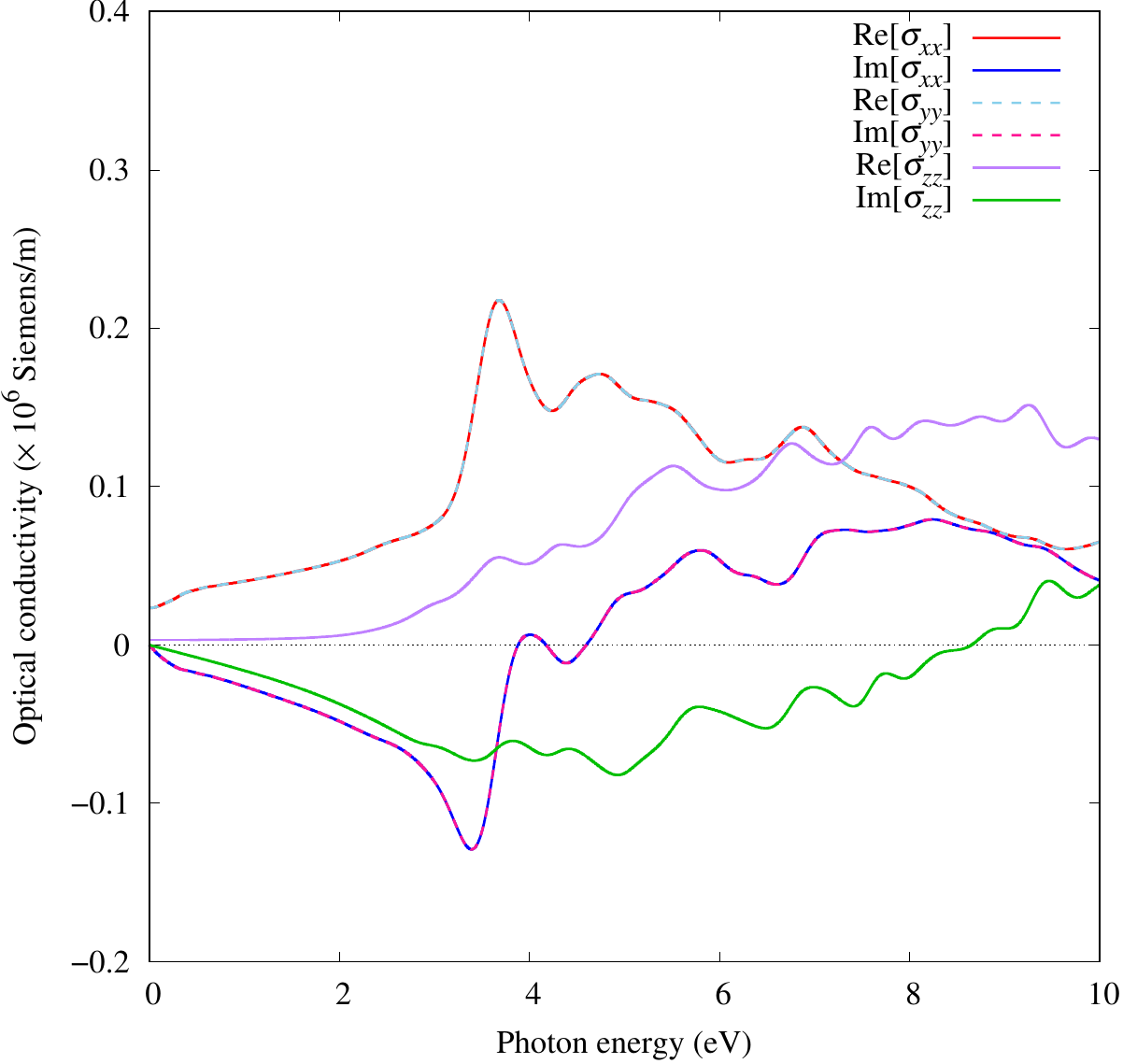}
          \subcaption{\textit{$\theta$}-Au$_2$S}
  \end{minipage}
  \begin{minipage}[b]{0.20\linewidth}
    \includegraphics[width=\linewidth]{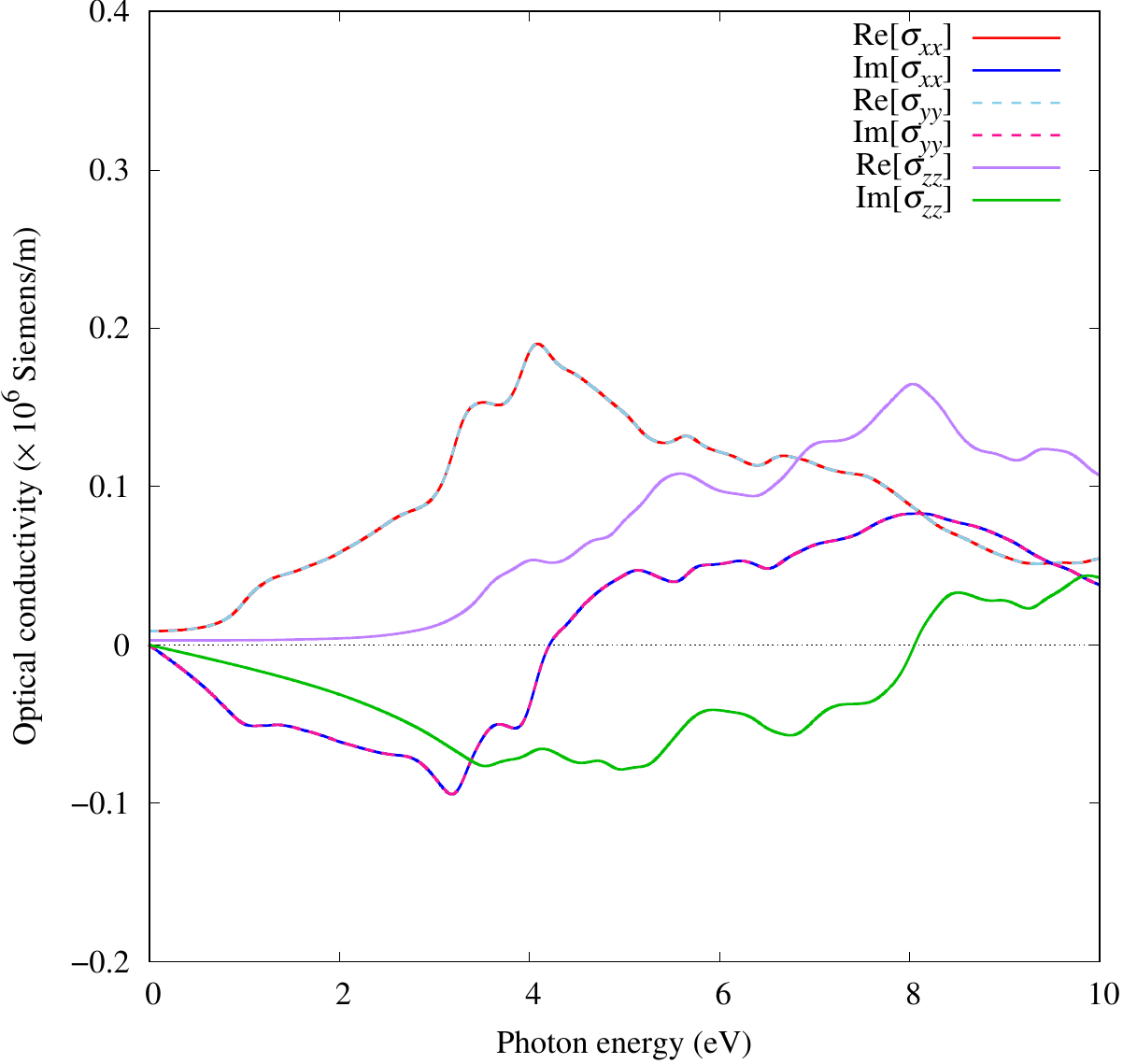}
          \subcaption{\textit{$\eta$}-Au$_2$S}
  \end{minipage}
  \begin{minipage}[b]{0.20\linewidth}
    \centering
    \includegraphics[width=\linewidth]{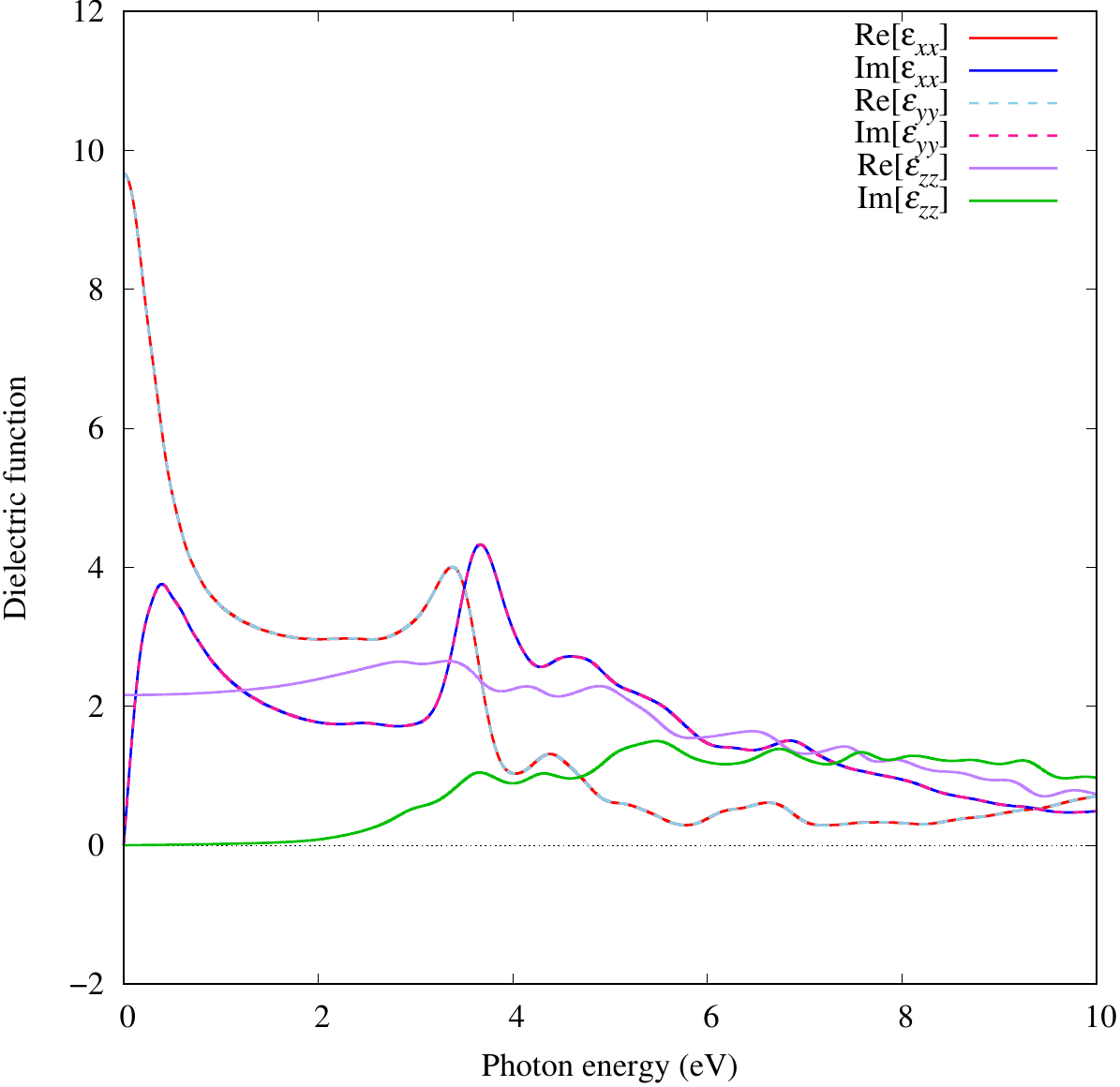}
          \subcaption{\textit{$\theta$}-Au$_2$S}
  \end{minipage}
  \begin{minipage}[b]{0.20\linewidth}
    \includegraphics[width=\linewidth]{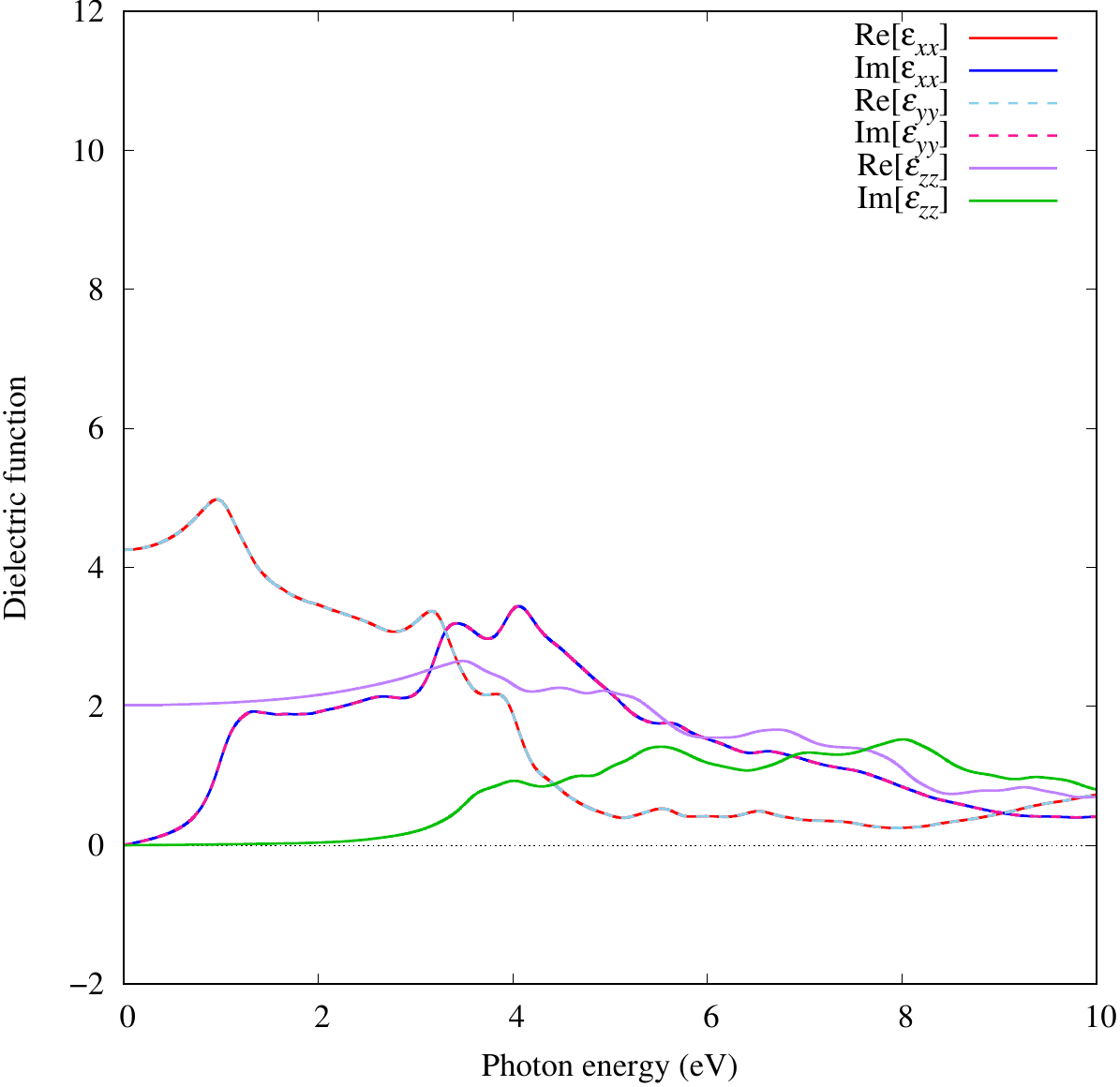}
          \subcaption{\textit{$\eta$}-Au$_2$S}
  \end{minipage}
  
  \vspace{5pt}
  \begin{minipage}[b]{0.20\linewidth}
    \centering
    \includegraphics[width=\linewidth]{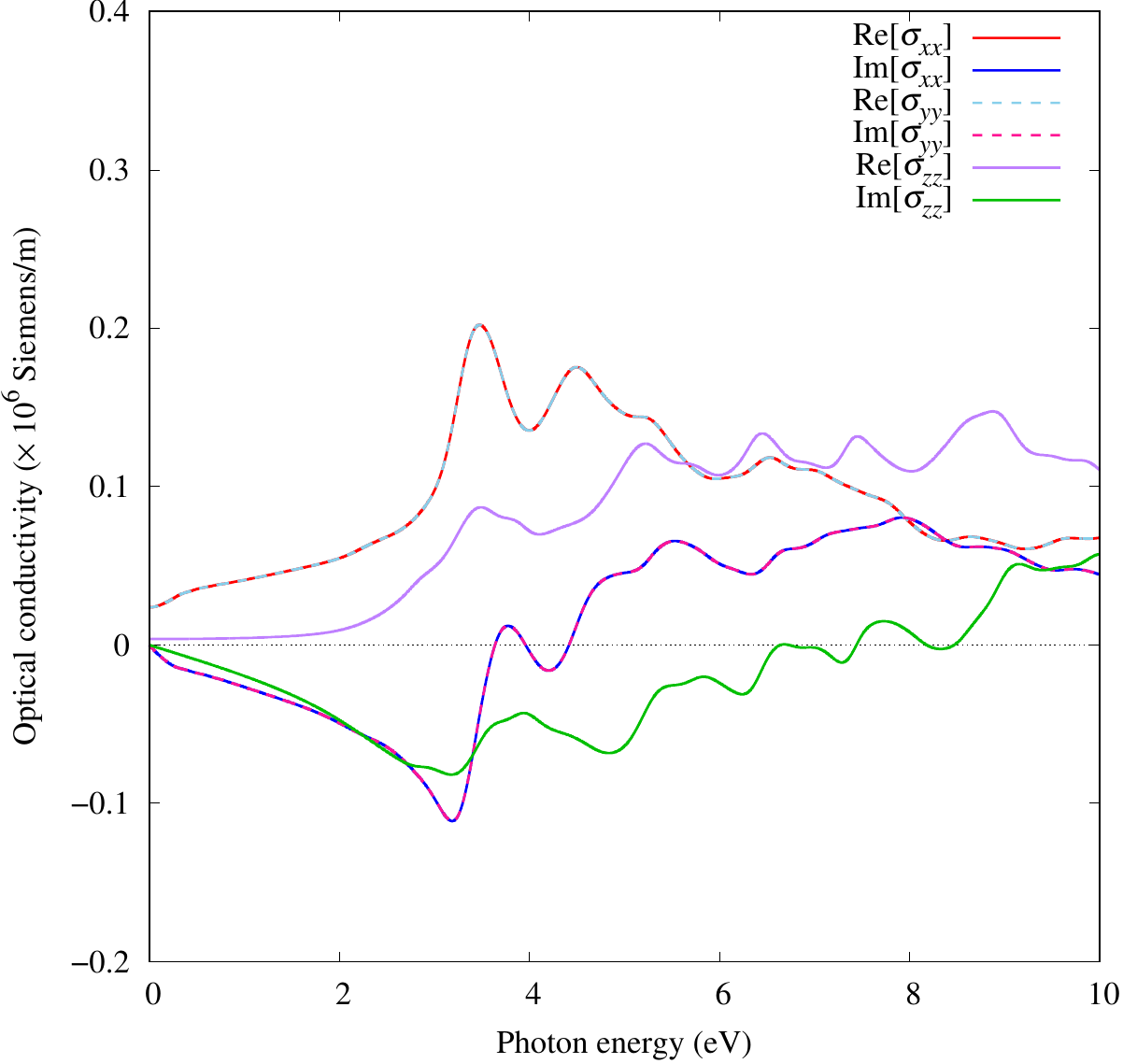}
          \subcaption{\textit{$\theta$}-Au$_2$Se}
  \end{minipage}
  \begin{minipage}[b]{0.20\linewidth}
    \includegraphics[width=\linewidth]{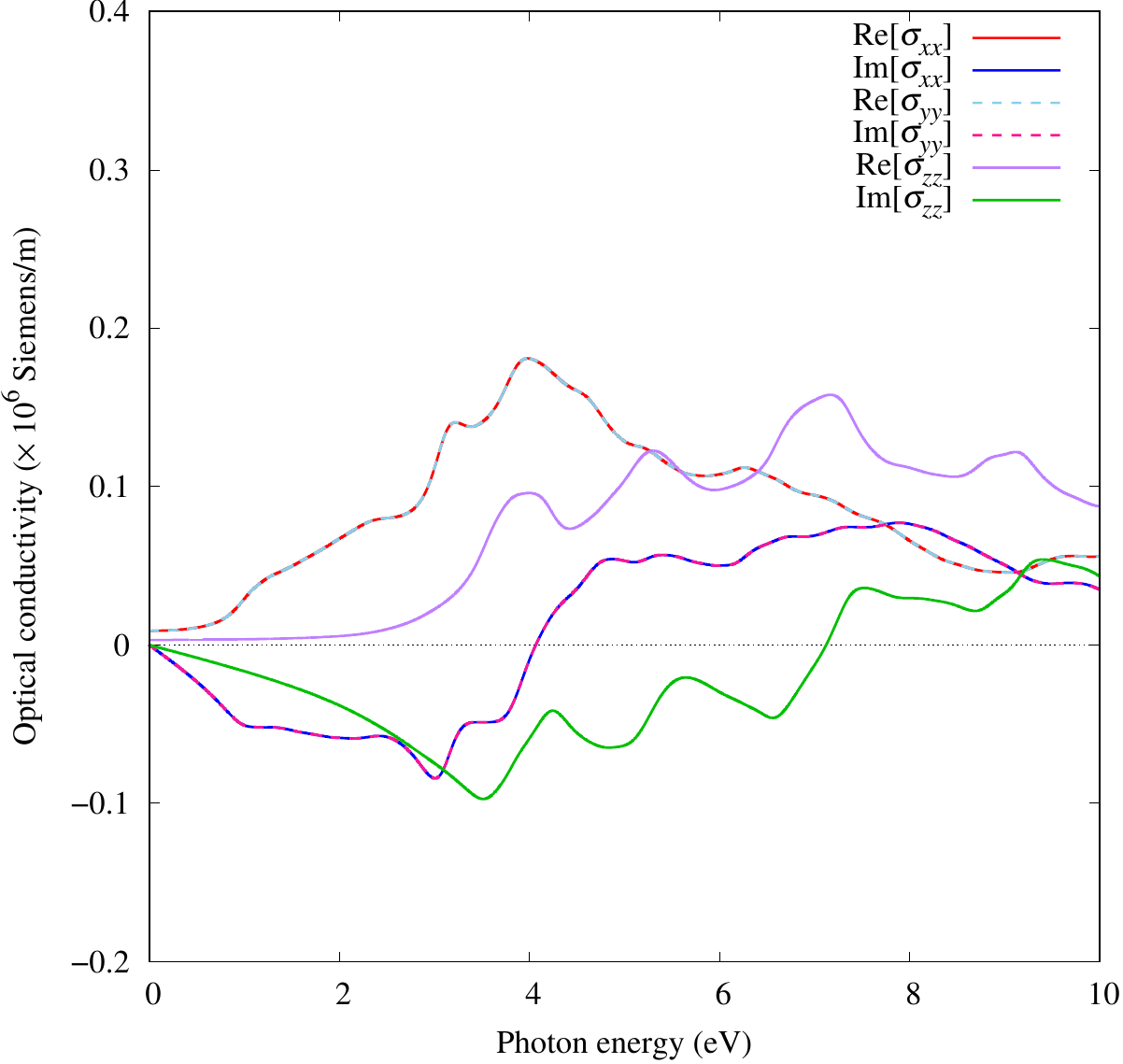}
          \subcaption{\textit{$\eta$}-Au$_2$Se}
  \end{minipage}
  \begin{minipage}[b]{0.20\linewidth}
    \centering
    \includegraphics[width=\linewidth]{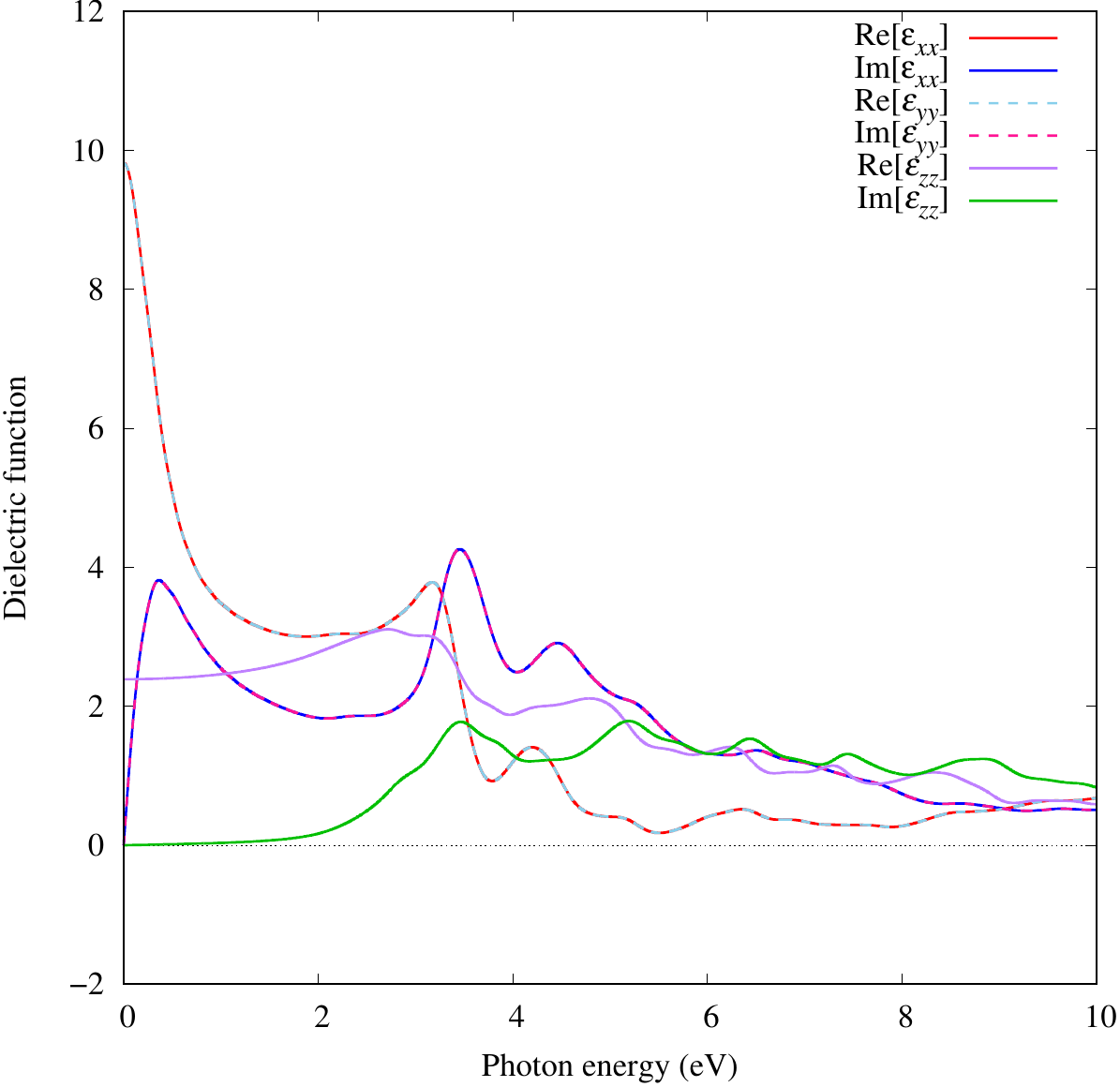}
          \subcaption{\textit{$\theta$}-Au$_2$Se}
  \end{minipage}
  \begin{minipage}[b]{0.20\linewidth}
    \includegraphics[width=\linewidth]{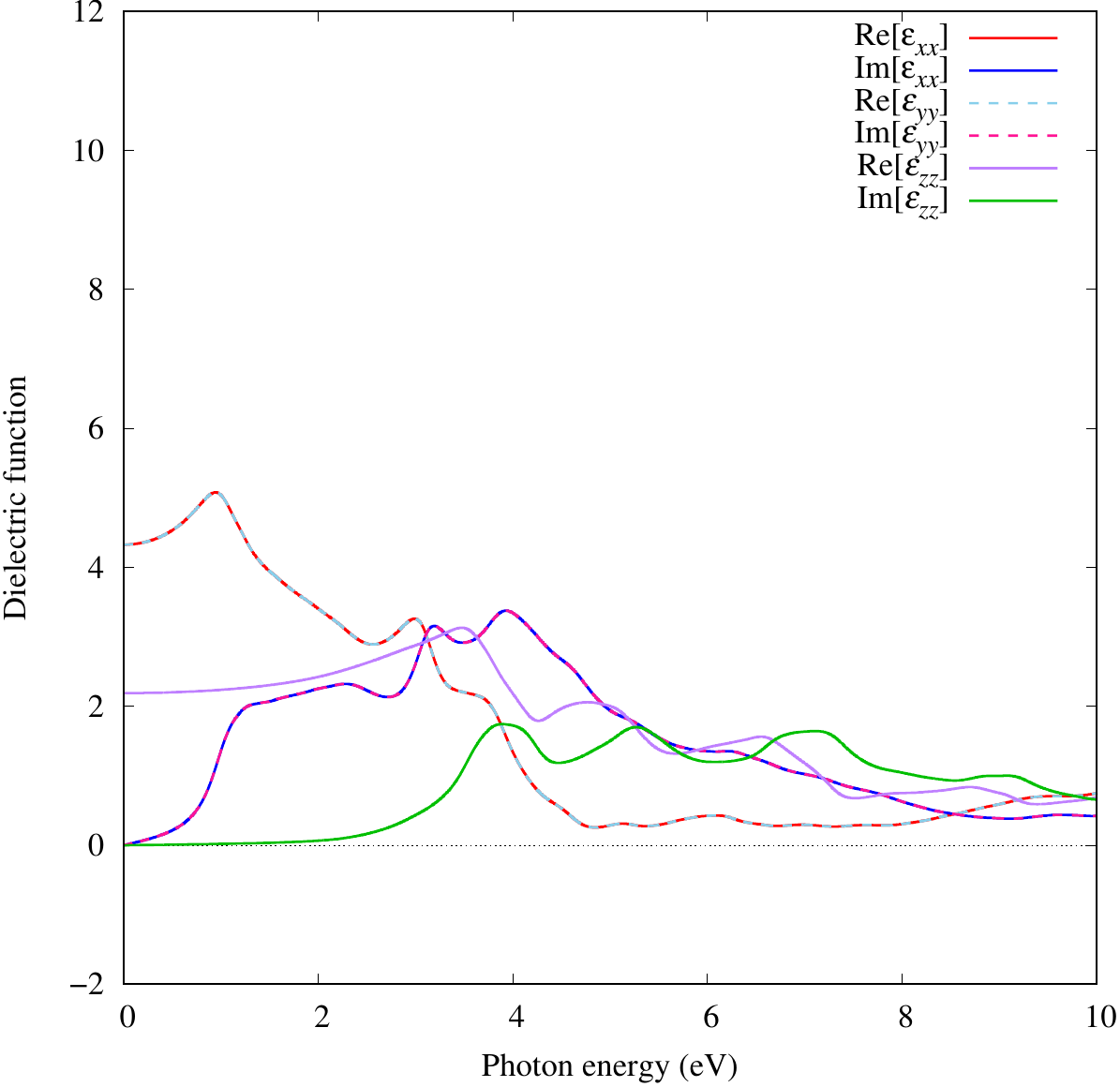}
          \subcaption{\textit{$\eta$}-Au$_2$Se}
  \end{minipage}
    
  \vspace{5pt}
  \begin{minipage}[b]{0.20\linewidth}
    \centering
    \includegraphics[width=\linewidth]{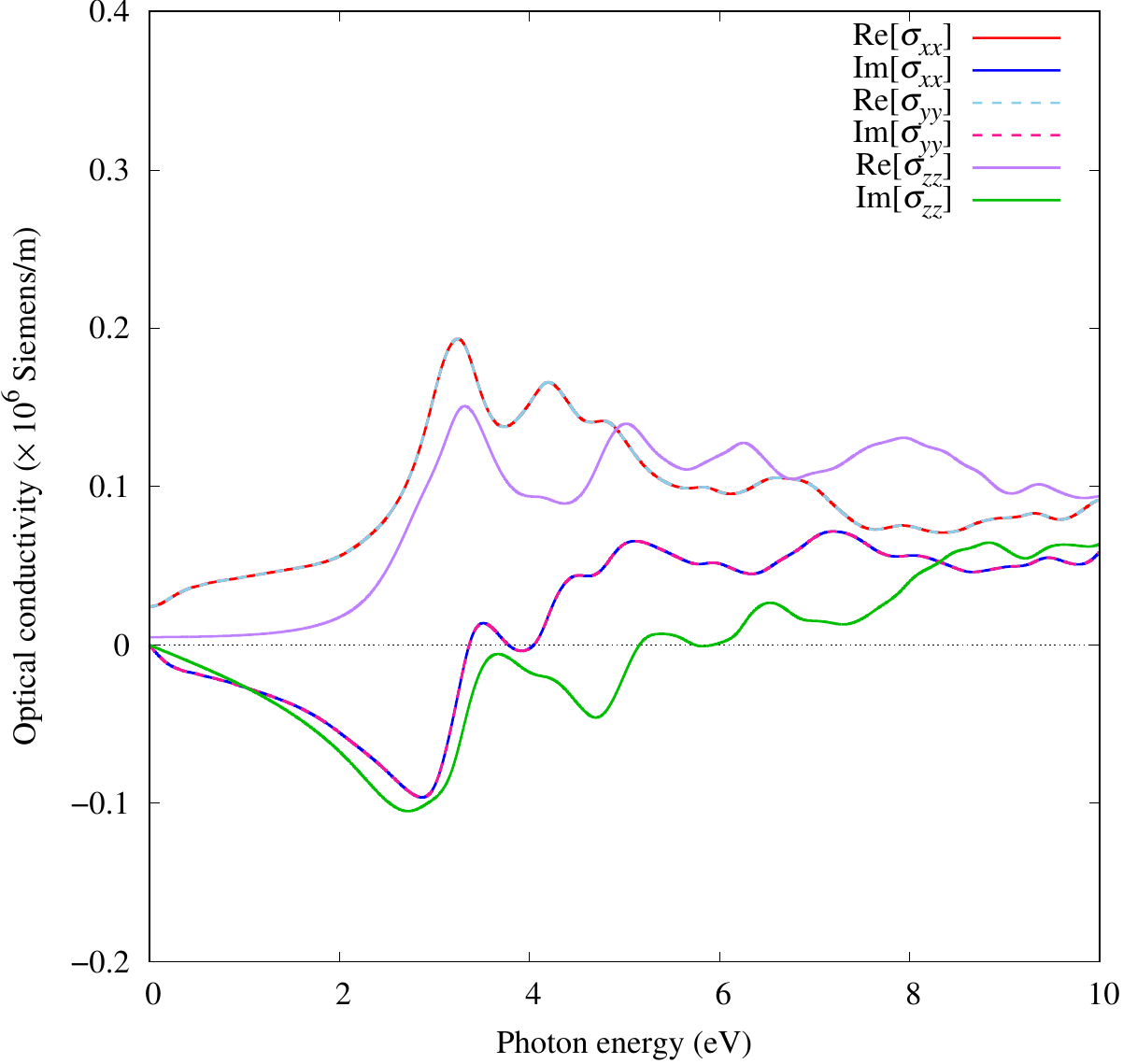}
          \subcaption{\textit{$\theta$}-Au$_2$Te}
  \end{minipage}
  \begin{minipage}[b]{0.20\linewidth}
    \includegraphics[width=\linewidth]{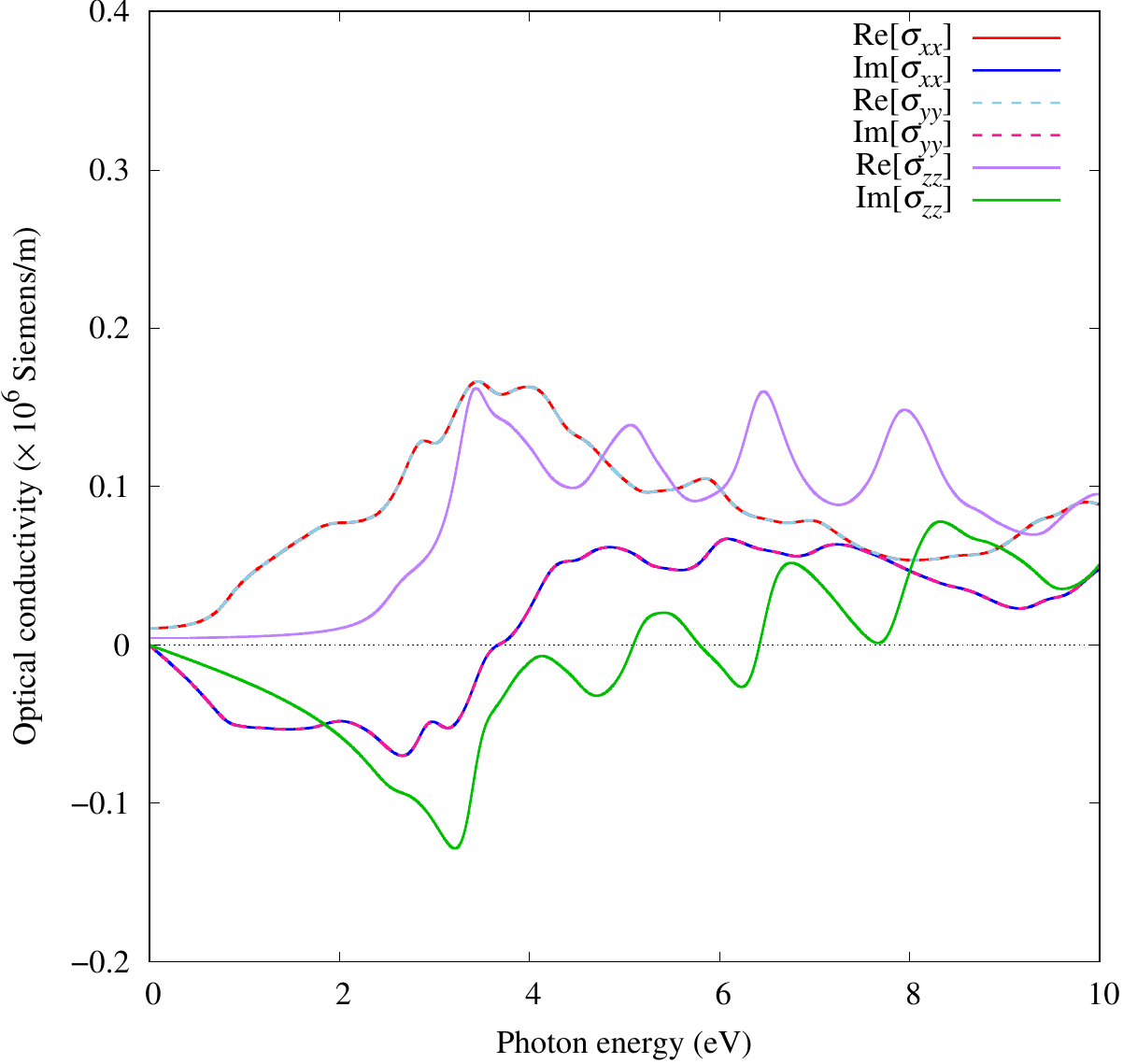}
          \subcaption{\textit{$\eta$}-Au$_2$Te}
  \end{minipage}
  \begin{minipage}[b]{0.20\linewidth}
    \centering
    \includegraphics[width=\linewidth]{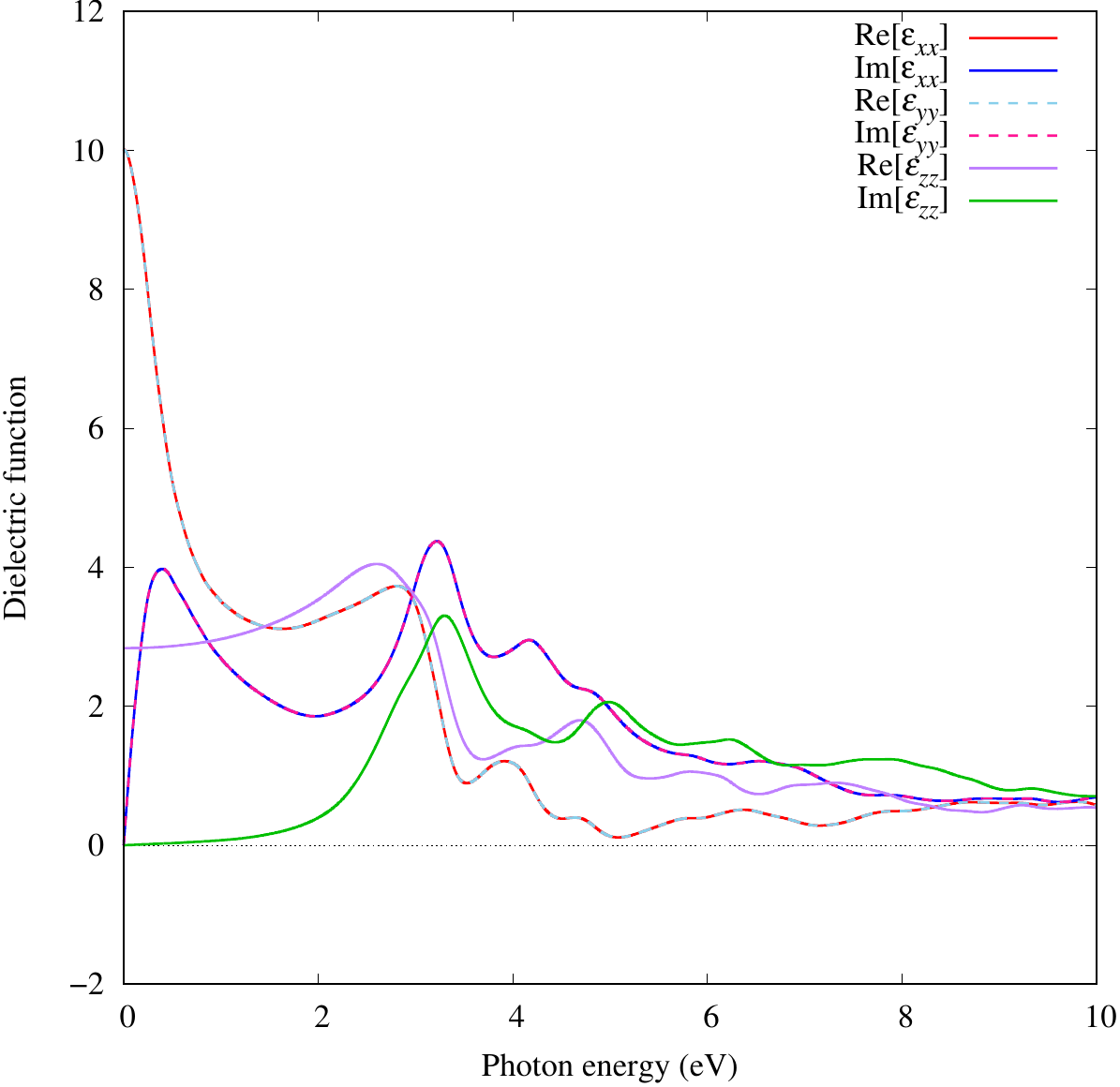}
          \subcaption{\textit{$\theta$}-Au$_2$Te}
  \end{minipage}
  \begin{minipage}[b]{0.20\linewidth}
    \includegraphics[width=\linewidth]{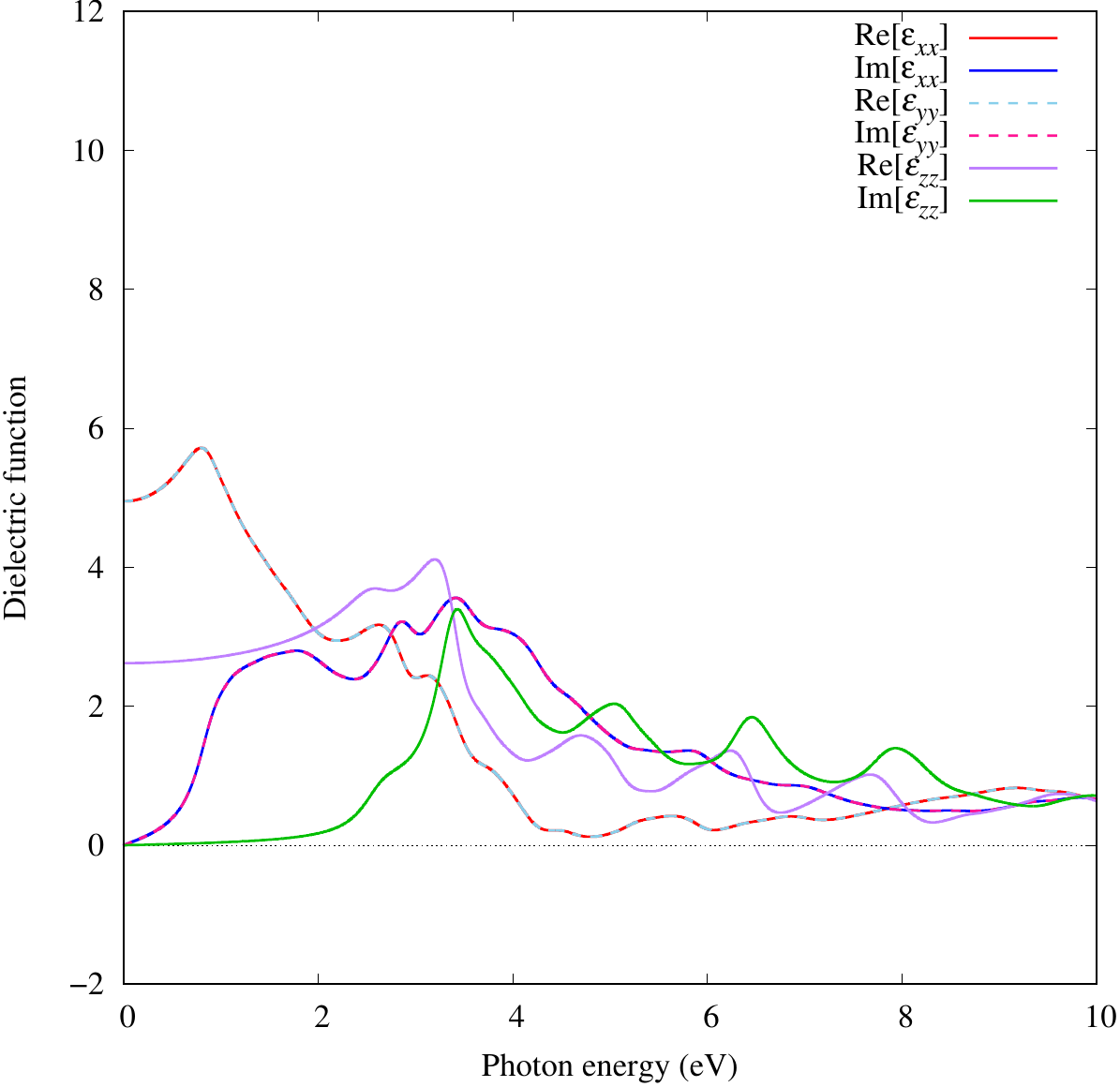}
          \subcaption{\textit{$\eta$}-Au$_2$Te}
  \end{minipage}
    \begin{minipage}[b]{0.20\linewidth}
    \centering
    \includegraphics[width=\linewidth]{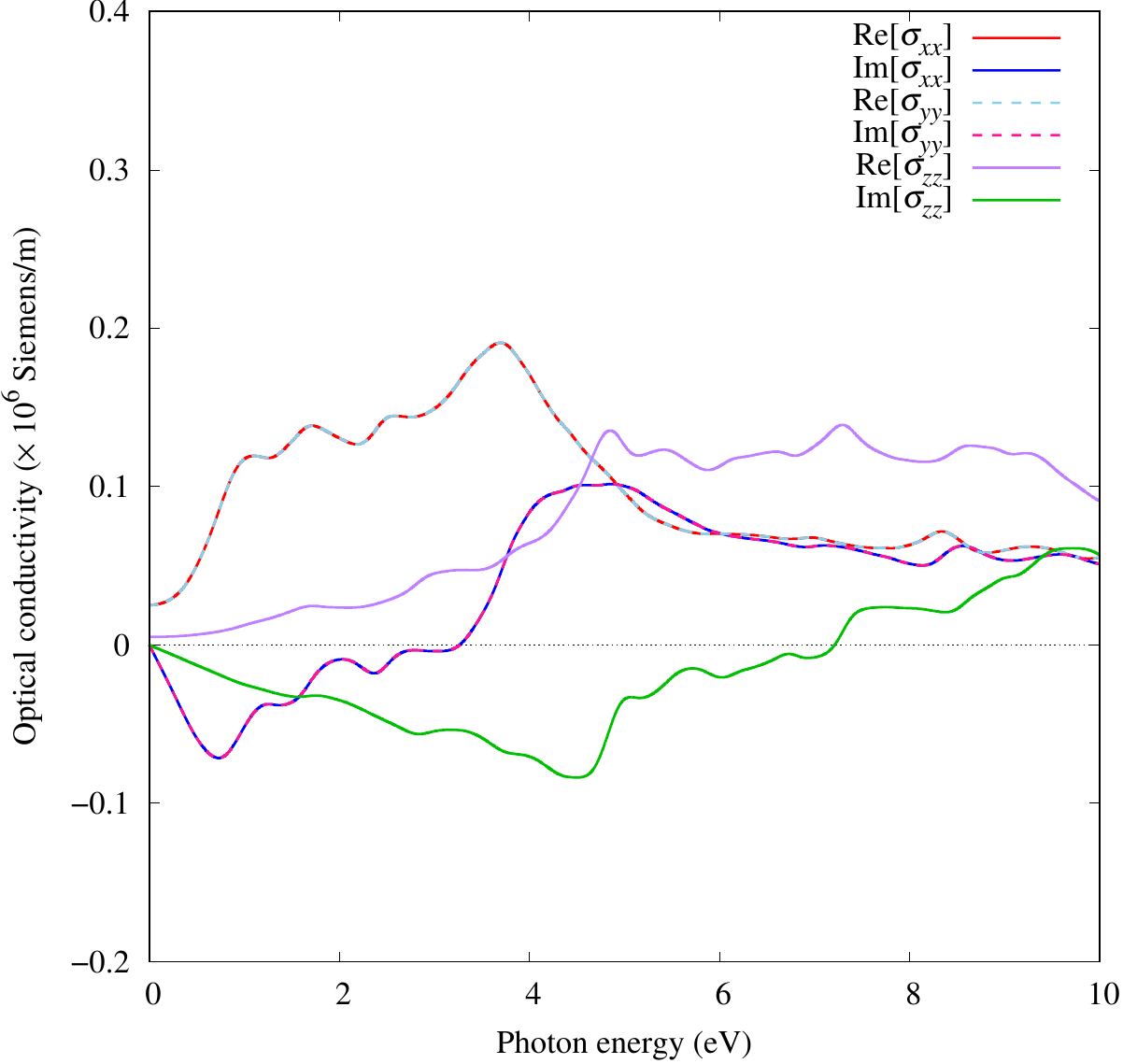}
          \subcaption{\textit{$\theta$}-Au$_2$Si}
  \end{minipage}
  \begin{minipage}[b]{0.20\linewidth}
    \includegraphics[width=\linewidth]{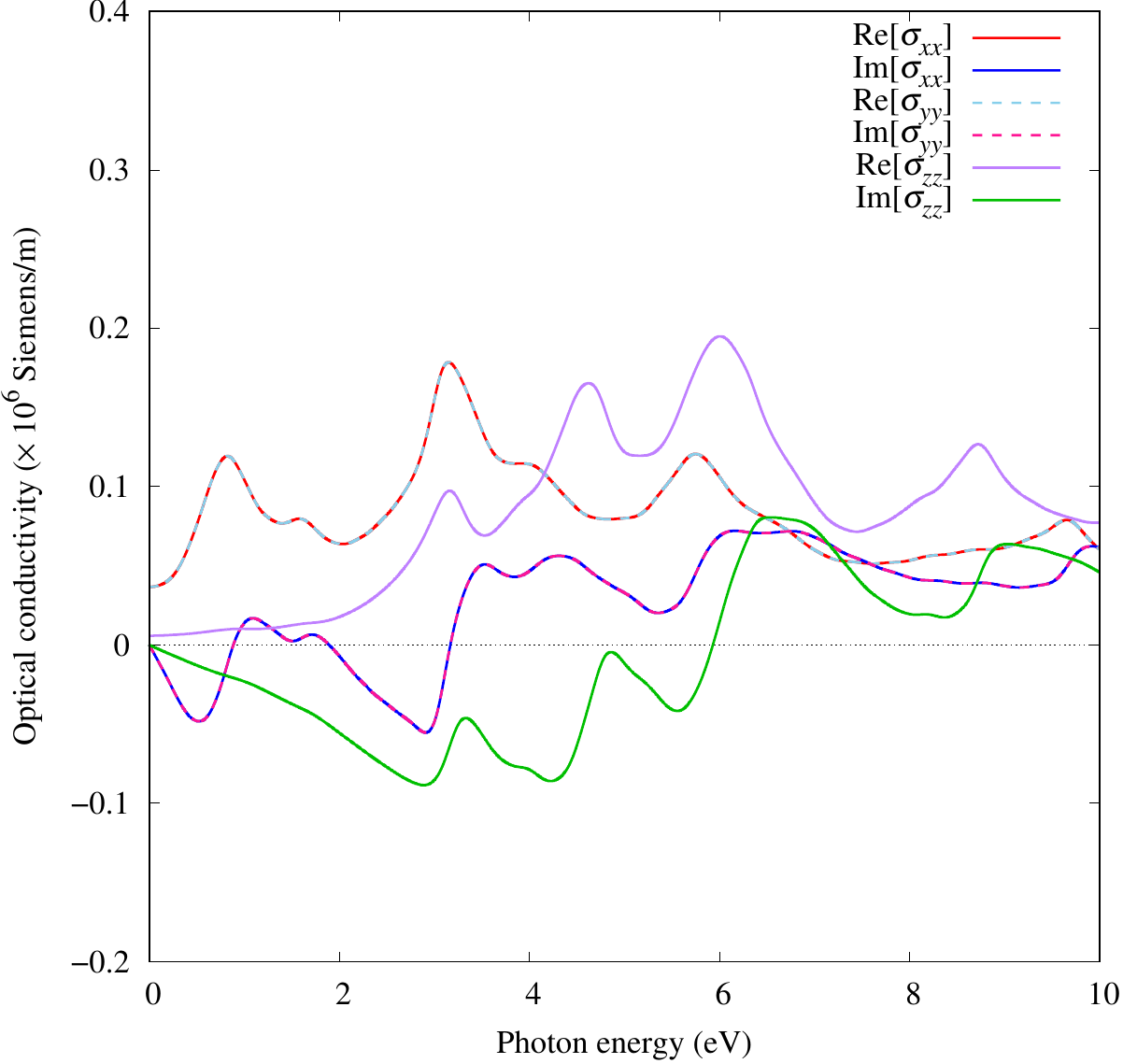}
          \subcaption{\textit{$\eta$}-Au$_2$Si}
  \end{minipage}
  \begin{minipage}[b]{0.20\linewidth}
    \centering
    \includegraphics[width=\linewidth]{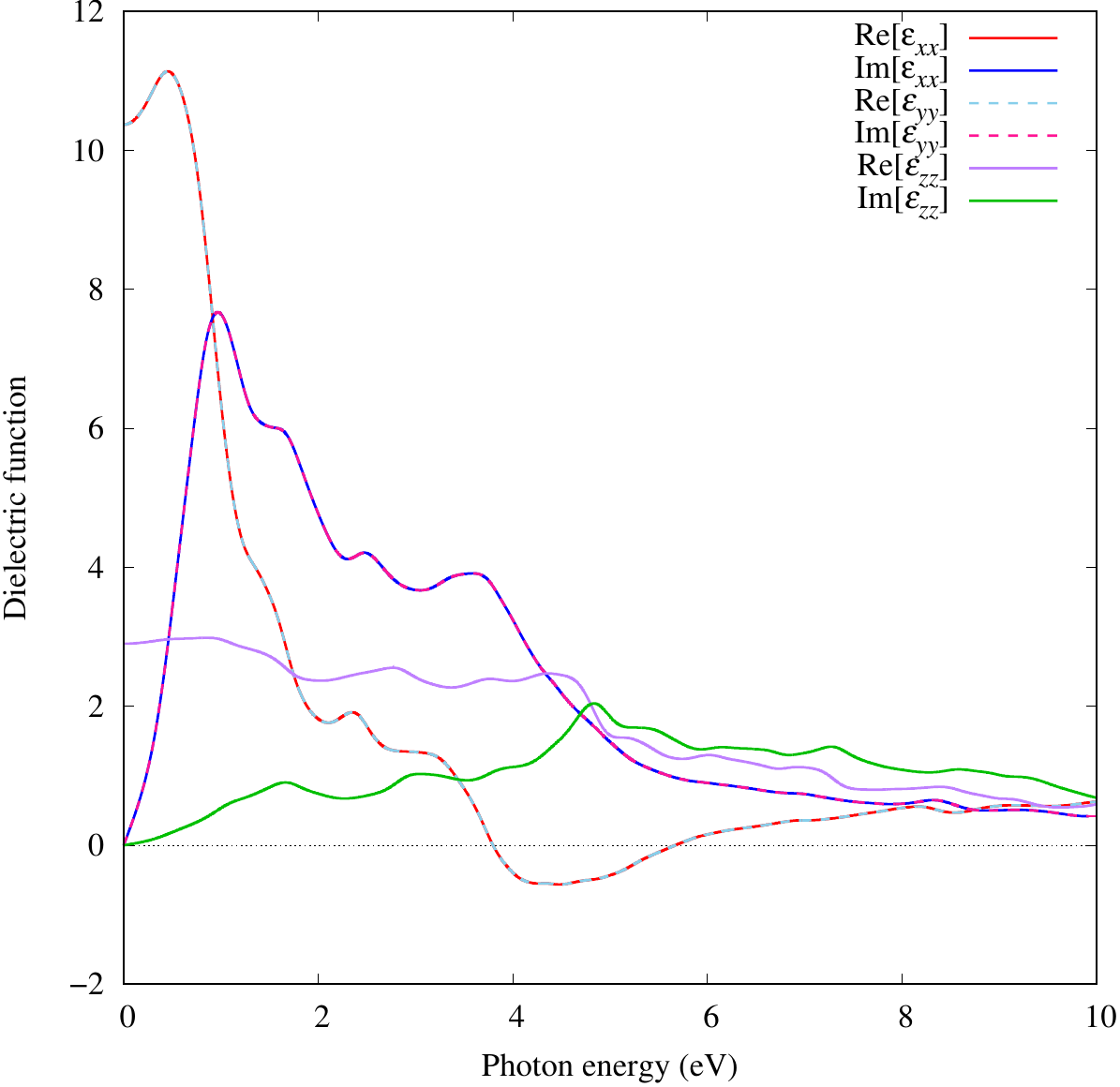}
          \subcaption{\textit{$\theta$}-Au$_2$Si}
  \end{minipage}
  \begin{minipage}[b]{0.20\linewidth}
    \includegraphics[width=\linewidth]{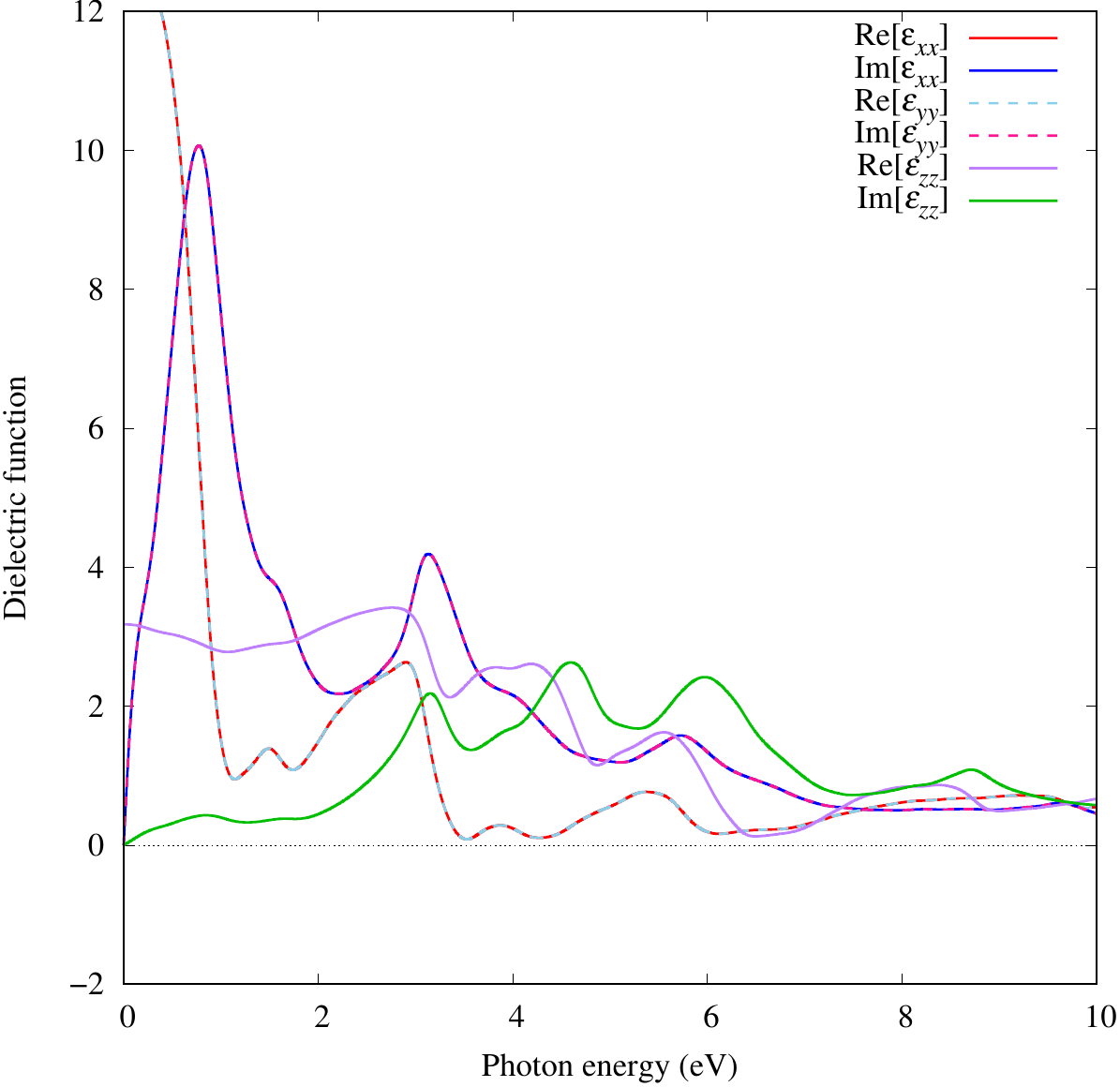}
          \subcaption{\textit{$\eta$}-Au$_2$Si}
  \end{minipage}
  
  \vspace{5pt}
  \begin{minipage}[b]{0.20\linewidth}
    \centering
    \includegraphics[width=\linewidth]{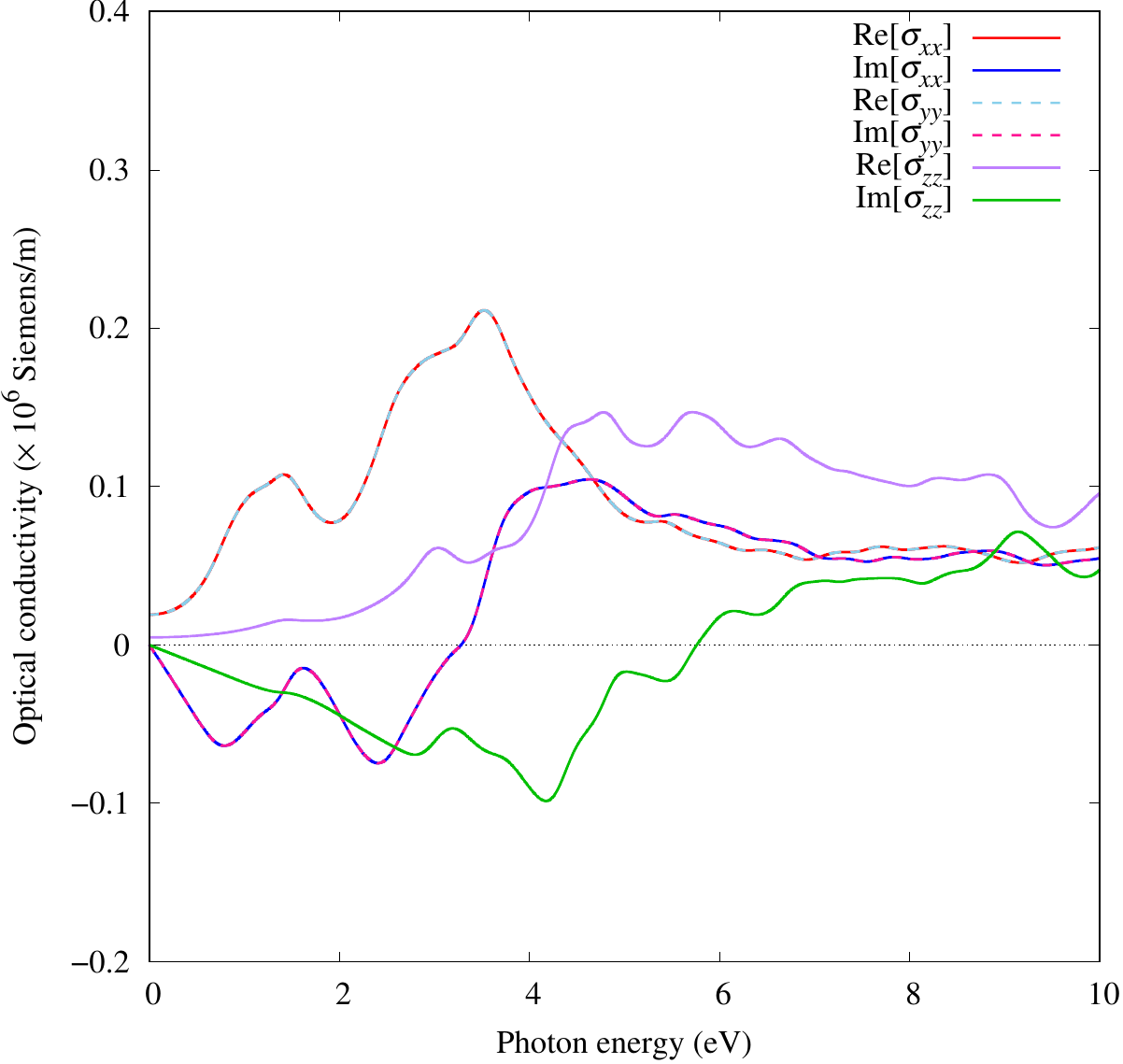}
          \subcaption{\textit{$\theta$}-Au$_2$Ge}
  \end{minipage}
  \begin{minipage}[b]{0.20\linewidth}
    \includegraphics[width=\linewidth]{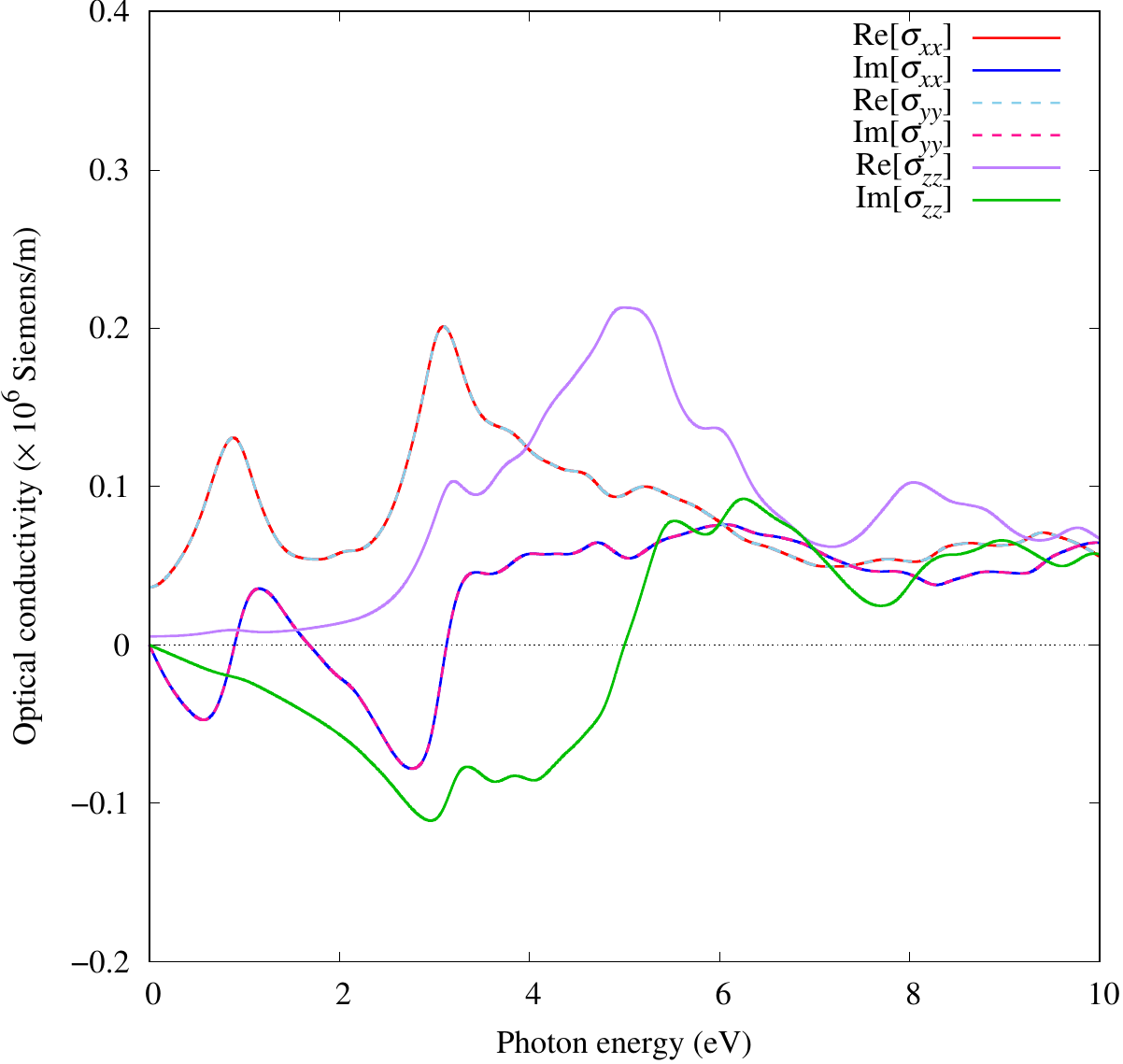}
          \subcaption{\textit{$\eta$}-Au$_2$Ge}
  \end{minipage}
  \begin{minipage}[b]{0.20\linewidth}
    \centering
    \includegraphics[width=\linewidth]{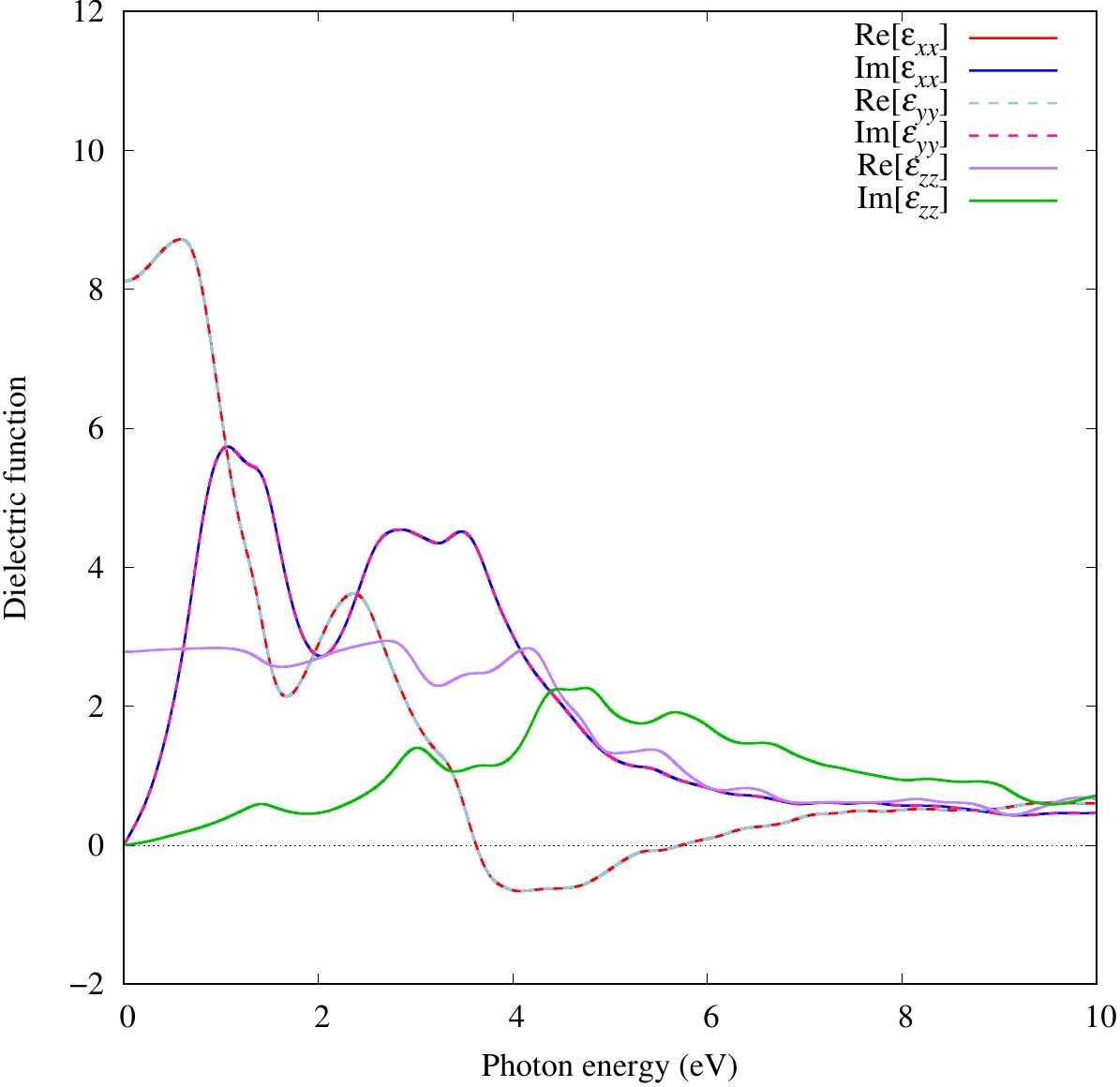}
          \subcaption{\textit{$\theta$}-Au$_2$Ge}
  \end{minipage}
  \begin{minipage}[b]{0.20\linewidth}
    \includegraphics[width=\linewidth]{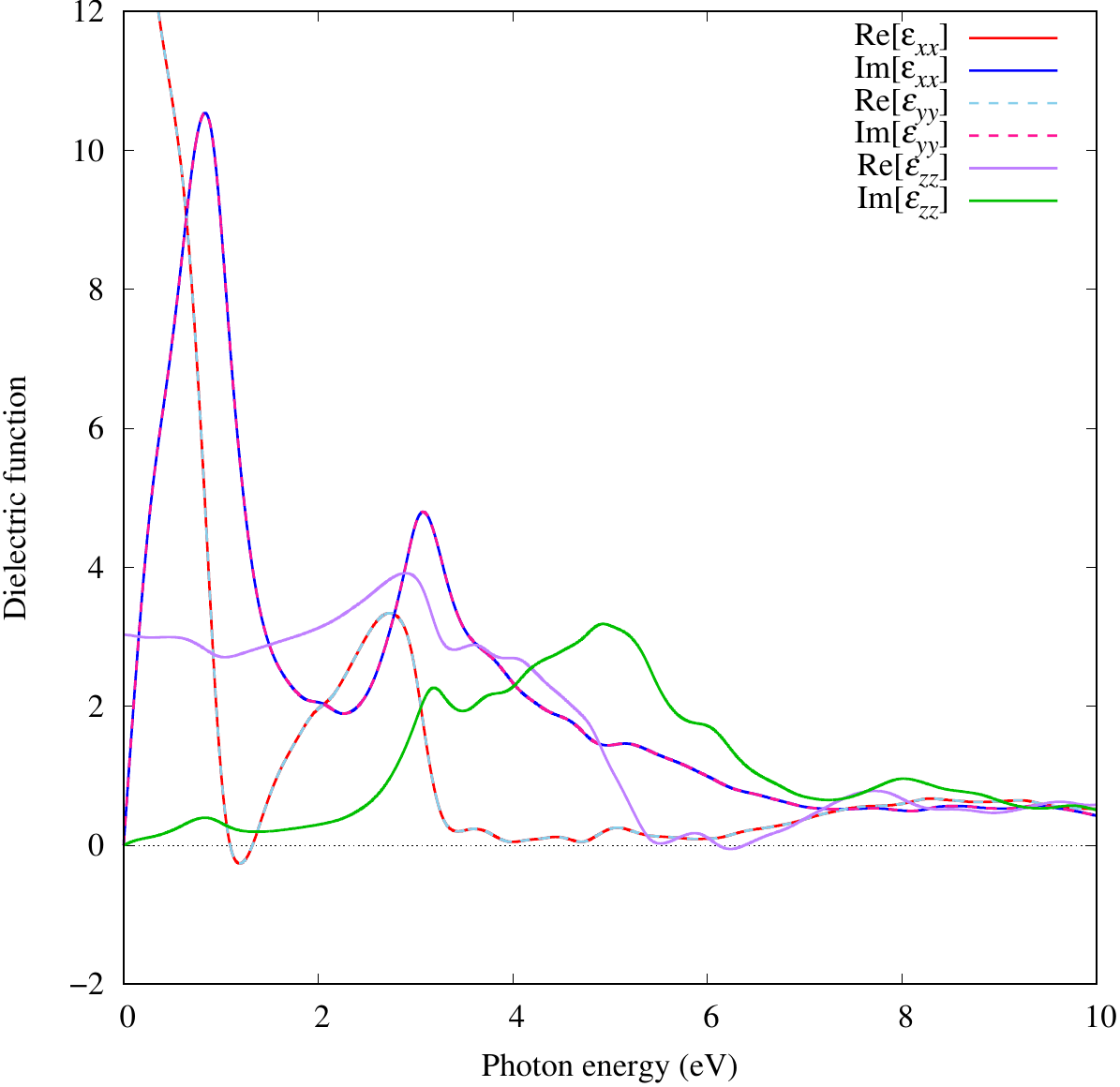}
          \subcaption{\textit{$\eta$}-Au$_2$Ge}
  \end{minipage}

        \caption{Optical conductivities (left two columns) and dielectric functions (right two columns) of Au$_2$X (X=S, Se, Te, Si, Ge) monolayers.}\label{SI_fig:Au2X_cddf}
\end{figure*}

\end{document}